\documentclass[aps,prx,superscriptaddress,twocolumn,showpacs, longbibliography]{revtex4}

\usepackage{amsfonts}
\usepackage[pdftex]{graphicx}
\usepackage{epstopdf}
\usepackage{amsmath}
\usepackage{amssymb}
\usepackage{verbatim}
\usepackage[ulem=normalem]{changes}
\usepackage{booktabs}
\usepackage{xcolor}

\usepackage{soul}

\newcommand{\bx     }{\mbox{\boldmath$x$}}

\newcommand{\bq     }{\mbox{\boldmath$q$}}

\usepackage{color}

\begin{document}

\title{Statistics of Infima and Stopping Times  of  Entropy production \\ 
and Applications to Active Molecular Processes}
\author{Izaak Neri}
\affiliation{Max Planck Institute for the Physics of Complex Systems, \\ N\"othnitzer Str. 38, 01187 Dresden, Germany.}
\affiliation{Max Planck Institute of Molecular Cell Biology and Genetics,\\ 
Pfotenhauerstra\ss e 108, 01307 Dresden, Germany.}

\author{\'{E}dgar Rold\'{a}n}
\affiliation{Max Planck Institute for the Physics of Complex Systems, \\ N\"othnitzer Str. 38, 01187 Dresden, Germany.} 
\affiliation{GISC-Grupo Interdisciplinar de Sistemas Complejos,  Madrid, Spain}

\author{Frank J\"{u}licher}
\affiliation{Max Planck Institute for the Physics of Complex Systems, \\ N\"othnitzer Str. 38, 01187 Dresden, Germany.}

\begin{abstract}
We study the statistics  of infima, stopping times and  passage probabilities 
of entropy production in nonequilibrium steady states, and show that they are universal.    We consider two 
examples of stopping times:  first-passage times of entropy production and waiting times of stochastic processes,
which are the times when a system reaches for the first time a given state.   
Our main results are: (i) the distribution of the  global infimum of entropy production is 
exponential with mean equal to minus  Boltzmann's constant;  (ii)    
 we find the exact expressions for the passage probabilities of entropy production to reach a given 
value;  (iii) we derive a fluctuation theorem for  stopping-time distributions of entropy production.  These results have interesting implications for stochastic processes that can be discussed in simple colloidal systems and in active molecular processes.  In particular, we show that the timing and statistics 
of discrete chemical transitions of molecular processes, such as, the steps of molecular motors, are governed by the statistics of entropy production.  We also show that the extreme-value statistics of active molecular processes are governed by entropy production,  for example, 
the infimum of entropy production of a motor can be related to the maximal excursion of a motor
against  the direction of an external force.    Using this relation, we   make predictions for the distribution of the maximum backtrack depth   of   RNA polymerases, which follows from our universal results for entropy-production infima.
\end{abstract}

\pacs{05.40.-a, 05.70.-a, 02.50.-r, 05.70.Ln} 

\maketitle

\section{Introduction and statement of the main results}
The total entropy $S_{\rm tot}(t)$ produced by a mesoscopic process in a finite time interval $[0,t]$ is stochastic, and  can for a single realization be negative due to  fluctuations. The second law of thermodynamics  implies that its average, taken over many realizations of the process, increases in time, $\langle S_{\rm tot}(t)\rangle\geq 0$.    
 In the 19th century   Maxwell already formulated the
idea of  a stochastic entropy \cite{maxwell1878tait}, and in the last decades    definitions of entropy production of nonequilibrium processes were established  using the theory of stochastic processes~\cite{bochkov1977general, evans1993probability, evans1994equilibrium, gallavotti1995dynamical, sekimoto1998langevin, crooks1998nonequilibrium,  crooks1999entropy, lebowitz1999gallavotti, maes1999fluctuation, maes2003time, gaspard2004time, jiang2004mathematical, seifert2005entropy, PhysRevLett.98.080602, sekimoto2010stochastic,  PhysRevE.85.051113, PhysRevE.85.031129, murashita2014nonequilibrium}.     

Little is known beyond the second law about the statistics of entropy-production fluctuations.  The best insights, so far, in fluctuations of  entropy production are provided by fluctuation theorems. 
They express a fundamental asymmetry of the fluctuations of 
entropy production: it is exponentially more likely to produce  a positive amount of entropy than to reduce entropy by the same but negative amount.  
An example is the detailed fluctuation theorem, which can be written as  $p_{\rm S}( S_{\rm tot};t)/p_{\rm S}(- S_{\rm tot};t) = e^{ S_{\rm tot}/k_{\rm B}}$, 
where $k_{\rm B}$ is Boltzmann's constant.
Here $p_{\rm S}(S_{\rm tot};t)$ is the probability density describing the
distribution of the entropy production $S_{\rm tot}$ at a given time $t$. 
 The detailed fluctuation theorem  is universal and holds for a broad class of physical processes in steady state  \cite{evans1993probability, evans1994equilibrium, gallavotti1995dynamical, kurchan1998fluctuation, crooks1999entropy, lebowitz1999gallavotti, maes1999fluctuation, seifert2005entropy, speck2007jarzynski,  PhysRevE.80.011117, aron2010symmetries}.  
Moreover, the detailed fluctuation theorem  has been tested in several experiments \cite{ciliberto1998experimental, wang2002experimental, gomez2010steady, garnier2005nonequilibrium, feitosa2004fluidized, tietz2006measurement, hayashi2010fluctuation, saira2012test}, for reviews see \cite{bustamante2005nonequilibrium, jarzynski2011equalities, seifert2012stochastic}.

In addition to fluctuation theorems, an important question is to understand the extreme-value statistics of entropy production.   In particular, because entropy must on average increase, it is interesting to understand the statistics of  records of negative entropy production during a given time interval $[0,t]$.  
 To address this question, here we  introduce the {\it infimum} of entropy production, for a single realization, $S_{\rm inf}(t) \equiv {\rm inf}_{0\leq \tau\leq t}\:S_{\rm tot}(\tau)$,  which is the negative record of entropy production over a time interval~$[0,t]$.   
 
\begin{figure}
\includegraphics[width=7cm]{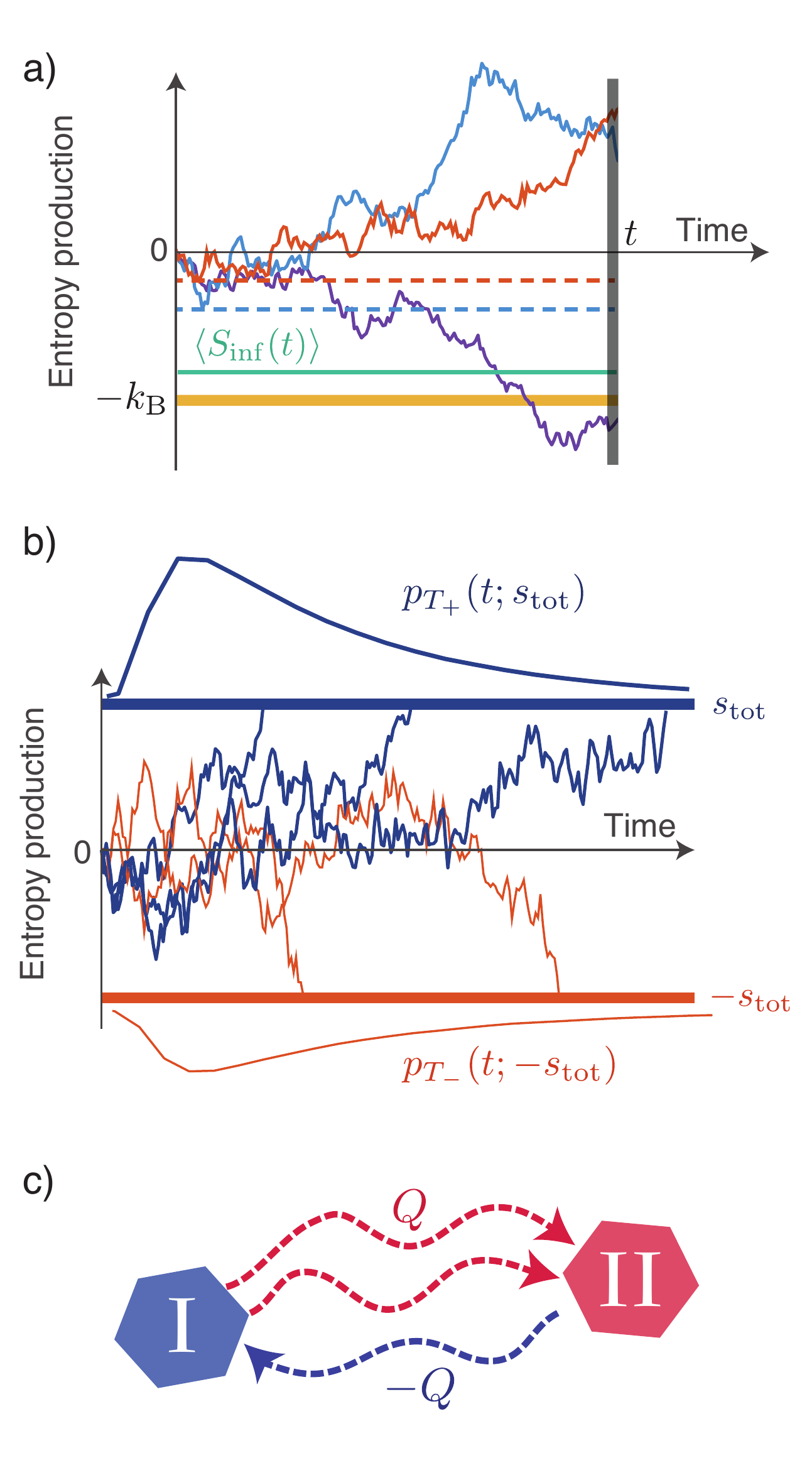} 
\caption{Illustration of  three  key results  of the paper. \\a)~Schematic representation of the \textit{Infimum law} for entropy production. Several stochastic trajectories of entropy production are shown (solid lines), and their infima  are indicated (dashed lines). The infima law implies that the  average  infimum of the entropy production (green solid line) is larger or equal than $-k_{\rm B}$  (orange line). b)~\textit{ First-passage-time fluctuation theorem} for entropy production  with two absorbing boundaries. Examples of trajectories of  stochastic entropy production as a function of time, which first reach a positive threshold $s_{\rm tot}$ (blue thick curves), and which first reach a negative threshold $-s_{\rm tot}$ (red thin curves). The  probability distribution $p_{T_+}(t;s_{\rm tot})$ to first reach the positive threshold at time $t$, and the probability distribution 
$p_{T_-}(t;-s_{\rm tot})$ to first reach the negative threshold at time $t$, are related by Eq.~\eqref{eq:FPTime}.  c)~\textit{Waiting-time fluctuations}:  the statistics of the waiting times between two states {\rm I} and {\rm II} are the same for forward and backward trajectories that absorb or dissipate a certain amount of heat $Q$ in isothermal conditions.}\label{fig1}
\end{figure}

  In this paper we derive universal equalities and inequalities on the statistics of entropy production infima. 
   We show that the mean of the infimum of  the stochastic entropy production, 
in a given  time interval $[0,t]$, is bounded from below by minus the Boltzmann constant:
\begin{eqnarray}
\big\langle S_{\rm inf}(t)\big\rangle \geq -k_{\rm B}\quad. \label{eq:infx}
\end{eqnarray}
This {\it  infimum law} for entropy production is illustrated in Fig.~\ref{fig1}a) and expresses a fundamental bound on how much entropy can be reduced in a finite time.  
  The infimum law follows from  a universal  bound for the cumulative distribution of entropy-production  infima: 
\begin{eqnarray}
\mathsf{Pr}\left(S_{\rm inf}(t)\geq -s\right)  
&\geq & 1-e^{-s/k_{\rm B}}\quad. \label{eq:infxx}
\end{eqnarray} 
Here $\mathsf{Pr}\left(\cdot\right)$   denotes the probability of an event, and  the left-hand side is the cumulative distribution of entropy production with   $s\geq 0$.   Remarkably, as we show in this paper, the infimum law, given by Eq.~(\ref{eq:infx}), is universal and holds in general for classical and stationary stochastic processes.  

The global infimum of entropy production, $S^{\infty}_{\rm inf} \equiv \lim_{t\rightarrow \infty}S_{\rm inf}(t)$, is the lowest value that entropy production will ever reach in one realization of the process;    note that the global infimum is always smaller or equal to the local infimum, $S^{\infty}_{\rm inf}\leq S_{\rm inf}(t)$.
 We show that the distribution of the global infimum of entropy production is exponential 
 \begin{eqnarray}
p_{S^\infty_{\rm inf}}(-s) =\frac{e^{\, -s/k_{\rm B}}}{k_{\rm B}}\quad, \label{eq:infxxxx}
\end{eqnarray}
where $s\geq 0$,
and the mean value of the global infimum equals to minus the Boltzmann constant:
\begin{eqnarray}
\big\langle  S^\infty_{\rm inf}\big\rangle = -k_{\rm B}\quad. \label{eq:infxxx}
\end{eqnarray} 
The shape of the distribution of the global infimum implies that   the infimum lies  with 50 percent  probability within $-k_{\rm B}\ln 2\leq S_{\rm inf}^{\infty}\leq 0$, and its standard deviation equals  the Boltzmann constant.  
 Whereas Eqs.~(\ref{eq:infx}) and (\ref{eq:infxx}) hold generally in steady states, the equalities given by Eqs.~(\ref{eq:infxxxx}) and (\ref{eq:infxxx})  are  shown to be true  for continuous stochastic processes.

Related to the global infimum are the passage probabilities  $\mathsf{P}^{(2)}_+$ ($\mathsf{P}^{(2)}_-$) for entropy production 
to reach a threshold $s^+_{\rm tot}$ ($-s^-_{\rm tot} $) without having reached $-s^-_{\rm tot}$ ($s^+_{\rm tot} $) before.  This corresponds to  the  stochastic process $S_{\rm tot}(t)$  with two absorbing boundaries, a positive absorbing boundary at $S_{\rm tot}(t)=s^+_{\rm tot}$ and a negative absorbing boundary at $S_{\rm tot}(t)=-s^-_{\rm tot}$.     If the process $S_{\rm tot}(t)$ is continuous and $\langle S_{\rm tot}(t)\rangle\neq 0$, we find 
\begin{eqnarray}
\mathsf{P}^{(2)}_+ &=&\frac{e^{s^-_{\rm tot}/k_{\rm B}}-1}{e^{s^-_{\rm tot}/k_{\rm B}}-e^{-s^+_{\rm tot}/k_{\rm B}}} \label{eq:pass1xxx}\quad,\\ 
\mathsf{P}^{(2)}_- &=&\frac{1-e^{-s^+_{\rm tot}/k_{\rm B}}}{e^{s^-_{\rm tot}/k_{\rm B}}-e^{-s^+_{\rm tot}/k_{\rm B}}}\quad.\label{eq:pass2xxx}
\end{eqnarray}
Interestingly, the relations (\ref{eq:pass1xxx}) and (\ref{eq:pass2xxx}) relate  entropy-production fluctuations between two asymmetric values $s^+_{\rm tot} \neq  s^-_{\rm tot}$.   The asymptotic value of the passage probability  $\mathsf{P}^{(2)}_+$ for  $s^+_{\rm tot} = +\infty$  is the probability that entropy never reaches the value $-s^-_{\rm tot}$.    It is equal to the probability that the global infimum is larger or equal than $-s^-_{\rm tot}$.    The relations for the passage probabilities given by Eqs.~(\ref{eq:pass1xxx}) and (\ref{eq:pass2xxx}) thus imply Eqs.~(\ref{eq:infxxxx}) and (\ref{eq:infxxx}) for the global infimum.  Notably, the infima and passage statistics of entropy production are independent of the strength of the non-equilibrium driving, i.e., the mean entropy-production rate.

We also discuss stopping times.   A stopping time is the time at which a stochastic trajectory satisfies for the first time a certain criterion.    
We discuss here $s_{\rm tot}$-stopping times $T_+$, for which entropy production at the stopping time equals to $S_{\rm tot}(T_+) = s_{\rm tot}$, with $s_{\rm tot}>0$.  An example is the first-passage time of entropy production, at which entropy production reaches $s_{\rm tot}$ for the first time.
This value of entropy $S_{\rm tot}(T_+)$ is  a new record of entropy production, and first-passage times of entropy production are thus times at which a given record is reached.    Analogously, we define  a  ($-s_{\rm tot}$)-stopping time $T_-$ associated to $T_+$ at which entropy production equals to $S_{\rm tot}(T_-) = -s_{\rm tot}$ at the stopping time.    For example, if $T_+$ is the first-passage time of entropy production to first reach $s_{\rm tot}$, then $T_-$ is the  first-passage time of entropy production to first reach $-s_{\rm tot}$.
 Remarkably, we find that the mean stopping time $\langle T_+ \rangle$  equals to the  mean stopping time  $\langle T_- \rangle$: 
\begin{eqnarray}
\langle T_+ \rangle= \langle T_- \rangle\quad.
\end{eqnarray}
A similar equality holds for all the higher order moments of stopping times of entropy production.  These results follow from the
{\it stopping-time fluctuation theorem}
\begin{eqnarray}
\frac{p_{T_+}(t;s_{\rm tot})}{p_{T_-}(t;-s_{\rm tot})} = e^{ s_{\rm tot}/k_{\rm B}}\quad, \label{eq:FPTime}
\end{eqnarray} 
which we derive in this paper for  classical and continuous stochastic  processes in steady state.  
Here  $p_{T_+}(t; s_{\rm tot})$  is the probability density for the stopping time $T_+$, and $p_{T_-}(t; -s_{\rm tot})$ 
is the probability density for the stopping time $T_-$.
The stopping-time fluctuation theorem (\ref{eq:FPTime}) is illustrated in Fig.~\ref{fig1}b) for the example where $T_+$ and $T_-$ are first-passage times of entropy production with two absorbing boundaries. 
 
Other examples of  stopping times are waiting times,  defined as the time a stochastic trajectory  takes while changing from  an initial state I to a final state II, see  Fig.~\ref{fig1}c).     We show in this paper that for a nonequilibrium and stationary isothermal  process the ratio of  waiting-time distributions corresponding to forward trajectories (I $\rightarrow$ {\rm II}) and backward trajectories ({\rm II} $\rightarrow$ I) obey
\begin{eqnarray}
\frac{p_{T^{\rm I \rightarrow {\rm II}}_+}(t;-Q)}{p_{T^{\rm {\rm II} \rightarrow I}_-}(t; Q)} = e^{-Q/k_{\rm B} \mathsf{T}_{\rm env}} \label{eq:waitingTime}\quad,  \label{eq:9} 
\end{eqnarray} 
for all trajectories between I and {\rm II} that exchange the amount   $Q$  of heat with an equilibrated environment at  temperature $\mathsf{T}_{\rm env}$; if $Q>0$, then the system absorbs heat from the environment.   Here $p_{T^{\rm I \rightarrow {\rm II}}_+}(t; -Q)$ denotes the probability density  for the waiting time $T^{\rm I \rightarrow {\rm II}}_+$ to reach the state {\rm II} while absorbing the heat $Q$.   Equation (\ref{eq:9}) is a generalization of the local detailed-balance condition for transition rates $k^{{\rm I}\rightarrow {\rm {\rm II}}}_{+}/k^{{\rm {\rm II}}\rightarrow {\rm I}}_- = e^{-Q/k_{\rm B}\mathsf{T}_{\rm env}}$ \cite{de2013non, julicher1997modeling, PhysRevE.85.041125} to waiting-time distributions.    Indeed, transition rates are given by $k = \int^\infty_0{\rm dt}\: t^{-1}\:p_{\rm T}(t)$.  Notably, Eq.~(\ref{eq:9}) implies  a symmetry relation on the normalized waiting-time distributions 
 \begin{eqnarray}
 \frac{p_{T^{{\rm I}\rightarrow {\rm  {\rm II}}}_+}(t| -Q)}{\mathsf{P}^{{\rm I}\rightarrow {\rm  {\rm II}}}_+ } = \frac{p_{T^{{\rm {\rm II}}\rightarrow {\rm  I}}_-}(t| Q)}{\mathsf{P}^{{\rm {\rm II}}\rightarrow {\rm  I}}_-}  \label{eq:waitingTimex}\quad,  \label{eq:10}
 \end{eqnarray}
 where  $\mathsf{P} = \int^\infty_0{\rm dt} \, p_T(t)$.  
Therefore, the mean waiting times $\langle T \rangle =  \int^\infty_0{\rm dt}\, t \,p_{\rm T}(t)/\mathsf{P}$
 for the forward and backward transitions are the same, $\langle T^{{\rm I}\rightarrow {\rm  {\rm II}}}_+ \rangle = \langle T^{{\rm {\rm II}}\rightarrow {\rm  I}}_- \rangle$. 

We derive all these results  on  infima, passage probabilities and stopping times of entropy production in a new  unified formalism that uses   the theory of martingales \cite{doob1953stochastic, doob2012measure}, and apply our results to the dynamics of colloidal particles in periodic potentials and molecular motors, which  transduce chemical energy into mechanical work.       
The paper is structured as follows:  In Sec.~\ref{sec:2}, we briefly review the formalism of stochastic thermodynamics.  
  In Sec.~\ref{sec:3},  we discuss the connection between martingale processes and entropy production.  In Sec.~\ref{sec:4}, Sec.~\ref{sec:5}, and Sec.~\ref{sec:6} we derive, respectively, the infimum law (\ref{eq:infx}) and the bound (\ref{eq:infxx}); the statistics of the global infimum of entropy production (\ref{eq:infxxxx})-(\ref{eq:infxxx}) and the equalities for the passage probabilities (\ref{eq:pass1xxx})-(\ref{eq:pass2xxx});    fluctuation theorems for stopping times of entropy production, which include  first-passage times of entropy production (\ref{eq:FPTime}) and waiting times of stochastic processes (\ref{eq:9}).   
We apply our results in Sec.\ref{sec:7} to a  drifted colloidal particle moving in a periodic potential.   In Sec.~\ref{sec:8}, we apply our results to discrete molecular  processes such as the stepping statistics of molecular motors or the dynamics of  enzymatic reactions. 
The paper concludes  with a discussion in  Sec.~\ref{sec:9}.

\section{Stochastic thermodynamics and entropy production}   \label{sec:2}
We first briefly review  the basic concepts of stochastic entropy production based on path probabilities in discrete time.  We then present   a measure-theoretic formalism of stochastic thermodynamics, which defines entropy production in discrete and continuous time.    Using measure theory we avoid problems with the normalization of path probabilities.   
\subsection{Entropy production for processes in discrete time}
We consider the dynamics of a mesoscopic system in a nonequilibrium steady state, and describe its dynamics  with the coarse-grained state variables  $\omega(t) = (\bq(t), \tilde{\bq}(t))$ at time $t$.   
The variables $\bq(t)$ represent  $n$ 
degrees of freedom that are even under time reversal, and the variables  $\tilde{\bq}(t)$ represent  $\tilde{n}$ 
degrees of freedom that are  odd under time reversal \cite{parity}.  Notably, the  variables $\bq(t)$ and $\tilde{\bq}(t)$   represent the dynamics of  collective modes in a system of interacting particles, for instance, $\bq(t)$ describes the position of a colloidal particle in a fluid and $\tilde{\bq}(t)$ its effective momentum. 

In a given time window $[0,t]$, the  coordinates $\omega(t)$ trace a path in phase space $\omega_0^t =\{\omega(\tau)\}_{0\leq \tau\leq t}$.  We associate with each trajectory $\omega_0^t$  a probability density $\mathcal{P}(\omega_0^t;p_{\rm init})$, which captures the limited  information provided by the coarse-grained variables $\omega$, and the fact that the exact microstate is not known; the distribution $p_{\rm init}$ is the probability density of the initial state $\omega(0)$ .  
  The  entropy production  associated with a path $\omega_0^t$ of a stationary  process is  given by~\cite{crooks1998nonequilibrium, crooks1999entropy, maes2003time}  
\begin{equation}
 S_{\rm tot} \left(t\right) \equiv k_{\rm B} \ln \frac{\mathcal{P}\left(\omega_0^t;p_{\rm ss}\right)}{\mathcal{P}\left(\Theta_t \omega_0^t;\tilde{p}_{\rm ss}\right)}\quad, \label{eq:s1}
\end{equation}
where   $\Theta_t \omega_0^t =  \{\tilde{\omega}(\tau)\}_{\tau=0}^t$   is the time-reversed  trajectory with   $\tilde{\omega}(\tau) = (\mathbf{q} (t-\tau),-\tilde{\bq}(t-\tau))$; $p_{\rm ss}$ the steady-state distribution in the forward dynamics; and $\tilde{p}_{\rm ss}$ the steady-state distribution in the backward dynamics.    Equation (\ref{eq:s1})  is well-defined for discrete-time processes for which $\omega^t_0$  is a discrete sequence of states.  The ensemble average of entropy production of  a stationary process can be expressed as an integral:
\begin{equation}
\langle S_{\rm tot}(t)\rangle = k_{\rm B} \int \mathcal{D}\omega_0^t\, \mathcal{P}(\omega_0^t;p_{\rm ss})\,\ln \frac{\mathcal{P}(\omega_0^t;p_{\rm ss})}{\mathcal{P}(\Theta \omega_0^t;\tilde{p}_{\rm ss})}\quad.  \label{eq:av}
\end{equation}  
Entropy production is therefore  the observable that quantifies time   irreversibility of mesoscopic trajectories \cite{maes2003origin}.   
In fact, by measuring entropy production an observer can determine within a minimal  time 
whether a movie  of a stochastic process is run forwards or backwards  \cite{roldan2015decision}.  

Microscopic reversibility implies that a mesoscopic system in contact with an  equilibrated environment  satisfies local detailed balance \cite{bergmann1955new,  crooks1999entropy, crooks2000path, maes2003time}.   Local detailed balance 
manifests itself in a condition on the  path  probabilities conditioned on the initial state, and reads
\begin{equation}
\frac{\mathcal{P}\left(\omega_0^t|\omega(0)\right)}{\mathcal{P}\left(\Theta_t \omega_0^t |\tilde{\omega}(t)\right)} = e^{S_{\rm env}(t)/k_{\rm B}}\quad,\label{eq:SsysMesoscopicxx} 
 \end{equation} 
where $S_{\rm env}(t)$ is the entropy change in the environment.  
If local detailed balance holds, then our definition of entropy production (\ref{eq:s1}) equals the total entropy change, i.e.,  the sum of the  system-entropy change  $\Delta S_{\rm sys}$ \cite{seifert2005entropy} and the  environment-entropy change $S_{\rm env}$:
\begin{equation}
 S_{\rm tot} (t) = \Delta S_{\rm sys} \left(t\right) +S_{\rm env} \left(t\right) \quad,
\end{equation}
with 
\begin{equation}
 \Delta S_{\rm sys} \left(t\right)= -k_{\rm B}\ln\frac{p_{\rm ss}\left(\omega(t)\right)}{p_{\rm ss}\left(\omega(0)\right)} \quad.
 \label{eq:SsysMesoscopic}
\end{equation}
Note that in Eq.~(\ref{eq:SsysMesoscopic}) we have used that $\tilde{p}_{\rm ss}\left(\tilde{\omega}(t)\right) = p_{\rm ss}(\omega(t))$ 
 \cite{maes2003time}.
  For a system in contact with 
one or several thermal baths,   the environment-entropy change is related to the heat exchanged between system and environment~\cite{thermo}.

\subsection{Entropy production for processes in continuous time}
 In discrete time, the expressions Eqs.~(\ref{eq:s1}) and  (\ref{eq:av}) for entropy production  are well-defined \cite{risken1984fokker}.  In continuous time,    the path-probability densities  $\mathcal{P}$ 
are not normalizable.  In order to avoid this problem, we use a  formalism based on measure  theory to define entropy production in continuous-time processes \cite{doob1953stochastic, liptser2013statistics, royden1988real, williams1991probability, tao2011introduction, doob2012measure}.  
Measure theory studies probabilities of events in terms of  a probability space $(\Omega, \mathcal{F}, \mathbb{P})$.    The set $\Omega$ of all trajectories $\omega$  is called the sample space; the set $\mathcal{F}$  of all measurable subsets $\Phi$ of $\Omega$ is
a $\sigma$-algebra; and the function $\mathbb{P}$ is a measure, which associates probabilities to subsets $\Phi$. 
In the following we identify the symbol   $\omega$ with  the full trajectory of state variables over all times, $\omega = \left\{\textbf{q}(\tau), \tilde{\bq}(\tau)\right\}_{\tau\in (-\infty,\infty)}$.

 The  concept of a probability measure $\mathbb{P}(\Phi)$ generalizes the path probability densities $\mathcal{P}(\omega)$.  
  The value $\mathbb{P}\left(\Phi\right)$ denotes the probability to observe a trajectory $\omega$ in  the  set $\Phi$, in other words $\mathbb{P}\left(\Phi\right) = \mathsf{Pr}\left(\omega\in\Phi\right)$.  
 An example of a measure is 
\begin{eqnarray}
\mathbb{P}\left( \Phi\right) = \displaystyle\int_{\bx \in\Phi} {\rm d}\lambda\:p(\bx) \quad . \label{eq:dens}
\end{eqnarray}  
with  $p(\bx)$ is a probability density of  
 elements $\bx$ in $\mathbb{R}^n$.
Here, $\lambda$ denotes the Lebesgue measure and the Lebesgue integral is
over the set $\Phi$. One can also 
define a probability density $\mathcal{R}(\omega)$ of a measure $\mathbb{P}(\Phi)= \int_{\omega \in \Phi} {\rm d}\mathbb{P}$  with respect to a second probability measure $\mathbb{Q}(\Phi)$ using  
 the Radon-Nikod\'{y}m theorem \cite{royden1988real} 
\begin{eqnarray}
\mathbb{P}\left(\Phi\right) 
&=& \int_{\omega \in \Phi} {\rm d}\mathbb{Q} \:\mathcal{R} (\omega)\quad , \label{eq:ro}
\end{eqnarray}
where the integral is over the probability space $\left(\Omega, \mathcal{F}, \mathbb{Q}\right)$ \cite{royden1988real, tao2011introduction}.  
The function $\mathcal{R} (\omega)$  is  called the Radon-Nikod\'{y}m derivative, which we denote by $\mathcal{R} (\omega) = \frac{{\rm d}\mathbb{P}}{{\rm d} \mathbb{Q}}(\omega)$.   In Eq.~(\ref{eq:ro}), the function  $\mathcal{R}(\omega)$ generalizes the probability density $p(\bx)$ to  spaces for which the Lebesgue measure does not exist, e.g., the Wiener space of trajectories of a Brownian particle.

We now consider   probability measures of steady-state processes.    A stationary  probability measure  is time-translation invariant  and satisfies $\mathbb{P} = \mathbb{P}\circ \mathsf{T}_t$, where $\mathsf{T}_t$ is the map  that translates a trajectory $\omega$ by  a time  $t$ as   $\bq(\tau)\rightarrow \bq(\tau+t)$ and $\tilde{\bq}(\tau)\rightarrow \tilde{\bq}(\tau+t)$.
A stochastic process $X(\omega;t)$  provides the value of an observable  $X$ at time $t$ for a given trajectory  $\omega$.
 We denote the average or expectation value of the stochastic variable $X(\omega;t)$  by $\langle X(\omega;t) \rangle_{\mathbb{P}}  = \int_{\omega\in\Omega}X(\omega;t)\:{\rm d}\mathbb{P}$.     In the following, the stochastic process $X(\omega;t)$ is sometimes simply denoted by  $X(t)$ and  its average by  $\langle X(t)\rangle$.

Entropy production $S_{\rm tot}(\omega;t)$ is an example of a stochastic process.  
An appropriate definition of entropy production, which generalizes Eq.~(\ref{eq:s1}) to include  continuous-time processes, can be written using  the Radon-Nikod\'{y}m derivative 
\begin{eqnarray}
S_{\rm tot}(\omega;t)  \equiv k_{\rm B} \ln \frac{\left.{\rm d}\mathbb{P}\right|_{\mathcal{F}(t)}}{\left.{\rm d}\left(\mathbb{P}\circ\Theta\right)\right|_{\mathcal{F}(t)}}(\omega)\quad \label{Eq:entropDef}
\end{eqnarray}
of the measure   $\left.\mathbb{P}\right|_{\mathcal{F}(t)}$ with respect to the time-reversed measure $\left.\left(\mathbb{P}\circ\Theta\right)\right|_{\mathcal{F}(t)}$ \cite{maes2000definition}.
Here  $\left.\mathbb{P}\right|_{\mathcal{F}(t)}$  denotes the restriction of the measure $\mathbb{P}$
over those events in the sub-$\sigma$-algebra $\mathcal{F}(t)\subset \mathcal{F}$ that is generated by trajectories $\omega^t_0$ in the time interval $[0,t]$.
 The time-reversed measure $\mathbb{P}\circ\Theta$ is defined using the
 time-reversal map  $\Theta$, which time reverses  trajectories $\omega$ as $\bq(t)\rightarrow \bq(-t)$ and $\tilde{\bq}(t)\rightarrow -\tilde{\bq}(-t)$.   Note that Eq.~(\ref{Eq:entropDef}) is well-defined for continuous-time processes that may contain jumps.

\section{Martingale Theory  for  Entropy Production}\label{sec:3} 
A fundamental, but still unexplored, property of entropy production is that in steady state  its exponential $e^{-S_{\rm tot}(t)/k_{\rm B}}$ is a positive and uniformly integrable {\it martingale} process. 
 A process is called martingale if its expected value at any time $t$  equals to its value at a previous time $\tau$, when the expected value is conditioned on observations up to the time $\tau$ (see Appendix \ref{app:a}).    The process $e^{-S_{\rm tot}(t)/k_{\rm B}}$ satisfies this property, and therefore obeys (see Appendix   \ref{app:c})
 \begin{eqnarray}
\langle \:e^{-S_{\rm tot}(t)/k_{\rm B}}\:|\:\omega^\tau_0\rangle  = e^{-S_{\rm tot}(\tau)/k_{\rm B}} \quad, \label{eq:Jarz}
\end{eqnarray}   
for $\tau<t$, and where the average is conditioned on a particular trajectory $\omega(t')$ from $t'=0$ up to time $\tau$.      From Eq.~(\ref{eq:Jarz})  it follows that martingale processes have a time-independent average.    Interestingly, for $e^{-S_{\rm tot}(t)/k_{\rm B}}$ this implies 
the integral fluctuation theorem.     Indeed, using Eq.~(\ref{eq:Jarz}) for $\tau=0$ and 
$S_{\rm tot}(0) = 0$, it follows  that $\langle e^{-S_{\rm tot}(t)/k_{\rm B}} \rangle = 1$, for arbitrary initial conditions  \cite{bochkov1977general, jarzynski1997nonequilibrium, crooks1999entropy, seifert2005entropy}.

On average the total entropy $S_{\rm tot}(t)$ always increases, and therefore it cannot be a martingale.   
However entropy production  is a  submartingale with the property
\begin{eqnarray}
\langle \:S_{\rm tot}(t)\:|\:\omega^\tau_0\:\rangle  \geq  S_{\rm tot}(\tau) \quad.  \label{eq:Jarzxxx}
\end{eqnarray}
Equation (\ref{eq:Jarzxxx}) follows from Eq.~(\ref{eq:Jarz}) and the fact that $e^{-S_{\rm tot}(t)/k_{\rm B}}$ is a convex function of $S_{\rm tot}(t)$.    From (\ref{eq:Jarzxxx}) it follows that the average entropy production is greater or equal  than zero for any initial condition.     Note that this statement is stronger than   $\langle S_{\rm tot}(t)\rangle \geq 0$, where the brackets denote the steady-state ensemble.

 A key property of martingales is {\it Doob's maximal inequality} (see Appendix \ref{app:a})  \cite{doob2012measure, doob1953stochastic}.  For $e^{-S_{\rm tot}(t)/k_{\rm B}}$   this inequality  provides a bound on the cumulative distribution of its supremum \cite{chetrite2011two}: 
 \begin{eqnarray}
{\rm Pr}\left({\rm sup}_{\tau\in [0,t]}\left\{e^{-S_{\rm tot}(\tau)/k_{\rm B}}\right\}\geq \lambda\right) \leq \frac{1}{\lambda}\Big\langle e^{-S_{\rm tot}(t)/k_{\rm B}}\Big\rangle \quad. \nonumber \\ \label{eq:districxx}
\end{eqnarray}  
Equation (\ref{eq:districxx}) is a stronger condition than the well-known Markov inequality  Eq.~(\ref{eq:markov}), and holds for steady-state  processes in discrete time and steady-state continuous-time processes with jumps.
  
Another key property of martingales is {\it Doob's optional sampling theorem}.   For entropy production, this theorem generalizes Eq.~(\ref{eq:Jarz})  to averages conditioned on stochastic stopping times $T<t$  (see  Appendix \ref{app:a}):
\begin{eqnarray}
\langle \:e^{-S_{\rm tot}(t)/k_{\rm B}}\:|\:S_{\rm tot}(T)\:\rangle  = e^{-S_{\rm tot}(T)/k_{\rm B}} \quad.\label{eq:Jarzx}
\end{eqnarray}
The stopping time $T=T(\omega)$ is the time at which a trajectory $\omega$ satisfies for the first time a certain criterion,  and therefore differs for each realization $\omega$.    
This is a generalization of  passage times.   Equation~(\ref{eq:Jarzx}) holds for steady-state  processes in discrete time and  for steady-state continuous-time processes with jumps.   Equation~(\ref{eq:Jarzx}) implies that the expected value of  $e^{-S_{\rm tot}(t)/k_{\rm B}}$, over all trajectories for which the value of entropy at the stochastic stopping time $T$ is  given by the value $s_{\rm tot}$, equals  $e^{-s_{\rm tot}/k_{\rm B}}$.

\section{The Infimum law}\label{sec:4}
Using the martingale property of $e^{-S_{\rm tot}(t)/k_{\rm B}}$  we now derive the  infimum law for entropy production, which holds for non-equilibrium processes in  a steady state.   From Eq.~(\ref{eq:districxx}) and the  integral fluctuation theorem, $\langle e^{-S_{\rm tot}(t)/k_{\rm B}}\rangle = 1$,  we find the following bound for the cumulative distribution of the  supremum of $e^{-S_{\rm tot}(t)/k_{\rm B}}$
 \begin{eqnarray}
{\rm Pr}\left({\rm sup}_{\tau\in [0,t]}\left\{e^{-S_{\rm tot}(\tau)/k_{\rm B}}\right\}\geq \lambda\right) \leq \frac{1}{\lambda} \quad, \nonumber \\ \label{eq:distric}
\end{eqnarray}  
for $\lambda\geq 0$.    Equation (\ref{eq:distric}) implies a  lower bound on the cumulative  distribution of the infimum of $S_{\rm tot}$ in a given time interval $[0,t]$:
\begin{eqnarray}
{\rm Pr}\left(\frac{S_{\rm inf}(t)}{k_{\rm B}}\geq -s\right) \geq 1-e^{-s}\quad,\label{cumula}
\end{eqnarray}
with $s\geq 0$ and   $S_{\rm inf}(t) = {\rm inf}_{\tau\in [0,t]}\left\{S_{\rm tot}(\tau)\right\}$.  The right hand side of Eq.~(\ref{cumula}) is the cumulative distribution of an exponential random variable $S$ with distribution function $p_{S}(s) = e^{-s}$.  From Eq.~(\ref{cumula})
it thus follows that the random variable $-S_{\rm inf}(t)/k_{\rm B}$  dominates stochastically over $S$, and this implies an inequality on the mean values of the corresponding random variables as we show in Appendix~\ref{app:dom}. 
From Eq.~(\ref{cumula}) we thus find the following  universal bound for the mean infimum of entropy production at time $t$:
\begin{eqnarray}
\langle  S_{\rm inf}(t) \rangle \geq -k_{\rm B} \quad.  \label{eq:inf}
\end{eqnarray}
The infimum law given by Eq.~(\ref{eq:inf}) holds for 
stationary stochastic processes  in discrete time and for stationary stochastic processes  
in continuous time for which  $e^{-S_{\rm tot}(t)/k_{\rm B}}$  is right continuous.

For the special case of isothermal processes the total entropy change $S_{\rm tot}(t) =  \Delta S_{\rm sys}(t) - Q(t)/\mathsf{T}_{\rm env} =  \Delta S_{\rm sys}(t) +  \left(W(t)-\Delta E(t)\right)/\mathsf{T}_{\rm env}$, with $Q(t)$ denoting the heat absorbed by the  system from the reservoir,  $W(t)$ the work done on the system  and $\Delta E(t)$ the internal energy change of the system.  We thus have a bound on the infimum of the dissipated part of the work  $W^{\rm diss}$, which reads 
\begin{eqnarray}
\langle W^{\rm diss}_{\rm inf}(t) \rangle  \geq -k_{\rm B}\mathsf{T}_{\rm env}\quad.
\end{eqnarray}
Here we have defined  the dissipated work $W^{\rm diss}(t)= W(t)-\Delta F(t)$ with $\Delta F(t) = E(t)-\mathsf{T}_{\rm env}\:\Delta S_{\rm sys}(t)$. For a process that is isothermal and for which all states have the same energy and entropy we have 
$\Delta F(t) =0$, and thus
\begin{eqnarray}
\langle W_{\rm inf}(t) \rangle  \geq -k_{\rm B}\mathsf{T}_{\rm env}, \quad \langle Q_{\rm sup}(t) \rangle  \leq k_{\rm B}\mathsf{T}_{\rm env}\quad,  \label{eq:q}
\end{eqnarray}
with $W_{\rm inf}(t)$ the infimum of the work done on the system, and $Q_{\rm sup}(t)$ the supremum of the heat absorbed by the system over a time $t$.
Equation (\ref{eq:q}) implies that on average  a homogeneous system in isothermal conditions can absorb no more than $k_{\rm B} \mathsf{T}_{\rm env}$ of energy from the thermal reservoir.

\section{Passage probabilities and global infimum of entropy production}\label{sec:5}
We  now derive, using the theory of martingales, general expressions for the passage probabilities and the global infimum of entropy production in 
continuous  steady-state processes without jumps. 
\subsection{Passage probabilities of entropy production with two asymmetric absorbing boundaries}
We consider the stochastic entropy production   $S_{\rm tot}(\omega;t)$ of a stationary probability measure $\mathbb{P}$ in a time interval $[0,T^{(2)}(\omega)]$, which starts at $t=0$ and ends at a stopping  time $T^{(2)}(\omega)$.  Here $T^{(2)}$ is the first-passage time at  which    $S_{\rm tot}(\omega;t)$ passes for the first time one of the two threshold values $-s^-_{\rm tot}<0$ or $s^+_{\rm tot}>0$ (see Fig.~1(b) for the particular case of $s^+_{\rm tot}=s^-_{\rm tot}$).

We define the passage probabilities $\mathsf{P}^{(2)}_+$  as the probability that the process first passes $s^+_{\rm tot}$ before passing $s^-_{\rm tot}$, and analogously, $\mathsf{P}^{(2)}_-$  as the probability that the process first passes $s^-_{\rm tot}$ before passing $s^+_{\rm tot}$.   These passage probabilities can be written as: 
\begin{eqnarray}
\mathsf{P}^{(2)}_+ &=& \mathbb{P}\left(\Phi_+ \right)\quad,\\ 
\mathsf{P}^{(2)}_- &=& \mathbb{P}\left(\Phi_- \right)\quad,
\end{eqnarray}
with  $\Phi_+$  the set of trajectories $\omega$ that pass first  the positive threshold  $s^+_{\rm tot}$, and $\Phi_-$ the set of trajectories $\omega$ that pass first  the negative threshold  $-s^-_{\rm tot}$:
\begin{eqnarray}
\Phi_+ &\equiv& \left\{\omega\in\Omega : S_{\rm tot}(\omega;T^{(2)}(\omega))=s^+_{\rm tot}\right\}\quad,  \label{eq:Phi+}\\
\Phi_- &\equiv& \left\{\omega\in\Omega : S_{\rm tot}(\omega;T^{(2)}(\omega))=-s^-_{\rm tot}\right\}\quad. \label{eq:Phi-}
\end{eqnarray}
Note that if $s^+_{\rm tot}$ is different from $s^-_{\rm tot}$ then $\Phi_+$ and $\Phi_-$ are  not  each other's time reversal.  Therefore the probabilities of these sets   are in general not related by local detailed balance. 
We also define the conjugate  probabilities  $\tilde{\mathsf{P}}^{(2)}_+$  and  $\tilde{\mathsf{P}}^{(2)}_-$ of the sets $\Phi_+$ and $\Phi_-$ under the time-reversed dynamics: 
\begin{eqnarray}
\tilde{\mathsf{P}}^{(2)}_+ &=& \left(\mathbb{P}\circ \Theta\right)\left(\Phi_+ \right)\quad,\\ 
\tilde{\mathsf{P}}^{(2)}_- &=& \left(\mathbb{P}\circ \Theta\right)\left(\Phi_- \right)\quad.
\end{eqnarray}

For a steady-state process out-of-equilibrium, i.e., $\langle S_{\rm tot}(t)\rangle>0$, $S_{\rm tot}(t)$  passes in a finite time one of the two boundaries with probability one.  We thus  have: 
\begin{eqnarray}
\mathsf{P}^{(2)}_+ + \mathsf{P}^{(2)}_- &=& 1  \quad, \label{eq:E16x}\\ 
\tilde{\mathsf{P}}^{(2)}_+ + \tilde{\mathsf{P}}^{(2)}_- &=& 1 \label{eq:E17x}\quad.
\end{eqnarray}

In addition, we  derive, using Doob's optional sampling theorem, the following  two identities: 
\begin{eqnarray}
\frac{\mathsf{P}^{(2)}_+}{\tilde{\mathsf{P}}^{(2)}_+} &=& e^{s^+_{\rm tot}/k_{\rm B}}\quad,\label{eq:E16} \\ 
\frac{\mathsf{P}^{(2)}_-}{\tilde{\mathsf{P}}^{(2)}_-} &=& e^{-s^-_{\rm tot}/k_{\rm B}}\quad. \label{eq:E17}
\end{eqnarray}
 Eq.~(\ref{eq:E16}) follows from  the  equalities:
\begin{eqnarray}
\tilde{\mathsf{P}}^{(2)}_+&=&\int_{\omega\in\Phi_+} {\rm d}\left(\mathbb{P}\circ\Theta\right)  \label{eq:E11}
 \\ 
&=& \int_{\omega\in\Phi_+} e^{-S_{\rm tot}(\omega;+\infty)/k_{\rm B}}\:{\rm d}\mathbb{P} \label{eq:E12}
 \\ 
&=& \int_{\omega\in\Phi_+} e^{-S_{\rm tot}(\omega;T^{(2)}(\omega))/k_{\rm B}}\:{\rm d}\mathbb{P} \label{eq:E13}
 \\ 
&=&e^{-s^+_{\rm tot}/k_{\rm B}} \int_{\omega\in\Phi_+} \:{\rm d}\mathbb{P} \label{eq:E14}
\\ 
&=& e^{-s^+_{\rm tot}/k_{\rm B}} \: \mathbb{P}\left(\Phi_+\right) \\
&=& e^{-s^+_{\rm tot}/k_{\rm B}}\:  \mathsf{P}^{(2)}_+
 \quad.\label{eq:E15}
\end{eqnarray}
   In Eq.~(\ref{eq:E12}) we transform an integral over the measure~$\mathbb{P}\:\circ\:\Theta$  to an integral over the measure $\mathbb{P}$, using the definition of entropy production, given by Eq.~(\ref{Eq:entropDef}) and $ e^{-S_{\rm tot}(\omega;+\infty)/k_{\rm B}} = \lim_{t\rightarrow +\infty}e^{-S_{\rm tot}(\omega;t)/k_{\rm B}}$, (see Appendix B).   In  Eq.~(\ref{eq:E13})  we replace $e^{-S_{\rm tot}(\omega;+\infty)/k_{\rm B}}$ by its value at the stopping time, $e^{-S_{\rm tot}(\omega;T^{(2)}(\omega))/k_{\rm B}}$,  using  
   Doob's optional sampling theorem, given by Eq.~(\ref{eq:Jarzx}).
    Finally, in Eq.~(\ref{eq:E14}) we use the fact that for continuous processes $S_{\rm tot}(\omega;T^{(2)}(\omega)) = s^+_{\rm tot}$, for all realizations  of the process $\omega$ in the set  $\Phi_+$.

From Eqs.~(\ref{eq:E16x})-(\ref{eq:E17}) we find the following explicit expressions for the passage probabilities:
\begin{eqnarray}
\mathsf{P}^{(2)}_+ &=&\frac{e^{s^-_{\rm tot}/k_{\rm B}}-1}{e^{s^-_{\rm tot}/k_{\rm B}}-e^{-s^+_{\rm tot}/k_{\rm B}}}  \label{eq:pass1xx}\quad,\\ 
\mathsf{P}^{(2)}_- &=&\frac{1-e^{-s^+_{\rm tot}/k_{\rm B}}}{e^{s^-_{\rm tot}/k_{\rm B}}-e^{-s^+_{\rm tot}/k_{\rm B}}}\quad.\label{eq:pass2xx}
\end{eqnarray}   
For the  case of    symmetric boundaries    $s_{\rm tot} =  s^+_{\rm tot} = s^-_{\rm tot}$ we have
\begin{eqnarray}
\mathsf{P}^{(2)}_{+} = \frac{e^{s_{\rm tot}/k_{\rm B}}}{1+e^{s_{\rm tot}/k_{\rm B}}}\quad, \label{eq:Integralx}\\ 
\mathsf{P}^{(2)}_{-}  =  \frac{1}{1+e^{s_{\rm tot}/k_{\rm B}}}\quad. \label{eq:Integralxx} 
\end{eqnarray}
We can also discuss  the limits where one the two thresholds move to infinity, whereas the other threshold remains finite.  This corresponds to a  process with one absorbing boundary.    If the lower threshold  $s^-_{\rm tot}\gg k_{\rm B}$, the process ends with probability one in the positive threshold,
$\mathsf{P}^{(2)}_+ = 1$ and $\mathsf{P}^{(2)}_- = 0$,  in accordance with the second law of thermodynamics.   
If however the upper threshold becomes large, $s^+_{\rm tot}\gg k_{\rm B}$,  entropy production can still reach the positive threshold, since on average entropy always increases, but  with a probability that depends on $s^-_{\rm tot}$.  In this case the passage probabilities are given by 
\begin{eqnarray}
\mathsf{P}^{(2)}_+ &\simeq& 1-e^{-s^-_{\rm tot}/k_{\rm B}}\quad, \\ 
\mathsf{P}^{(2)}_- &\simeq& e^{-s^-_{\rm tot}/k_{\rm B}} \quad.
\end{eqnarray}

From these limits we can also determine the passage probabilities $\mathsf{P}^{(1)}_+$ and $\mathsf{P}^{(1)}_-$ of entropy production with one absorbing boundary.  They denote,  respectively, the probability to reach a positive boundary $s_{\rm tot}$ or a negative boundary $-s_{\rm tot}$:
\begin{eqnarray}
\mathsf{P}^{(1)}_{+}&=& 1\quad, \label{eq:IntegralxT}\\ 
\mathsf{P}^{(1)}_{-} &=&  e^{-s_{\rm tot}/k_{\rm B}}\quad. \label{eq:IntegralxxT} 
\end{eqnarray}

The above arguments also hold  for sets $\Phi_{+,{\rm I}}$ and $\Phi_{-,{\rm I}}$ of trajectories $\omega$, which are conditioned on an initial coarse grained state I.  
They are defined as the subsets of, respectively, $\Phi_+$ and $\Phi_-$  with the additional constraint that the initial state falls in the coarse-grained state I,  $\omega(0)\in{\rm I}$.   With these definitions 
Eqs.~(\ref{eq:pass1xx})-(\ref{eq:pass2xx}) can be generalized to passage probabilities of entropy production conditioned on the initial state, see Appendix \ref{app:pass}.  Note that this generalization holds for coarse-grained states that are invariant with respect to time-reversal, i.e., ${\rm I} = \Theta({\rm I})$.  

\begin{figure}
\includegraphics[width=9cm]{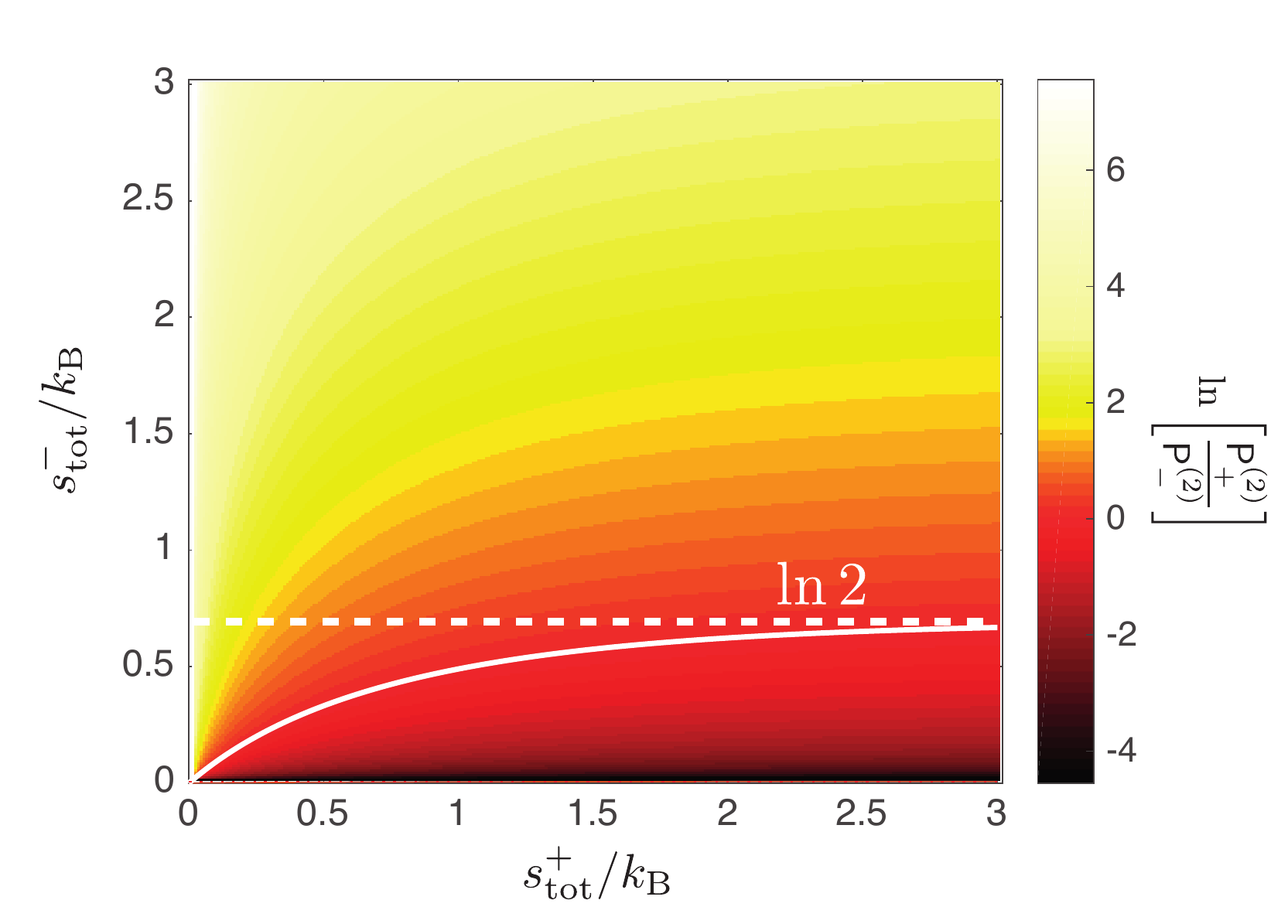} 
\caption{Entropy-production passage-probability ratio  $\ln \left[\mathsf{P}^{(2)}_+/\mathsf{P}^{(2)}_-\right]$ as a function of the thresholds $s^-_{\rm tot}$ and $s^+_{\rm tot}$, obtained from Eqs.~(\ref{eq:pass1xx})-(\ref{eq:pass2xx}).  The solid white line is the curve for which $\mathsf{P}^{(2)}_+ = \mathsf{P}^{(2)}_- = 1/2$, and follows from Eq.~(\ref{eq:fixedRatio}).  This curve converges asymptotically to the value $s^-_{\rm tot} = k_{\rm B} \ln 2$ for $s^+_{\rm tot} \rightarrow +\infty$ (white dashed line).  }\label{fig:fpixx}
\end{figure}

In Fig.~\ref{fig:fpixx} we illustrate the expressions of the passage probabilities, given by Eqs.~(\ref{eq:pass1xx}) and (\ref{eq:pass2xx}), by plotting $\ln(\mathsf{P}^{(2)}_+/\mathsf{P}^{(2)}_-)$
  as a function of the thresholds $s^-_{\rm tot}$ and $s^+_{\rm tot}$.   A characteristic feature of this figure are the lines of constant ratio $\mathsf{P}^{(2)}_+/\mathsf{P}^{(2)}_-$, which are given by: 
\begin{eqnarray}
s^-_{\rm tot} = k_{\rm B}\:\ln\left(1+\frac{\mathsf{P}^{(2)}_+}{\mathsf{P}^{(2)}_-}\left(1- e^{-s^+_{\rm tot}/k_{\rm B}}\right)\right)\quad.  \label{eq:fixedRatio}
\end{eqnarray}
In situations for which $\mathsf{P}^{(2)}_+ = \mathsf{P}^{(2)}_- = 1/2$, the stopping process is unbiased, and the probability to reach the threshold $s^+_{\rm tot}$ equals the probability to reach the threshold $-s^-_{\rm tot}$.  Since entropy production is a stochastic process with positive drift, the passage probabilities can be equal only if the  negative threshold  lies closer to the origin than the positive threshold, $s^-_{\rm tot}<s^+_{\rm tot}$.    Additionally, it follows from Eq.~(\ref{eq:fixedRatio}) that for $\mathsf{P}^{(2)}_+ = \mathsf{P}^{(2)}_-$  the negative threshold obeys
$s^-_{\rm tot}  \leq k_{\rm B}\:\ln 2$, as is illustrated in Fig.~\ref{fig:fpixx}.    This bound on $s^-_{\rm tot}$ can also be discussed for passage probabilities   $\mathsf{P}^{(2)}_-\neq 1/2$, for which  the lower threshold must satisfy   $s^-_{\rm tot} \leq -k_{\rm B}\:\ln \mathsf{P}^{(2)}_-$.     

The discussion of stopping events of entropy production with two boundaries is an example of thermodynamics of symmetry breaking.   Thermodynamics of symmetry breaking is usually discussed in finite times  \cite{roldan2014universal}.  Here we find   the analogous relation $\pm s^{\pm}_{\rm tot}\geq k_{\rm B}\ln  \mathsf{P}^{(2)}_{\pm}$, which is  valid for stopping times.

\subsection{Global infimum of entropy production}
The global infimum of entropy production $S^{\infty}_{\rm inf}$ is the lowest value of entropy production over all times $t\geq 0$ during one realization of the process.  The global infimum can be defined in terms of the local infimum $S_{\rm inf}(t)$ as
\begin{eqnarray}
S^{\infty}_{\rm inf} \equiv \lim_{t\rightarrow \infty}S_{\rm inf}(t)\quad.
\end{eqnarray}
Therefore, the global infimum is always negative and smaller or equal than  local infima $S_{\rm inf}(t)$.  The statistics of the global infimum follow from the expressions for the passage probabilities (\ref{eq:pass1xx})-(\ref{eq:pass2xx}).   This can be most easily understood 
in terms of the cumulative distribution of the global infimum
\begin{eqnarray}
\mathsf{Pr}(S^{\infty}_{\rm inf}\geq -s) = \mathsf{Pr}(S_{\rm tot}(t)\geq -s, \forall t\geq 0)\quad.  \label{eq:cumGlobal}
\end{eqnarray}
The  right-hand side of Eq.~(\ref{eq:cumGlobal}) is the  survival probability for entropy production in a process with one absorbing boundary located at  $-s$, with $s\geq 0$.   Therefore, the survival probability is the passage probability 
$\mathsf{P}^{(2)}_+$ with $s^+_{\rm tot}=+\infty$ and 
 $s^-_{\rm tot}=s$.    This implies $\mathsf{Pr}(S^{\infty}_{\rm inf}\geq -s) = 1-e^{-s/k_{\rm B}}$.  The corresponding  distribution of the global infimum is
\begin{eqnarray}
p_{S^\infty_{\rm inf}}(-s) =\frac{e^{-s/k_{\rm B}}}{k_{\rm B}}\quad. \label{eq:infmain}
\end{eqnarray}
The mean of the global infimum therefore is
\begin{eqnarray}
\langle S^{\infty}_{\rm inf}\rangle  = -k_{\rm B}\quad. \label{eq:meanmain}
\end{eqnarray}
These properties of the global infimum hold for continuous processes in steady state. 
 The infimum law, given by Eq.~(\ref{eq:inf}),  becomes thus an equality at large times.   Since $S_{\rm inf}(t)\geq S^{\infty}_{\rm inf}$, the equalities on the global infimum, given by Eqs.~(\ref{eq:infmain}) and  (\ref{eq:meanmain}), valid for continuous processes, imply the inequalities for the local infima, given by Eqs.~(\ref{cumula}) and (\ref{eq:inf}), for continuous processes.  Note however that Eqs.~(\ref{cumula}) and (\ref{eq:inf}) are also valid for processes in discrete time and processes in continuous time with jumps.

Remarkably, the distribution of the  global infimum  of entropy production is  universal.   For any continuous steady-state process  the distribution of the global infimum is an exponential with mean equal to $-k_{\rm B}$.

\section{Stopping times of entropy production} \label{sec:6}  
In this section we derive  fluctuation theorems for stopping times of entropy production using the martingale property of $e^{-S_{\rm tot}(t)/k_{\rm B}}$.    The stopping-time fluctuation theorem entails   fluctuation theorems for  first-passage-times of entropy production and for waiting times of a stochastic process.

\subsection{Stopping-time fluctuation theorem}\label{sec:IVA}
We consider the statistics of   $s_{\rm tot}$-stopping times  $T_+=T(\omega)$ for which entropy production at the stopping time takes the value $s_{\rm tot}$, i.e.,~ $S_{\rm tot}(T_+) = s_{\rm tot}$ ($s_{\rm tot}>0$).
An example of such an $s_{\rm tot}$-stopping time is the first-passage time $T^{(1)}_{+}$,  which determines the time at which  entropy production $S_{\rm tot}(t)$  reaches for the first time the value  $s_{\rm tot}>0$.   
Another example is given by the first-passage time $T^{(2)}_{+}$, which is the time 
at which  entropy  production $S_{\rm tot}(t)$ passes for the first  time a threshold value $s_{\rm tot}$, 
given that it has not reached  $-s_{\rm tot}$ before. The latter process is therefore equivalent to a first-passage problem  with two absorbing boundaries.  
 More generally,  $s_{\rm tot}$-stopping times $T^{(n)}_{+}$ can be defined by multiple threshold crossings  and the condition $S_{\rm tot}(T^{(n)}_+)= s_{\rm tot}$, with $n$ the order of threshold crossings.

\begin{figure}
\includegraphics[width=6cm]{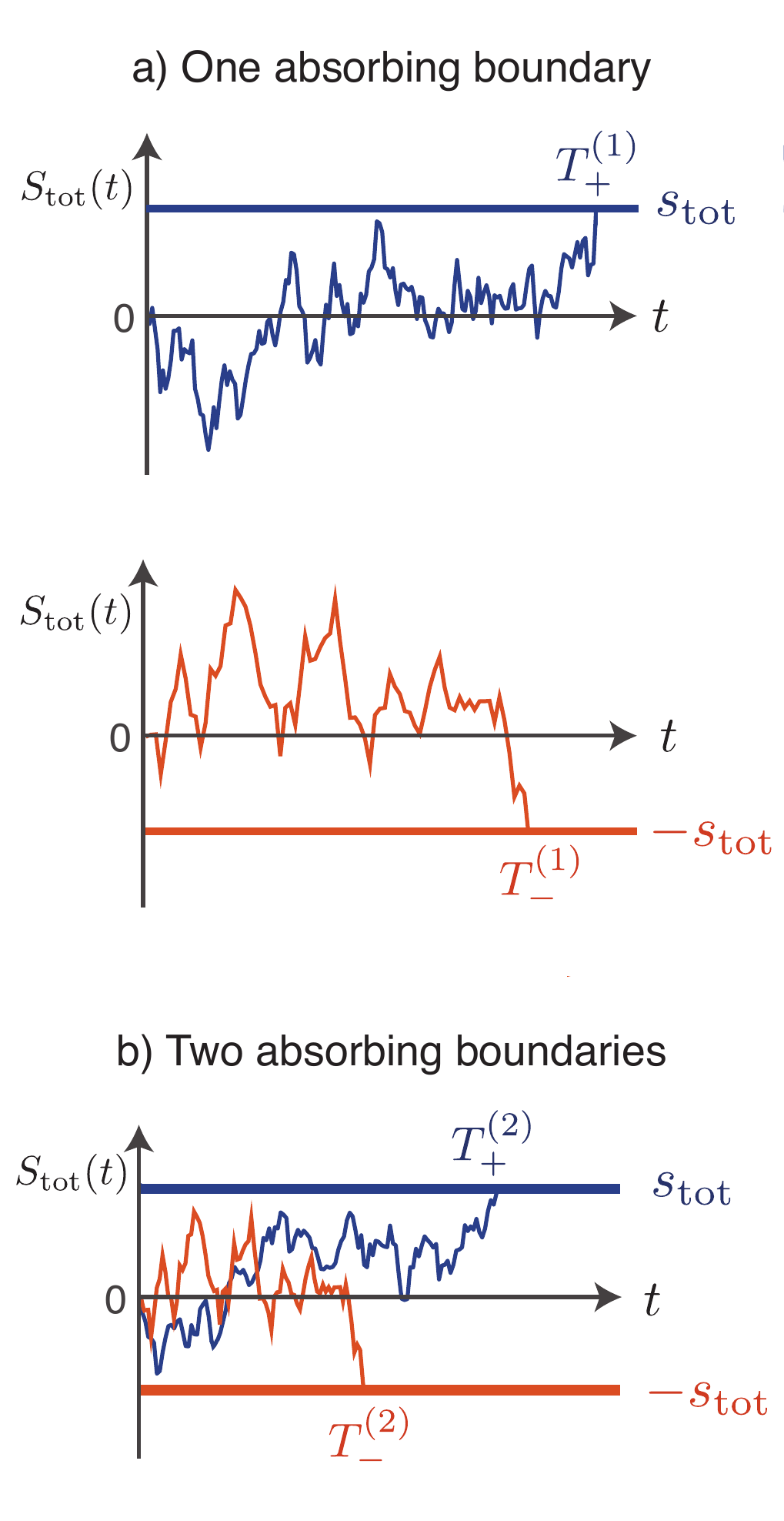} 
\caption{Illustration of the first-passage times for entropy production.  a) First-passage time $T^{(1)}_{+}$ for entropy production with  one positive absorbing boundary $s_{\rm tot}$ (top) and  first-passage time $T^{(1)}_{-}$ for entropy production with one negative absorbing boundary $-s_{\rm tot}$ (bottom).  b) First-passage times $T^{(2)}_{+}$ and $T^{(2)}_{-}$ for entropy production with two absorbing boundaries at $\pm s_{\rm tot}$.  At the time  $T^{(2)}_{+}$ entropy production passes for the first time the threshold $s_{\rm tot}$ without having reached $-s_{\rm tot}$ before.   At the time  $T^{(2)}_{-}$ entropy production passes for the first time the threshold $-s_{\rm tot}$ without having reached $s_{\rm tot}$ before. }\label{fig:fpi}
\end{figure}

We derive the following general fluctuation theorem for $s_{\rm tot}$-stopping times $T_+$  (see Appendix \ref{sec:proofApp}): 
\begin{eqnarray}
\frac{\mathbb{P}\left(\Phi_{T_+\leq t}\right)}{\mathbb{P}\left(\Theta_{T_+}\left(\Phi_{T_+\leq t}\right)\right)} = e^{s_{\rm tot}/k_{\rm B}}\quad, \label{eq:stopp}
\end{eqnarray}
where  $\mathbb{P}\left(\Phi_{T_+\leq t}\right)$ is the 
probability to observe a trajectory $\omega$  that satisfies the stopping-time criterion at a time $T_+\leq t$, and $\Phi_{T_+\leq t}$ denotes the set of these trajectories.
The set $\Theta_{T_+}\left(\Phi_{T_+\leq t}\right)$ describes the time-reversed trajectories of $\Phi_{T_+\leq t}$.  It is generated by applying the 
time-reversal map $\Theta_{T_+}$ to all the elements of the original set.  The map $\Theta_{T_+}= \mathsf{T}_{T_+}\circ\Theta$ time reverses trajectories $\omega$  with respect to the reference time $T_+(\omega)/2$, and thus  $X(\Theta_{T_+}(\omega); \tau) = X(\omega; T_+(\omega)-\tau)$ for all stochastic processes $X$ that are even under time reversal.   The  fluctuation theorem for $s_{\rm tot}$-stopping times  Eq.~(\ref{eq:stopp})    is  valid for continuous and stationary stochastic processes.   In our derivation, given in Appendix \ref{sec:proofApp}, we use the martingale property of $e^{-S_{\rm tot}(t)/k}$.  
  
The probability density $p_{T_+}$ of the $s_{\rm tot}$-stopping time $T_+$ is given by:
\begin{eqnarray}
p_{T_+}(t;s_{\rm tot}) = \frac{{\rm d}}{{\rm d}t} \mathbb{P}\left(\Phi_{T_+\leq t}\right) \label{eq:pt}\quad.  
\end{eqnarray} 
Entropy production is   odd under time reversal $S_{\rm tot}(\Theta_{T_+}(\omega);T_+(\omega)) = -S_{\rm tot}(\omega;T_+(\omega))$ as shown in  Appendix \ref{app:b3}.   Therefore,  we can associate to the $s_{\rm tot}$-stopping time $T_+$ a ($-s_{\rm tot}$)-stopping time $T_-$ with the property 
 \begin{eqnarray}
 \Theta_{T_+}\left(\Phi_{T_+\leq t}\right) = \Phi_{T_-\leq t}\quad. \label{eq:defTheta}
 \end{eqnarray}
 For example, the $(-s_{\rm tot})$-stopping times $T_-$ associated to the 
 first-passage time $T^{(1)}_{+}$  is the first-passage time $T^{(1)}_{-}$ 
  when entropy production first reaches $-s_{\rm tot}$, see Fig.\ref{fig:fpi} a).   Analogously, the    
  $(-s_{\rm tot})$-stopping times $T_-$ associated to the  
  the first-passage time
$T^{(2)}_{+}$ is the first-passage time  $T^{(2)}_{-}$  when entropy production  first reaches  $-s_{\rm tot}$ without having reached $s_{\rm tot}$ before, see Fig.\ref{fig:fpi} b).

      We can thus identify the distribution of $T_{-}$ with the measure of time-reversed trajectories:
 \begin{eqnarray}
p_{ T_-}(t;-s_{\rm tot}) = \frac{{\rm d}}{{\rm d}t} \mathbb{P}\left(\Theta_{T_+} \left(\Phi_{T_+\leq t}\right)\right)\quad.  \label{eq:-sp} 
\end{eqnarray}
This equation can be applied to all pairs of stopping times $T_+$ and $T_-$ related by 
 Eq.~(\ref{eq:defTheta}).   From Eqs.~(\ref{eq:stopp}), (\ref{eq:pt}) and~(\ref{eq:-sp}) follows the stopping-time fluctuation theorem for entropy production
 \begin{eqnarray}
\frac{p_{ T_+}(t;s_{\rm tot})}{p_{ T_-}(t;-s_{\rm tot})} = e^{s_{\rm tot}/k_{\rm B}}\quad.\label{eq:fpt}
 \end{eqnarray}  

The stopping-time fluctuation theorem for entropy production, given by Eq.(\ref{eq:fpt}), 
generalizes  the results derived in  \cite{roldan2015decision} for  first-passage times.  
Equation (\ref{eq:fpt}) implies two interesting results for stopping times of entropy production  that are outlined below.

\subsection{Symmetry of the normalized stopping-time distributions}
The stopping-time fluctuation relation Eq.~(\ref{eq:fpt})  implies an equality between  
the normalized stopping-time distributions $p_{ T_+}(t|s_{\rm tot})$ and $p_{ T_-}(t|-s_{\rm tot})$ which reads
\begin{eqnarray}
p_{ T_+}(t|s_{\rm tot}) &=& p_{ T_-}(t|-s_{\rm tot})\quad. \label{eq:synnp}
\end{eqnarray} 
The normalized distributions are  defined as: 
\begin{eqnarray}
p_{ T_+}(t|s_{\rm tot}) &= & \frac{p_{T_+}(t;s_{\rm tot})}{\int^\infty_0 {\rm d}t\:p_{T_+}(t;s_{\rm tot})} \quad,\\ 
p_{T_-}(t|-s_{\rm tot}) &= & \frac{p_{T_-}(t;-s_{\rm tot})}{\int^\infty_0 {\rm d}t\:p_{T_-}(t;-s_{\rm tot})}\quad.
\end{eqnarray}
The symmetric relation Eq.~(\ref{eq:synnp}) comes from the fact that the ratio of the stopping-time distributions in Eq.~(\ref{eq:fpt}) is time independent.

The stopping-time fluctuation theorem (\ref{eq:fpt}) thus implies that the mean stopping time, given that the process terminates at the positive boundary, equals to the mean  stopping time, given that the process terminates  at the negative boundary: 
\begin{eqnarray}
\langle T_+ \rangle\quad = \langle T_- \rangle\quad,  \label{eq:symmFP}
\end{eqnarray}
with $\langle T_+ \rangle = \int^{+\infty}_0{\rm d}t\:p_{T_+ }(t|s_{\rm tot})\:t$ and  $\langle T_-   \rangle= \int^{+\infty}_0{\rm d}t\:p_{T_-}(t|-s_{\rm tot})\:t$.
 This remarkable symmetry extends to all the moments of the  stopping-time   distributions.  A similar result has been found for waiting-time distributions in chemical kinetics \cite{qian2006generalized, wang2007detailed, PhysRevE.71.031902, ge2008waiting,  linden2008decay, jia2014cycle, qian2016entropy}, for the cycle-time distributions in Markov chains \cite{bauer2014affinity, jia2014cycle}, and for decision-time distributions in sequential hypothesis tests~\cite{roldan2015decision, dorpinghausinformation}.   These results could therefore be interpreted as a consequence of the fundamental relation  for the stopping-time fluctuations of entropy production given by Eq.~(\ref{eq:synnp}).  

\subsection{Passage probabilities for symmetric boundaries}
Equation (\ref{eq:fpt}) implies a relation for the  stopping probabilities of entropy production: 
\begin{eqnarray}
\frac{\mathsf{P}_+}{\mathsf{P}_-} = e^{s_{\rm tot}/k_{\rm B}}\quad. \label{eq:Integral}
\end{eqnarray}  
Stopping probabilities are the probabilities that the process satisfies the stopping criterion in a finite time.   
They are defined as 
\begin{eqnarray}
\mathsf{P}_+ &=& \int^\infty_0 {\rm d}t\:p_{T_+}(t;s_{\rm tot}) \quad,\\ 
\mathsf{P}_- &=&  \int^\infty_0 {\rm d}t\:p_{T_-}(t;-s_{\rm tot})\quad.
\end{eqnarray}  
 Equation (\ref{eq:Integral}) follows directly from integrating Eq.~(\ref{eq:fpt}) over time.    The relations (\ref{eq:IntegralxT}) and (\ref{eq:IntegralxxT}) for passage probabilities of first-passage times with one absorbing boundary, and the relations (\ref{eq:IntegralxT}) and (\ref{eq:IntegralxxT}) for passage probabilities of first-passage times with two absorbing boundaries, are examples of stopping times that satisfy   Eq.~(\ref{eq:Integral}).

\subsection{Fluctuation relation for waiting times}
An interesting question, closely related to entropy stopping times, is the following:  what is the waiting time $T^{{\rm I}\rightarrow {\rm {\rm II}}}$ a  process takes to travel from a state  ${\rm I}$ to 
a state ${\rm II}$.    We derive here exact relations characterizing    $(\pm s_{\rm tot})$-waiting times $T^{{\rm I}\rightarrow {\rm {\rm II}}}_{\pm}$.      The $(\pm s_{\rm tot})$-waiting time  $T^{{\rm I}\rightarrow {\rm {\rm II}}}_{\pm}$ denotes the time a process takes to travel  from a state  ${\rm I}$ to 
a state ${\rm II}$, and  produce a total positive or negative entropy $\pm s_{\rm tot}$ (see Appendix    \ref{app:D4}).

Following Kramers \cite{kramers1940brownian}, we define states as points  in phase space, i.e., ${\rm I} = \left\{\bq_{\rm I}\right\}$  and ${\rm II} = \left\{\bq_{{\rm II}}\right\}$.   In the Appendix   \ref{app:D4}, we also consider the more general case for which  states consist  of sets of points, which may also contain   odd-parity variables.   

We first derive a generalized fluctuation theorem which applies to trajectories starting from a given initial state~I  (see Appendix \ref{app:D3}): 
\begin{eqnarray}
\frac{\mathbb{P}\left[\Phi_{T_+\leq t}\cap \Gamma_{\rm I}\right]}{\mathbb{P}\left[\Theta_{T_+}\left(\Phi_{T_+\leq t}\cap \Gamma_{\rm I}\right)\right]}  = e^{s_{\rm tot}/k_{\rm B}}\quad, \label{eq:fluctIntermediateMain}
\end{eqnarray} 
with $\Gamma_{\rm I}$ the set of trajectories $\omega$ for which $\omega(0)\in{\rm I}$.    

We use Eq.~(\ref{eq:fluctIntermediateMain})  to derive a fluctuation theorem for waiting times (see Appendix \ref{app:D4}): 
\begin{eqnarray}
\frac{p_{T^{{\rm I}\rightarrow {\rm {\rm II}}}_+}\left(t\right)}{p_{T^{{\rm II} \rightarrow I}_-}\left(t\right)} = e^{s_{\rm env}/k_{\rm B}} \quad, \label{eq:fpttxla}
\end{eqnarray}
where $s_{\rm env} = s_{\rm tot} - k_{\rm B}\log \frac{p_{\rm ss}(\bq_{{\rm II}})}{p_{\rm ss}(\bq_{\rm I})}$ is the change in the environment entropy during the transition from state ${\rm I}$ to state ${\rm II}$.
Equation (\ref{eq:fpttxla}) relates the waiting-time distributions between two states with the environment-entropy  change along trajectories connecting both states.   

We normalize the distributions in Eq.~(\ref{eq:fpttxla}), and find a relation for the normalized waiting-time distributions
\begin{eqnarray}
p_{T^{{\rm I}\rightarrow {\rm {\rm II}}}_+}(t|s_{\rm env} ) =p_{T^{{\rm {\rm II}}\rightarrow {\rm I}}_-}(t|-s_{\rm env} )\quad,\label{eq:envRx}
\end{eqnarray}
and for the associated passage probabilities: 
\begin{eqnarray}
\frac{\mathsf{P}^{{\rm I}\rightarrow {\rm {\rm II}}}_+}{\mathsf{P}^{{\rm {\rm II}}\rightarrow {\rm I}}_-} = e^{s_{\rm env}/k} \quad. \label{eq:envRxx}
\end{eqnarray} 
Interestingly, the relations  Eqs.~(\ref{eq:fpttxla}), (\ref{eq:envRx}) and (\ref{eq:envRxx}) are similar to the waiting-time relations (\ref{eq:fpt}), (\ref{eq:synnp}) and (\ref{eq:Integral}) discussed above.   However, in Eq.~(\ref{eq:fpttxla}) and (\ref{eq:envRxx}) the environmental entropy production occurs, instead of the total entropy production, because the  trajectories are conditioned on passage through  initial and final states.     For isothermal processes, $s_{\rm env} = -Q/\mathsf{T}_{\rm env}$, with $Q$ the heat absorbed by the system   and  $\mathsf{T}_{\rm env}$ the temperature of the environment.

\section{Application to Simple Colloidal Systems}\label{sec:7}
\begin{figure}
\includegraphics[width=7.5cm]{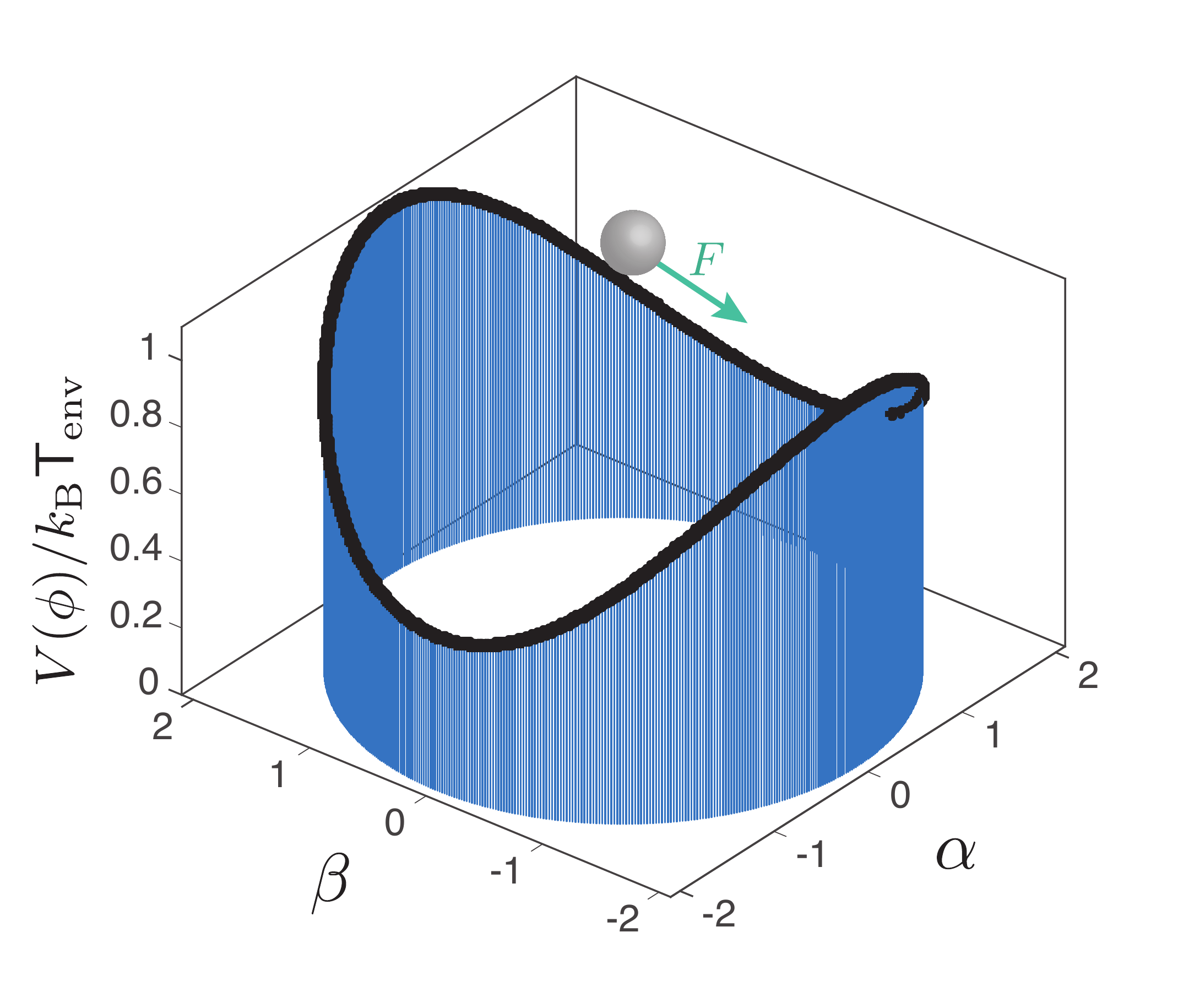} 
\caption{Illustration of a Smoluchowski--Feynman ratchet. A Brownian particle (gray sphere)  immersed in a thermal bath of temperature $\mathsf{T}_{\rm env}$  moves in a periodic potential $V(\phi)$ (black shaded curve)  with friction coefficient $\gamma$.   The coordinate $\phi$ is the azimuthal angle of the particle. When applying an external force $F=\gamma v$ in the azimuthal direction, the particle reaches a nonequilibrium steady state. In this example, $V(\phi) = k_{\rm B}\mathsf{T}_{\rm env}\ln [\cos(\phi)+2]$, $\alpha=R\cos (\phi)$ and $\beta=R\sin(\phi)$, with $R=2$. 
\label{fig:potential}}
\end{figure}

Infima of entropy production fluctuations and stopping times can be calculated for specific stochastic processes.  In this section we discuss these quantities for the dynamics of  a colloidal particle  with  diffusion coefficient $D$ that moves in a periodic potential $V$, with period $\ell$, and under the influence of a constant external force $F$ \cite{reimann2002brownian,hanggi2009artificial} (see Fig.~\ref{fig:potential} for a graphical illustration).   This process has been realized  in several experiments using colloidal particles trapped with  toroidal optical potentials~\cite{speck2007distribution,blickle2007characterizing,gomez2009experimental,gomez2011fluctuations}.     We discuss how our results can be tested in this type of experiments.

 We describe the dynamics of this colloidal particle in terms of a one-dimensional overdamped Brownian motion 
 with periodic boundary conditions.    The state of the particle at time $t$ 
is characterized by a phase variable $\phi(t)\in [0,\ell)$.  In the illustration of a ring geometry   in Fig.~\ref{fig:potential}, $\phi$ is the azimuthal angle and $\ell = 2\pi$.  
Equivalently, one can consider a stochastic process $X(t)$ given by the net distance traveled by the Brownian particle up to time $t$: $X(t)=\phi(t) + \ell  N(t)$, where $N(t)$ is the winding number or the net number of clockwise turns (or minus the number of counterclockwise turns) of the particle up to time $t$~\cite{saito2015waiting}. The time evolution of $X(t)$ is obeys the Langevin equation
   \begin{eqnarray}
 \frac{\text{d}X(t)}{\text{d}t} = -\frac{1}{\gamma}\frac{\partial V(X(t))}{\partial x}  + v  + \zeta(t)\quad, 
 \label{eq:LESF}
 \end{eqnarray}
where $\gamma$ is a friction coefficient, $v= F/\gamma$ is the drift velocity,  and  $\zeta$ is  a Gaussian white noise with zero mean $\langle\zeta(t)\rangle=0$ and with autocorrelation $\langle \zeta(t)\zeta(t')\rangle= 2D\delta(t-t')$.   If the Einstein relation holds, $D = k_{\rm B}\mathsf{T}_{\rm env}/\gamma$, with $\mathsf{T}_{\rm env}$ the temperature of the thermal reservoir.   Here  $V(x)$ is the periodic potential of period $\ell$, $V(x+\ell)=V(x)$, and $\frac{\partial V(X(t))}{\partial x} = \left.\frac{\partial V(x)}{\partial x}\right|_{X(t)}$. 

 The steady-state entropy production after a time $t$ is~\cite{reimann2002brownian,gomez2006test}:
 \begin{eqnarray}
 S_{\rm tot}(t) = k_{\rm B}\ln \frac{\displaystyle\int_{X(0)}^{X(0)+\ell} \text{d}y\, e^{(V(y)-Fy)/k_{\rm B}\mathsf{T}_{\rm env}}}{\displaystyle\int_{X(t)}^{X(t)+\ell} \text{d}y\, e^{(V(y)-Fy)/k_{\rm B}\mathsf{T}_{\rm env}}}\quad.
 \label{eq:s}
 \end{eqnarray}
For a drift-diffusion process, with $V(x)=0$, the total entropy production reads 
\begin{equation}
S_{\rm tot}(t) = k_{\rm B}\frac{v}{D}(X(t)-X(0))\quad.
\label{eq:SDD}
\end{equation}
Equation (\ref{eq:SDD}) implies that the first-passage and extreme-value statistics of entropy production in the drift-diffusion process follow from the statistics of the position $X(t)$ of a drifted  Brownian particle in the real line.   The drift-diffusion process is an example for which the infimum  and first-passage statistics of entropy production can be calculated analytically.     We also consider a Smoluchowski-Feynman ratchet whose dynamics is given by  Eq.~(\ref{eq:LESF}) with the  nonzero potential,
 \begin{eqnarray}
V(x) = V_0\: \ln [\cos(2\pi x/\ell) + 2]\quad,
\label{eq:pot}
\end{eqnarray}
as illustrated in Fig.~\ref{fig:potential}.    For the potential (\ref{eq:pot}) with $V_0 = k_{\rm B}\mathsf{T}_{\rm env}$ the stochastic entropy production in steady state, given by Eq.~\eqref{eq:s}, equals \cite{gomez2006test}
 \begin{equation}
 \frac{ S_{\rm tot}(t)}{k_{\rm B}} =f\,(X(t)-X(0)) - \ln \frac{\psi(X(t),f)}{\psi(X(0),f)}\quad,
 \label{eq:stota}
 \end{equation}
with $f=F\ell/(2\pi k_{\rm B}\mathsf{T}_{\rm env})$ and $\psi(x,f) = f^2[\cos(2\pi x/\ell) + 2] - f\sin(2\pi x/\ell) + 2$.    

In the following we present analytical results for the infima and passage statistics of entropy production in the drift-diffusion process, and we present simulation results for the Smoluchowski-Feynman ratchet with the potential (\ref{eq:pot}).   The simulation results are for parameters corresponding to experimental conditions in optical tweezers experiments \cite{gomez2009experimental, gomez2011fluctuations, blickle2007characterizing}, namely, for a polystyrene spherical Brownian particle of radius 1 $\mu${\rm m} immersed in water at room temperature.  

\begin{figure}
\includegraphics[width=7cm]{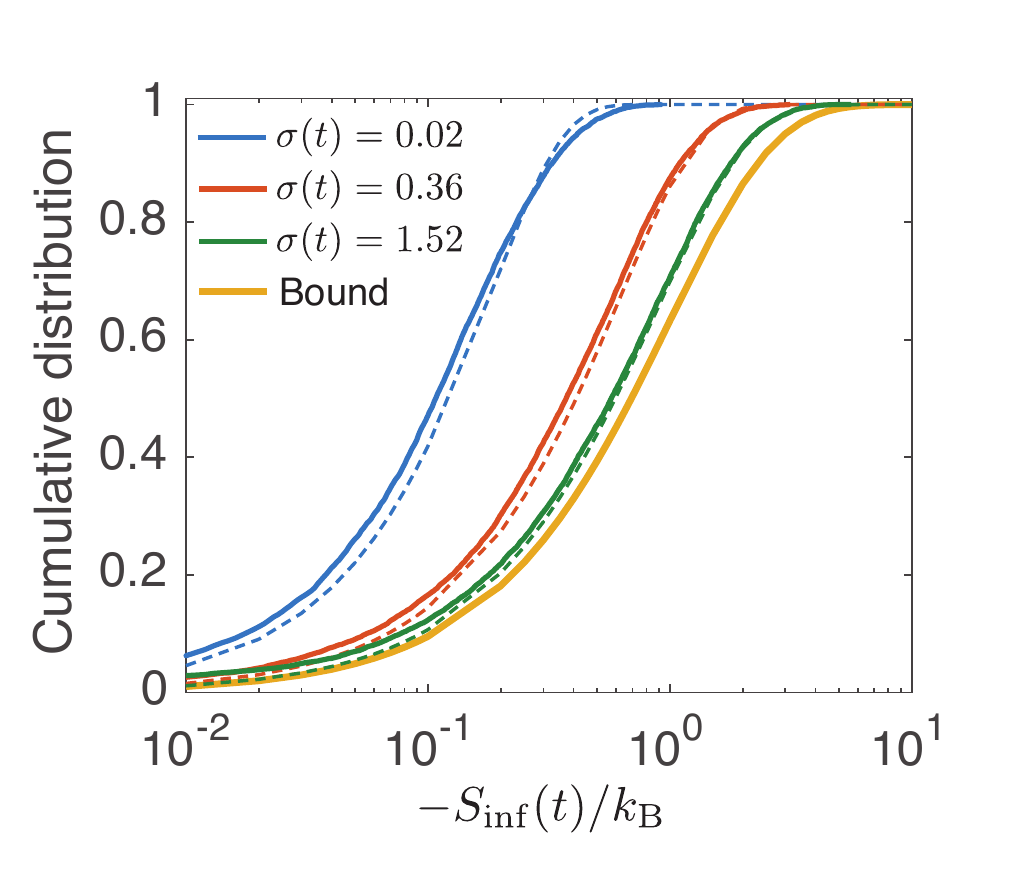} 
\caption{Cumulative distributions of the infimum of  entropy production  for a Smoluchowski--Feynman ratchet in steady state for different values of the mean entropy change $\sigma(t) = \langle S_{\rm tot}(t)\rangle/k_{\rm B}$. 
Simulation results for the Smoluchowski-Feynmann ratchet  with the potential (\ref{eq:pot}) (solid lines) are compared with analytical results 
for the drift-diffusion process given by Eq.~\eqref{eq:infCDFS} (dashed lines), and with the universal bound given by $1-e^{-s}$ (rightmost yellow curve). Simulation parameters: $10^4$ simulations with   $\mathsf{T}_{\rm env}=300\,\rm K$, $\ell=2\pi\,\rm nm$, $\gamma = 8.4\,\rm pN s /nm$, $V_0 = k_{\rm B} \mathsf{T}_{\rm env}$, simulation time step $\Delta t= 0.00127\,\rm s$ and total simulation time $t=0.4\,\rm s$.   The external force $F$, and the drift velocity $v$, are determined by the average entropy production $\sigma(t)$.
\label{fig:CDFINFSF}}
\end{figure}

\begin{figure}
\includegraphics[width=7.1cm]{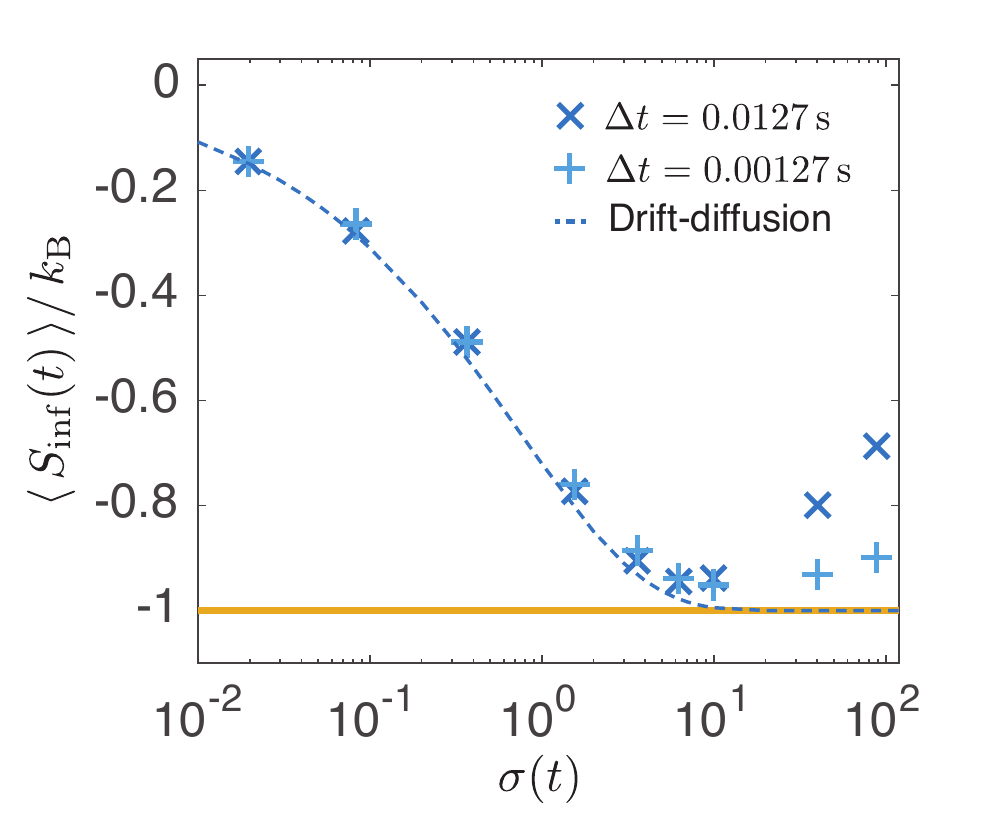} 
\caption{Mean of the local  entropy-production infimum $\langle S_{\rm inf}(t)\rangle $  as a function of the mean entropy change $\sigma(t) = \langle S_{\rm tot}(t)\rangle/k_{\rm B}$   for a Smoluchowski--Feynman ratchet in steady state.    Simulation results for the Smoluchowski-Feynmann ratchet  with the potential (\ref{eq:pot}) (symbols) are compared with analytical results 
for the drift-diffusion process given by Eq.~(\ref{eq:INFSDD}) (dashed lines), and with the universal infimum law given by $-k_{\rm B}$ (yellow thick bar). 
   Different symbols are obtained for  different simulation time steps $\Delta t$.  Simulation parameters: $10^4$ simulations with  $\mathsf{T}_{\rm env}=300\,\rm K$, $\ell=2\pi\,\rm nm$,  $\gamma = 8.4\,\rm pN s /nm$,  $V_0 = k_{\rm B} \mathsf{T}_{\rm env}$, and total simulation time $t=0.4\,\rm s$. 
\label{fig:MEANINFSF}}
\end{figure}

\subsection{Infimum statistics}
We now study the infimum properties of entropy production for the Smoluchowski--Feynman ratchet  in steady state.   In Fig.~\ref{fig:CDFINFSF}  we show that the cumulative distribution of the local entropy-production infimum $S_{\rm inf}(t)$ is bounded from below by $1-e^{-s}$, 
which confirms the universality of Eq.~(\ref{eq:infxx}).   We compare analytical results for the drift-diffusion process ($V(x)=0$, dashed lines) with numerical results for the Smoluchowski-Ferynman ratchet (with a potential $V(x)$ given by Eq.~(\ref{eq:pot}), solid lines), for different values of the mean entropy production $\sigma (t)=\langle S_{\rm tot}(t)\rangle/k_{\rm B}$.   The analytical expression for the cumulative distribution of $S_{\rm inf}$ is, for the drift-diffusion process, given by 
 (see Appendix~\ref{app:BPI}):
\begin{eqnarray}
\lefteqn{ \mathsf{Pr}\left( \frac{S_{\rm inf}(t)}{k_{\rm B}} \geq -s\right) }&& \label{eq:infCDFS}\\
&=& \frac{1}{2}\left[ \text{erfc}\left(\frac{-s-\sigma (t)}{2\sqrt{\sigma (t)}}  \right) - e^{-s}\text{erfc}\left(\frac{s-\sigma (t)}{2\sqrt{\sigma (t)}}  \right)   \right]\;,\nonumber  
\end{eqnarray}
where $s>0$, ${\rm erfc}(x) = \left(2/\sqrt{\pi}\right)\int^{+\infty}_xe^{-y^2}{\rm d}y$ is the complementary error function, and $ \sigma (t) $  is the average entropy production in steady state at time $t$, which for the drift-diffusion process is $\sigma (t) = (v^2/D) t$.    Interestingly, the bound saturates  
for large values of the average entropy production $\sigma (t)$, which illustrates the universal equality on the distribution of the global infimum of entropy production, given by Eq.~(\ref{eq:infxxxx}).   Remarkably, as shown in Fig.~\ref{fig:CDFINFSF} the  cumulative distribution for the infimum of entropy production of the Smoluchowski--Feynman ratchet  is  nearly identical for different shapes of the potential $V(x)$. This equivalence between the infimum cumulative distributions holds even for small values of  $\sigma(t)$  where the shape of the potential $V(x)$ affects  the entropy-production fluctuations.

Secondly, in Fig.~\ref{fig:MEANINFSF} we  illustrate the infimum law, $\langle\,S_{\rm inf}(t)\,\rangle\geq -k_{\rm B}$,  
for the Smoluchowski--Feynman ratchet.      We show the average local infimum $\langle\,S_{\rm inf}(t)\,\rangle$ as a function of the mean entropy production $\sigma(t)$:  we compare analytical results for the drift-diffusion process without potential (dashed lines) with numerical results for the Smoluchowski-Ferynman ratchet with a potential given by Eq.~(\ref{eq:pot}) (solid lines).      
The analytical expression for the drift-diffusion process is (see Appendix \ref{app:BPI}):
\begin{eqnarray}
\lefteqn{\left\langle  \frac{S_{\rm inf}(t)}{k_{\rm B}}\right\rangle} && \label{eq:INFSDD} \\
&=&-\text{erf}\left[\frac{\sqrt{\sigma(t)}}{2}\right] +\frac{\sigma(t)}{2}\text{erfc}\left[\frac{\sqrt{\sigma(t)}}{2}\right]- \sqrt{\frac{\sigma(t)}{\pi}}e^{-\sigma(t)/4} \,,\nonumber
\end{eqnarray}
where ${\rm erf}(x) = 1-{\rm erfc}(x) $ is the error function.    In the limit of large times, the global infimum $\langle\, S^{\rm \infty}_{\rm inf}\,\rangle =-k_{\rm B}$, in accordance to the universal equality (\ref{eq:infxxx}).   The results in Fig.~\ref{fig:MEANINFSF}    show that the mean local infimum  of entropy production has the same functional dependency on $\sigma(t)$ independent of the potential $V(x)$.     This  points towards a universal behaviour of the statistics of  local infima of entropy production.

\subsection{Passage probabilities and first-passage times}

\begin{figure}
\includegraphics[width=7.5cm]{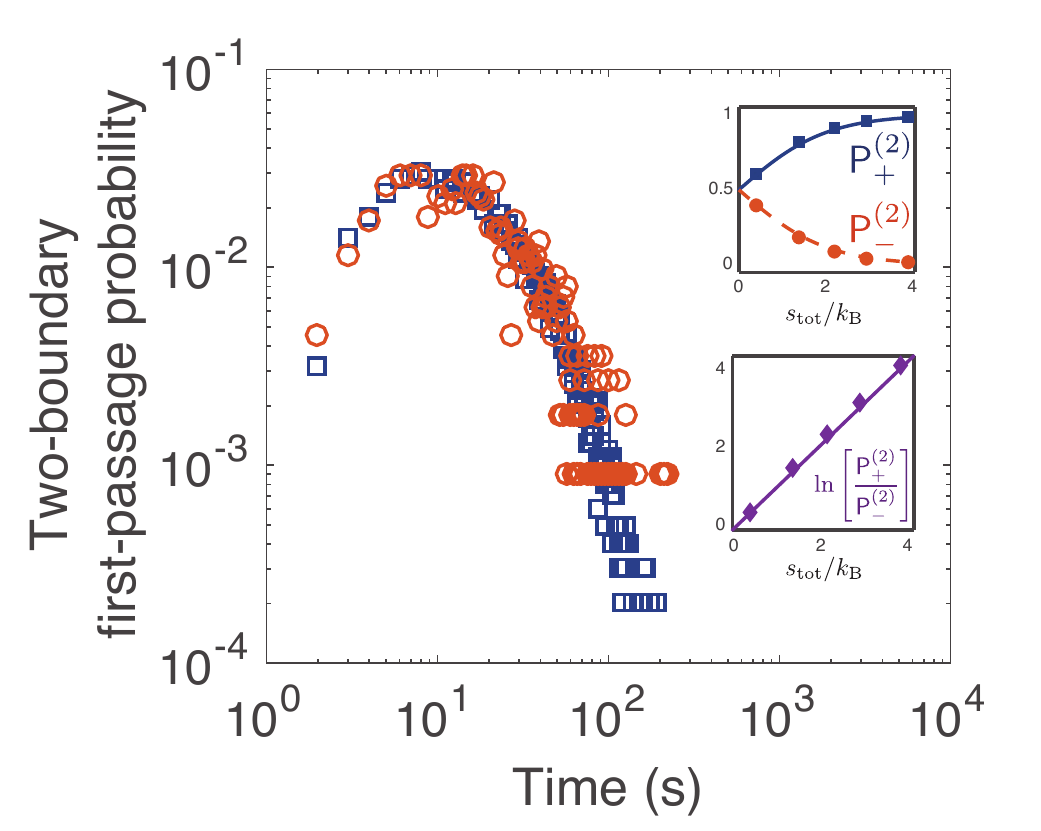} 
\caption{Empirical two-boundary first-passage-time distributions of entropy production to first reach a positive threshold $p_{T^{(2)}_+}(t;s_{\rm tot})$ (blue squares) and the rescaled distribution for the negative threshold  $p_{T^{(2)}_-}(t;- s_{\rm tot})e^{s_{\rm tot}/k_{\rm B}}$  (red circles) for the Smoluchowski-Feynman ratchet with the potential (\ref{eq:pot}) in steady state. The distributions are obtained from $10^4$ numerical simulations  and the threshold values are set to  $\pm s_{\rm tot}=\pm 2.2 k_{\rm B}$. The simulations are done using Euler numerical scheme with with the following parameters:  $F=4\,\rm pN$, $\mathsf{T}_{\rm env}=300\,\rm K$, $\ell=2\pi\,\rm nm$, $\gamma = 8.4\,\rm pN s /nm$,  $V_0 = k_{\rm B} \mathsf{T}_{\rm env}$,  and simulation time step $\Delta t= 0.0127\,\rm s$.  {\em Top inset}: Empirical passage probabilities of entropy production to first reach the positive threshold ($\mathsf{P}^{(2)}_+$, blue squares) and to first reach  the negative  threshold ($\mathsf{P}^{(2)}_-$, red circles)  as a function of the threshold value $s_{\rm tot}$. The analytical expressions for $\mathsf{P}^{(2)}_+$ (blue solid line), given by Eq.~\eqref{eq:Integralx}, and for $\mathsf{P}^{(2)}_-$  (red dashed line), given by Eq.~\eqref{eq:Integralxx}, are also shown.  {\em Bottom inset}: Logarithm of the ratio between the empirical passage probabilities $\mathsf{P}^{(2)}_+$ and $\mathsf{P}^{(2)}_-$ as a function of the threshold value $s_{\rm tot}$ (magenta diamonds). The solid line is a straight line of slope~1. 
\label{fig:twobounds}}
\end{figure}

\begin{figure}
\includegraphics[width=7.5cm]{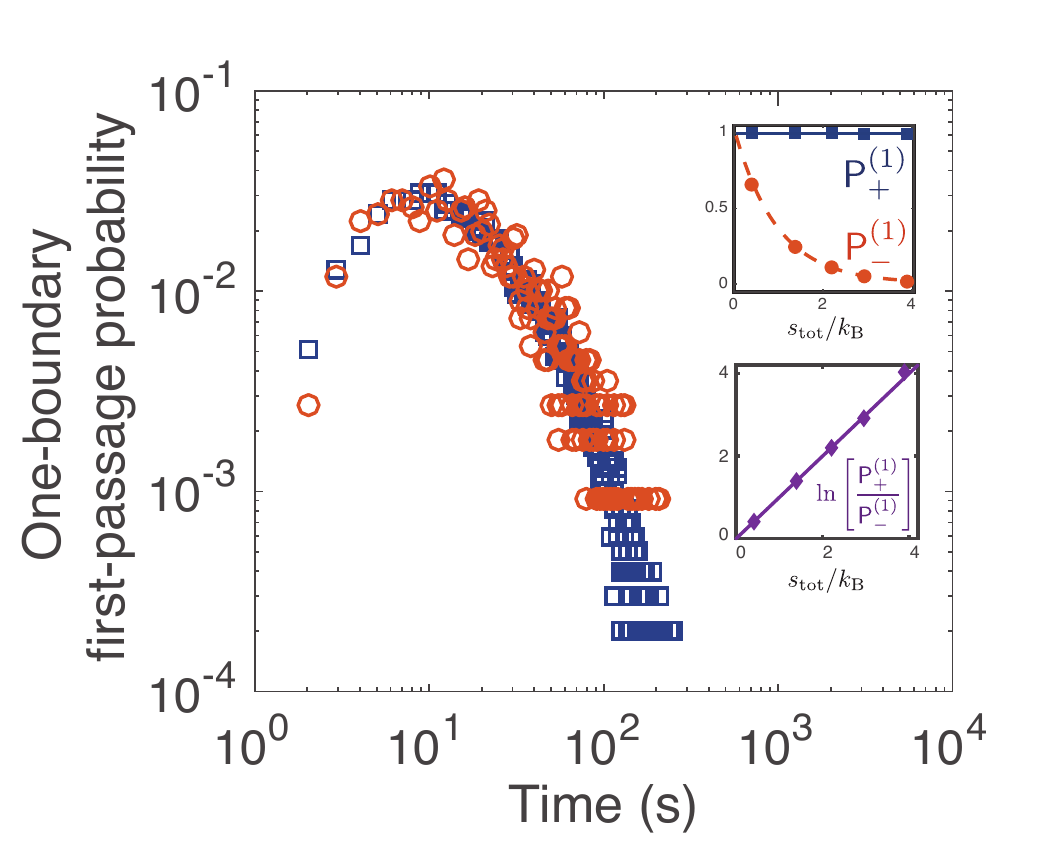} 
\caption{Empirical one-boundary first-passage-time distributions of entropy production to first reach a positive threshold $p_{T^{(1)}_+}(t;s_{\rm tot})$ (blue squares) and  the rescaled distribution for the negative threshold  $p_{T^{(1)}_-}(t;- s_{\rm tot})e^{s_{\rm tot}/k_{\rm B}}$  (red circles) for a Smoluchowski-Feynman ratchet with the potential (\ref{eq:pot}) in steady state. The estimate of $p_{T^{(1)}_+}(t;s_{\rm tot})$ ($p_{T^{(1)}_-}(t;-s_{\rm tot})$) is obtained measuring the time elapsed by the entropy production to first reach a single absorbing boundary in $s_{\rm tot}=  2.2 k_{\rm B}$ ($-s_{\rm tot}=  -2.2 k_{\rm B}$)  in $10^4$ simulations. The simulations are done with the same parameters as in Fig.~\ref{fig:twobounds}, and the empirical probabilties are calculated over a total simulation time of $\tau_{\rm max } = 20\,{\rm s}$. {\em Top inset}: Empirical passage probabilities of entropy production in the positive-threshold simulations ($\mathsf{P}^{(1)}_+$, blue squares) and in  the negative-threshold simulations ($\mathsf{P}^{(1)}_-$, red circles)  as a function of the value of the thresholds. The expressions $\mathsf{P}^{(1)}_+$ given by Eq.~\eqref{eq:IntegralxT} (blue solid line) and $\mathsf{P}^{(1)}_-$ given by Eq.~\eqref{eq:IntegralxxT} (red dashed line) are also shown.  {\em Bottom inset}: Logarithm of the ratio between $\mathsf{P}^{(1)}_+$ and $\mathsf{P}^{(1)}_-$ as a function of the threshold value (magenta diamonds). The solid line is a straight line of slope~1. 
\label{fig:onebound}}
\end{figure}

\begin{figure}
\includegraphics[width=0.4\textwidth]{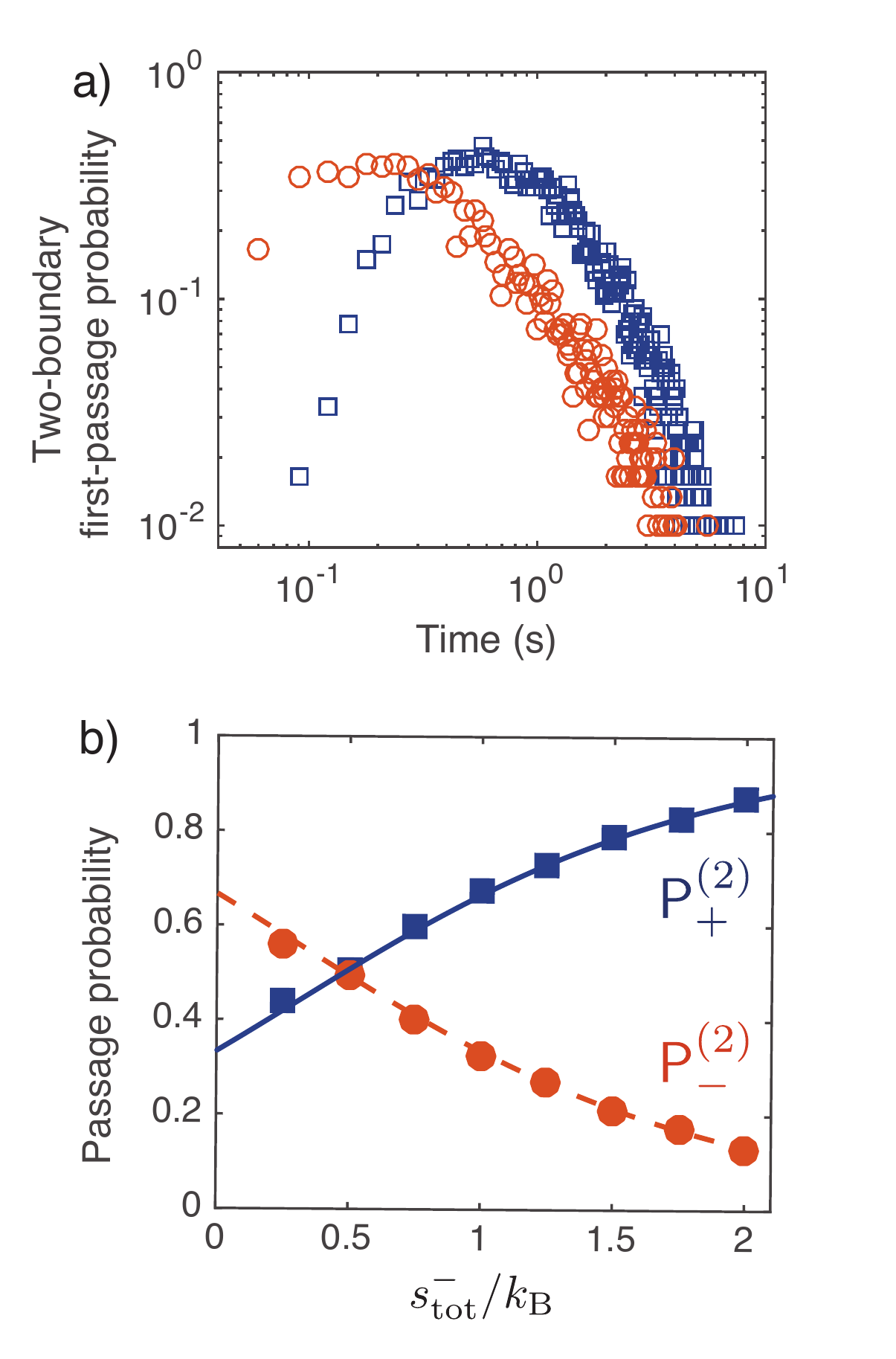} 
\caption{ First-passage statistics of entropy production with two absorbing and asymmetric  boundaries at values $s^+_{\rm tot}=2s^-_{\rm tot}$ for the Smoluchowski-Feynman ratchet with the potential (\ref{eq:pot}).   a) The first-passage-time distribution $p_{T^{(2)}_+}(t;s^+_{\rm tot})$ (blue open squares),  for entropy production to first reach $s^+_{\rm tot}=2k_{\rm B}$ given that it has not reached $-s^-_{\rm tot}=-k_{\rm B}$ before, and the first-passage-time distribution  $p_{T^{(2)}_-}(t;-s^-_{\rm tot})$  (red open circles), to first reach $-s^-_{\rm tot}$  given that it has not reached $s^+_{\rm tot}$ before.  The  data is from $10^4$ simulations of the Smoluchowski--Feynman ratchet using the same simulation parameters as in Fig.~\ref{fig:twobounds}. b) Passage probabilities for entropy production $\mathsf{P}^{(2)}_+$ (blue filled squares), to first reach the positive threshold  $s^+_{\rm tot}$ given that it has not reached  $-s^-_{\rm tot}$ before, and  passage probability for  entropy production $\mathsf{P}^{(2)}_-$    (red filled circles), to first reach the positive threshold  $-s^-_{\rm tot}$ given that it has not reached  $s^+_{\rm tot}$ before.   We show passage probabilities as a function of the negative threshold value $s^-_{\rm tot}$,  and with a positive threshold given by  $s^+_{\rm tot}=2s^-_{\rm tot}$.     The curves represent the analytical expressions for the passage proabilities  given by Eqs.~(\ref{eq:pass1xx})-(\ref{eq:pass2xx})  ($\mathsf{P}^{(2)}_+$ blue solid curve, $\mathsf{P}^{(2)}_-$ red dashed curve).   
\label{fig:asym}}
\end{figure}

\begin{figure}
\includegraphics[width=0.4\textwidth]{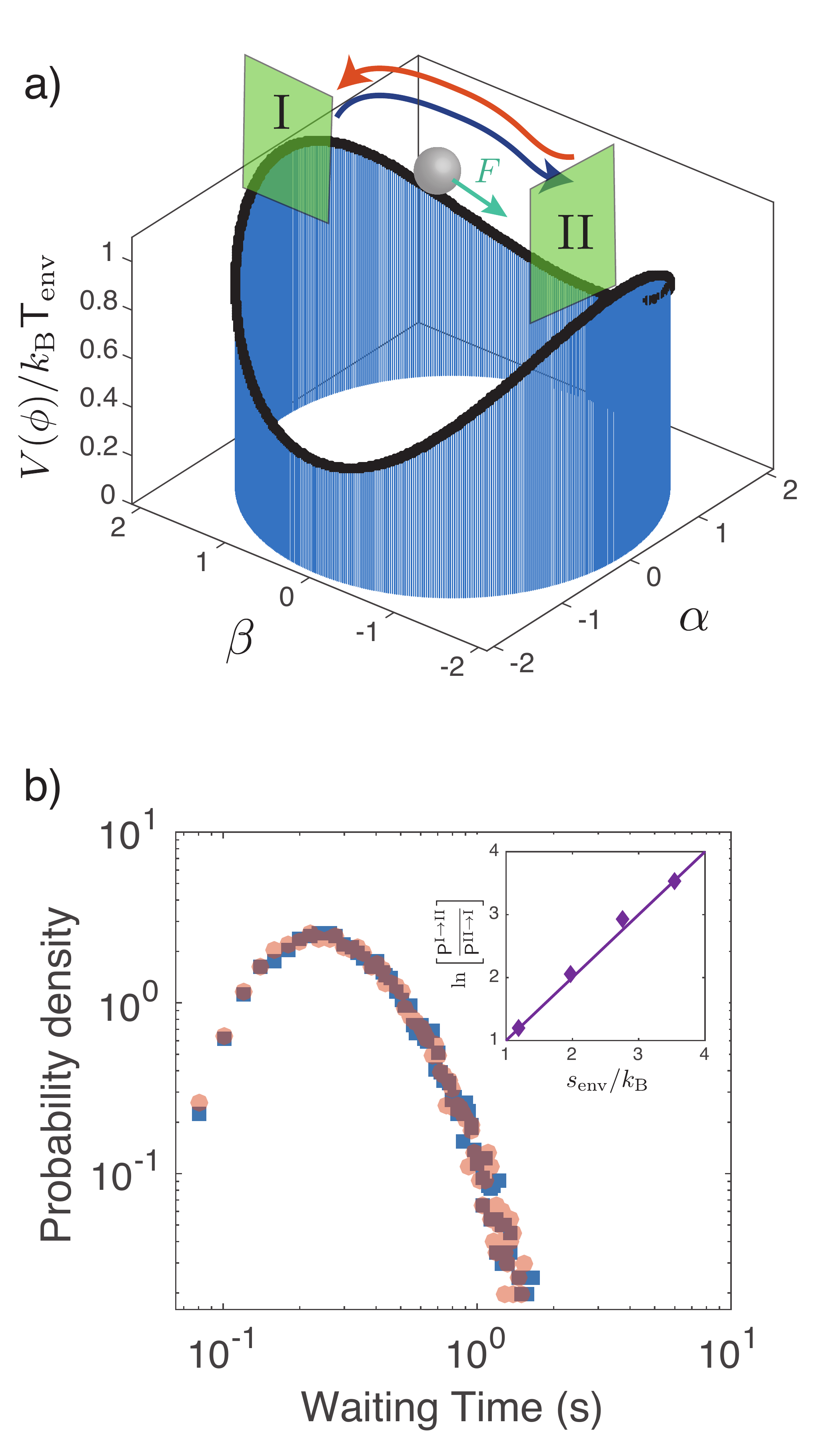} 
\caption{Waiting-time statisticsbetween the states $X_{\rm I}$ and $X_{\rm II}$ as illustrated in panel a)  for the Smoluchowski-Feyman ratchet with the potential (\ref{eq:pot}).   In panel b) 
 simulation results are shown for the normalized waiting-time distributions of transitions $\text{I}\to\text{II}$ (blue squares) and of reversed transitions $\text{II}\to\text{I}$ (red circles).   The simulations are done with the same paramerers as in Fig. 7, and the distributions are obtained from $10^4$ trajectories. Inset: Logarithm of the ratio between $\mathsf{P}^{\text{I}\to\text{II}}_+$ and $\mathsf{P}^{\text{II}\to\text{I}}_-$ as a function of the environment-entropy change in the transition $\text{I}\to\text{II}$ (magenta diamonds). The transition probability $\mathsf{P}^{\text{I}\to\text{II}}_+ (\mathsf{P}^{\text{II}\to\text{I}}_-)$ is the fraction of the trajectories starting from $X_{\rm I}$ $(X_{\rm II})$  at $t=0$ that reach $X_{\rm II}$   $(X_{\rm I})$ at a later time $t>0$ without returning to  $X_{\rm I}$ $(X_{\rm II})$  before.   The solid line is a straight line of slope 1.   
   The data in the inset is obtained from $10^5$ simulations starting from the state I, and  $10^5$ simulations starting from the state II.  
\label{fig:WT}}
\end{figure}
\subsubsection{Symmetric boundaries}
We illustrate our universal results on passage probabilities and first-passage times for the Smoluchowski-Feynman ratchet. 

 For the drift-diffusion process ($V(x)=0$),  we recover 
the first-passage-time fluctuation theorem for entropy production, given by  Eq.~\eqref{eq:FPTime}, from the analytical expressions of the first-passage-time distributions for the position of a Brownian particle (see Appendix \ref{app:F}).

We also compute, using numerical simulations,  the first-passage-time statistics for entropy production of the Smoluchowski-Feynman ratchet in steady state and with potential $V(x)$ given by Eq.~(\ref{eq:pot}).
First we study the first-passage times $T^{(2)}_{\pm}$ for entropy production   with two absorbing boundaries  at the threshold values  $s_{\rm tot}$ and $-s_{\rm tot}$ (with $s_{\rm tot}>0$).
 Figure~\ref{fig:twobounds} shows the empirical first-passage-time distribution $p_{T^{(2)}_+}(t; s_{\rm tot})$ to first reach the positive threshold (blue squares) together with the first-passage time distribution $p_{T^{(2)}_-}(t;- s_{\rm tot})$ to first reach the negative threshold (red circles, the distribution is rescaled by $e^{ s_{\rm tot}/k_{\rm B}}$). Since both distributions coincide we confirm the validity of the first-passage-time fluctuation theorem given by Eq.~\eqref{eq:FPTime}.  Moreover, the functional dependency of the empirical passage probabilities     $\mathsf{P}^{(2)}_+$ and  $\mathsf{P}^{(2)}_-$ on the threshold value $s_{\rm tot}$ obeys the expressions given by Eqs.~\eqref{eq:Integralx} and~\eqref{eq:Integralxx} (see top inset in Fig.~\ref{fig:twobounds});  analytical expressions for $\mathsf{P}^{(2)}_+$ and  $\mathsf{P}^{(2)}_-$ can be obtained using the method in \cite{Gichman, lindsey1977complete}.
   As a result, the integral first-passage-time fluctuation theorem given by Eq.~\eqref{eq:Integral} is also fulfilled in this example (see bottom inset in Fig.~\ref{fig:twobounds}).    
 lindsey1977complete

As a second case  we consider two  one-boundary first-passage problems for entropy production.  
We study the first-passage times  $T^{(1)}_{+}$ for entropy production   with one absorbing boundary  at the threshold value  $s_{\rm tot}$, and the corresponding 
first-passage time  $T^{(1)}_{-}$ at the negative threshold value 
$-s_{\rm tot}$.   We obtain numerical estimates of the distribution   $p_{T^{(1)}_{+}}(t; s_{\rm tot})$ and its conjugate distribution $p_{T^{(1)}_{-}}(t;- s_{\rm tot})$. Figure~\ref{fig:onebound} shows empirical estimates of  these   first-passage-time distributions  and confirms the validity of the first-passage-time fluctuation theorem given by Eq.~\eqref{eq:FPTime}.    In the top inset  of Fig.~\ref{fig:onebound} we show that the 
 the passage probabilities $\mathsf{P}^{(1)}_{+}$ and $\mathsf{P}^{(1)}_{-}$  are given by the analytical expressions in Eqs.~(\ref{eq:IntegralxT})-(\ref{eq:IntegralxxT}), and in the bottom inset, we verify the integral first-passage-time fluctuation theorem~given by Eq.~\eqref{eq:Integral} for one boundary first-passage processes.

\subsubsection{Asymmetric boundaries}
We now discuss the passage statistics for entropy production with asymmetric boundaries.  
In Appendix \ref{app:F} we discuss the drift-diffusion process, whereas here we discuss the 
the Smoluchowski-Feynman rachet with a potential, given by Eq.~(\ref{eq:pot}).  In Fig.~\ref{fig:asym}(a) we show  the  distributions of the first-passage times $p_{T^{(2)}_+}(t;s^+_{\rm tot})$ and $p_{T^{(2)}_-}(t;-s^-_{\rm tot})$ with two boundaries located at    $s^+_{\rm tot} = 2 k_{\rm B}$ and  $-s^-_{\rm tot} = -k_{\rm B}$.   Interestingly, the ratio of the two distributions $p_{T^{(2)}_+}(t;s^+_{\rm tot})/p_{T^{(2)}_-}(t;-s^-_{\rm tot})$ is  time dependent.  Therefore  the  first-passage-time fluctuation relation, given by Eq.~(\ref{eq:FPTime}), does not extend to asymmetric boundaries.  Consequently, for asymmetric boundaries the mean first-passage times $\langle T^{(2)}_+\rangle$ and   $\langle T^{(2)}_- \rangle$ to first reach, respectively, the positive and negative boundary, are in general different.    

  In Fig.~\ref{fig:asym}(b) we show numerical results for the entropy-production passage probabilities $\mathsf{P}^{(2)}_+$ and   $\mathsf{P}^{(2)}_-$  as a function of the value of the negative threshold $s^-_{\rm tot}$.    Simulation results are in accordance with the  universal expressions given by Eqs.~(\ref{eq:pass1xxx})-(\ref{eq:pass2xxx}).

\subsection{Fluctuation theorem in waiting times}
We illustrate the waiting-time fluctuation theorem, given by Eq.~(\ref{eq:9})  (or Eq.~(\ref{eq:fpttxla})), on the Smoluchowski-Feynman ratchet.  We compute, using numerical simulations, the  waiting times along forward trajectories ${\rm I}\rightarrow {\rm II}$ and backward trajectories ${\rm II}\rightarrow {\rm I}$ between two states characterised, respectively, by the coordinates $X = X_{\rm I}$ and  $X = X_{\rm II}$, as illustrated in Fig.~\ref{fig:WT}a).     In agreement with the fluctuation theorem for waiting times we find that the  normalized distribution $p_{T^{{\rm I}\rightarrow {\rm {\rm II}}}_+}(t|s_{\rm env} )$  is equal to  the normalized distribution $p_{T^{{\rm {\rm II}}\rightarrow {\rm I}}_-}(t|-s_{\rm env} )$ (see Fig.~\ref{fig:WT}b)).    Here the environment-entropy change is determined by the heat exchange between system and environment, i.e.,  $s_{\rm env} = -Q/\mathsf{T}_{\rm env} = F(X_{\rm II}-X_{\rm I})/\mathsf{T}_{\rm env}  +(V(X_{\rm I})-V(X_{\rm II}))/\mathsf{T}_{\rm env}$. 
  In the inset of Fig.~\ref{fig:WT}b)  we show simulation results  for the ratio  of passage probabilities, which is in agreement with our theoretical result Eq.~(\ref{eq:envRxx}), i.e.,   $\mathsf{P}^{{\rm {\rm I}}\rightarrow {\rm  II}}_+/\mathsf{P}^{{\rm {\rm II}}\rightarrow {\rm  I}}_- = e^{-Q/k_{\rm B}\mathsf{T}_{\rm env}}$.

\section{Applications  to  active molecular processes} \label{sec:8}
In contrast to the driven colloidal particles discussed in the last section, which is best viewed as a 
a continuous stochastic process,  many biochemical  processes are often described in terms of  transitions between discrete states.     Examples are the motion of a molecular motor on a substrate with discrete binding sites, or a chemical reaction  that turns reactants  into products.  
The statistics of waiting times of discrete processes can be obtained by a coarse-graining procedure of continuous processes.   We  
apply our theory to a chemical driven  hopping process with one degree of freedom  and to the dynamics of  RNA polymerases described by two degrees of freedom.

\begin{figure}
\includegraphics[width=7.5cm]{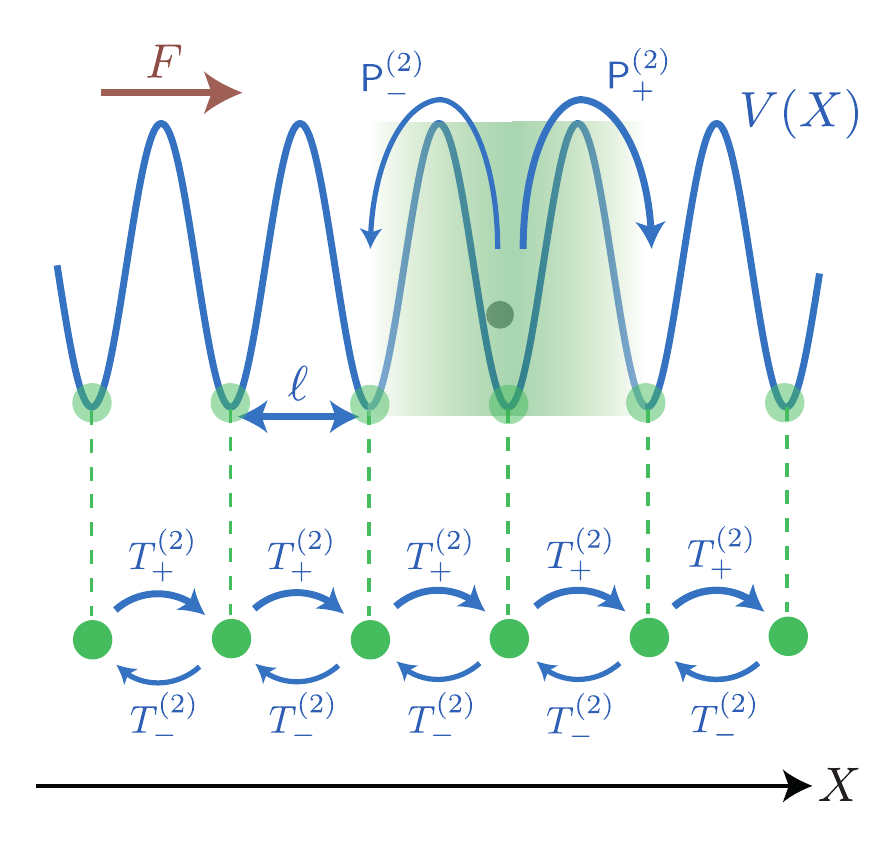} 
\caption{Coarse-graining procedure from a continuous Langevin process (top) to a discrete Markov process (bottom).  The horizonal axis denotes either a chemical coordinate, which quantifies the progress of a chemical reaction, or a position coordinate, which quantifies the position of a molecular motor.   In the Langevin description this coordinate is described by the position of  a Brownian particle (grey circle) moving in a periodic potential $V(X)$ of period $\ell$ (blue curve) and driven by an external bias $F$.  In the discrete process this coordinate is described by the state of  a Markov jump process with stepping rates $k_+$ and $k_-$, which are related to the waiting times and the passage probabilities through $k_+ = P^{(2)}_+\langle 1/T^{(2)}_+ \rangle$ and $k_- = P^{(2)}_-\langle 1/T^{(2)}_- \rangle$.   Our coarse-graining procedure corresponds to a partitioning of the coordinate axis into intervals (green shaded areas) centered to the minima of the potential (dashed lines).}\label{fig:chemical}
\end{figure}

\subsection{From continuous to discrete processes}
We  can apply  the theory developed above to systems which progress in discrete steps using a coarse-graining procedure \cite{keller2000mechanochemistry, hanggi1990reaction}.
  We consider a single continuous  mesoscopic degree of freedom $X$,
which can be used to describe a stochastic cyclic process.
The variable  $X$ can be interpreted as a generalized reaction coordinate or the position  coordinate of a molecular motor,  and  obeys  the Langevin equation (\ref{eq:LESF}) with the same noise correlations.   The effective potential $V(X)$ now describes the free energy profile along a chemical reaction or position coordinate  of a molecular motor.

 We coarse grain the continuous variable $X$  by considering  events when the particle passes discrete points at positions $X_M$.    The transition times between these points are examples of waiting times, similar to those in Kramers' theory \cite{kramers1940brownian}.  An example is shown in Fig.~\ref{fig:chemical}, for which the points $X_M$ are located at the minima of the potential $V$.   We introduce the transition times $T_{M\rightarrow M+1}$  to reach the final state $X_{M+1}$, for the first time, 
 starting from the initial state $X_M$  and allowing several passages through $X_M$.   Similarily, we define $T_{M+1 \rightarrow M}$  for the reverse process.    The entropy change associated with a transition is $s_{\rm tot} = F \ell/\mathsf{T}_{\rm env}$.    Entropy production is therefore related to position $X_M$ by: 
\begin{equation}
S_{\rm tot} (t) =    -N(t) \frac{F\ell }{\mathsf{T}_{\rm env}}\quad,
\end{equation}
where $N(t) = \left(X_{M}(0)-X_M(t)\right)/\ell$ is the number of steps in the negative direction minus the number of steps in the positive direction up to time $t$.    The transition times are thus first-passage times of entropy production:  $T_{M\rightarrow M+1} = T^{(2)}_+$ and $T_{M+1\rightarrow M} = T^{(2)}_-$.  The corresponding change in entropy is  $s_{\rm tot}= F\ell/\mathsf{T}_{\rm env}$, where we have used that the process is cyclic.    The probabilities for forward and backward hopping are the passage probabilities for entropy production $\mathsf{P}^{(2)}_+$ and $\mathsf{P}^{(2)}_-$.

 The passage probabilities and the  statistics of transition times  obey the universal equalities that we have derived in the  sections \ref{sec:4}, \ref{sec:5}  and \ref{sec:6}  of this paper.    They  can also be related to the usually discussed transition rates 
 $k_{\pm}\equiv \mathsf{P}^{(2)}_{\pm}\langle 1/T^{(2)}_{\pm}\rangle$, which satisfy the   condition of  local detailed balance $k_+/k_- = e^{F \ell/k_{\rm B}  \mathsf{T}_{\rm env}}$ \cite{de2013non, julicher1997modeling, PhysRevE.85.041125}, as follows from Eqs.~(\ref{eq:synnp}) and (\ref{eq:Integral}).

\subsection{Chemically-driven hopping processes described by one degree of freedom}
Enzymes or molecular motors that are driven out of equilibrium  by a chemical bias or  external force  are examples of such stepping processes.       The thermodynamic bias  of a chemical transition is often of the form $F\ell = \Delta G  - F_{\rm mech}\ell$ where $\Delta G$ denotes the chemical free energy change associated with the chemical transition  and $F_{\rm mech}$ is an externally applied force opposing motion  driven by positive $\Delta G$.   

We are interested in the statistics of chemical transition times $T_{M\rightarrow M'}$, which are the first-passage times of entropy production  $T^{(2)}_+$ and   $T^{(2)}_-$.    Equation  (\ref{eq:synnp}) implies  that the normalized distributions of    $T^{(2)}_+$ and   $T^{(2)}_-$  are identical. 
    This symmetry condition of the distributions of forward and backward transitions times   could be tested experimentally.   For instance, the relation $\langle T^{(2)}_+\rangle = \langle T^{(2)}_-\rangle$ on the  mean  dwell times  could be tested experimentally  for molecular motor that make reversible steps, for example, ${\rm F}_0{\rm F}_1$-ATP synthase \cite{toyabe2011thermodynamic} and RNA polymerase in the backtracking state \cite{galburt2007backtracking, depken2009origin}.

We can also discuss the extreme-value statistics of the number of steps $N$.  
We denote by  $N_{\rm max} (t)>0$ the  maximum number of steps against the bias.    The infimum law Eq.~(\ref{eq:infx}) implies an upper bound  for  the average of $N_{\rm max} (t)$ given by
\begin{equation}
 \,\langle N_{\rm max} (t) \rangle  \leq \frac{k_{\rm B}\mathsf{T}_{\rm env}}{\Delta G - F_{\rm mech}\ell}\quad,
\label{eq:tox}
\end{equation}
for $\Delta G > F_{\rm mech}\ell$.
  The right-hand side of Eq.~(\ref{eq:tox}) is the inverse of the P\'{e}clet number ${\rm Pe} = v \ell/D$.
Moreover, Eq.~(\ref{eq:infxx}) implies that the cumulative disitribution of $N_{\rm max}(t)$ is  bounded from above by an exponential 
\begin{eqnarray}
\mathsf{Pr}\left(N_{\rm max}(t) \geq  n\right) \leq e^{- n  \left(\Delta G - F_{\rm mech}\ell\right)/k_{\rm B}  \mathsf{T}_{\rm env}}\quad, \label{eq:toxx}
\end{eqnarray}
with $n\geq 0$.     Equation (\ref{eq:toxx}) states that the probability that a molecular motor 
makes more than $n$ backsteps  is smaller or equal than $e^{- n  \left(\Delta G - F_{\rm mech}\ell\right)/k_{\rm B}  \mathsf{T}_{\rm env}}$.    Therefore, our 
 results on infima of entropy production constrain the maximum excursion of a molecular motor against the direction of an external force.

\subsection{Dynamics of RNA polymerases: an example with two degrees of freedom}
\begin{figure}
\includegraphics[width=7cm]{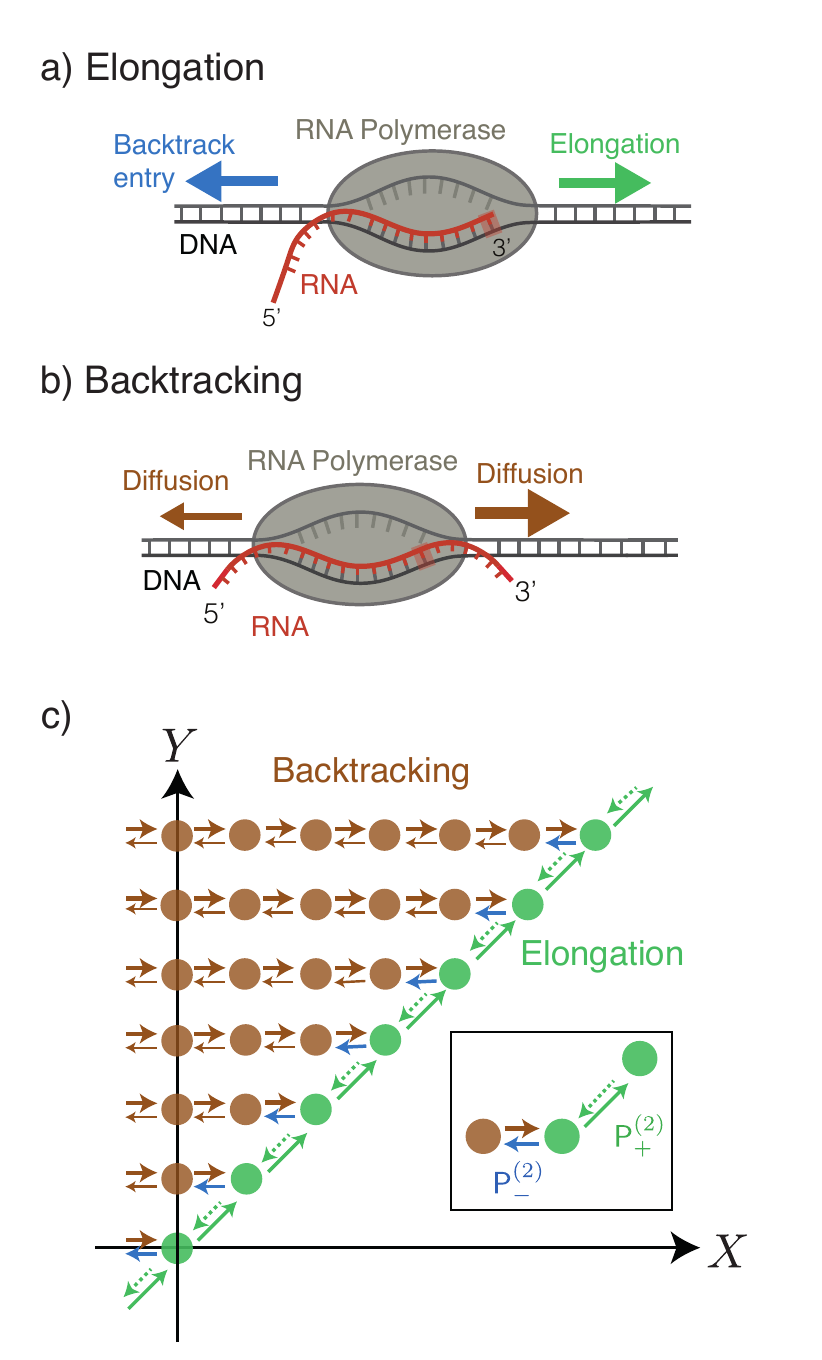} 
\caption{Stochastic model for transcription.  a) Sketch of an elongating RNA polymerase.  The polymerase can either incorporate a new nucleotide to its active site (red shaded square), or enter a bactracked state.   b) Sketch of a backtracking RNA polymerase.  The polymerase diffuses along the DNA template until its active site (red shaded area) aligns with the 3' end of the transcribed RNA.  c) Network representation of a Markov jump process describing the dynamics of RNA polymerases.  The $X$-coordinate gives the position of the polymerase along the DNA template, and the $Y$-coordinate gives the number of nucleotides transcribed.  Green nodes represent elongating states and brown nodes represent backtracked states.  The inset shows all possible transitions from or to an elongating state.  The green solid arrow denotes active translocation of the RNA polymerase, and the dashed green arrow denotes its time reversed transition.  Because the heat dissipated during the forward translocation is much larger than $k_{\rm B}\mathsf{T}_{\rm env}$, the rate of the backward transition is very small.   The blue arrow corresponds to the entry in a backtracked state, and the brown arrow is the exit from a backtracked state into the elongating state.  In the backtracked state motion is biased towards the elongating state.  }\label{fig:chemical2Dxx}  
\end{figure}

We now apply the general results of our theory to a more complex example of a   biomolecular process, which cannot be described in terms of a single degree of freedom, namely, the dynamics of RNA polymerases on a DNA template.   
RNA polymerases transcribe genetic information from DNA into RNA. During  transcription, RNA polymerases adopt two different states:  the elongating state and the backtracking state \cite{galburt2007backtracking}.    Elongation  is an active process where  
  RNA polymerases  move stepwise and unidirectionally along the DNA while polymerizing a complementary RNA driven by the chemical free energy of the incorporated nucleotides, as illustrated in Fig.~\ref{fig:chemical2Dxx}a).     In the elongating state the motion of RNA polymerase and the polymerization reaction are fuelled by the hydrolysis of ribonucleotide triphosphate (NTP), which provides  the free  energy $\Delta G_{\rm NTP}$ per nucleotide   \cite{julicher1998motion}.   
     Backtracking is a passive motion of the RNA polymerase on the DNA template that displaces the RNA 3' end from the active site  of the polymerase, and leaves the enzyme transcriptionally inactive \cite{komissarova1997transcriptional},  as illustrated in Fig.~\ref{fig:chemical2Dxx}b).    Transcription is thus an active polymerization process  that is interspersed by   pauses of passive stepping motion. 
  
\begin{figure}
\includegraphics[width=7.5cm]{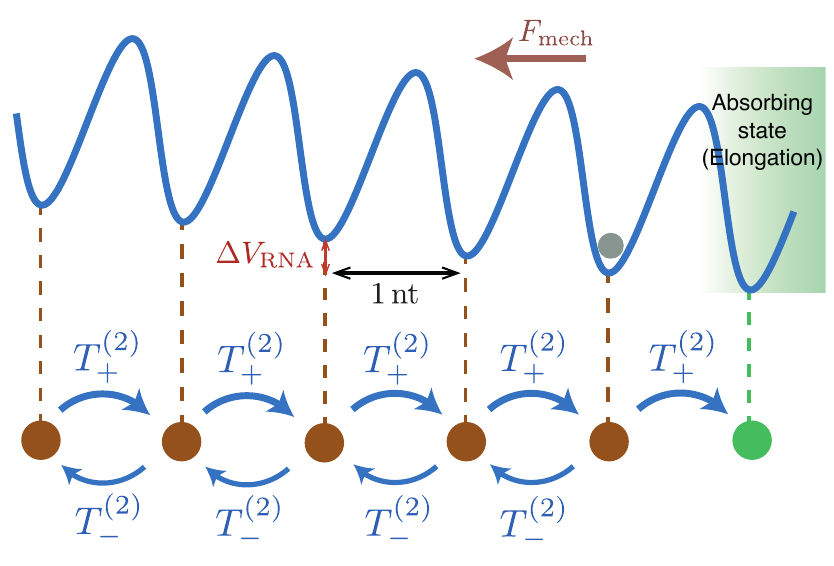} 
\caption{The dynamics of RNA polymerase during backtracking.  The motion of the polymerase is represented as a diffusion process in a tilted periodic potential with an absorbing boundary on the right (which corresponds to the transition from the backtracked state into the elongating state).    The period of the potential  is equal to the length of one nucleotide (nt). The potential is tilted towards the absorbing boundary due to an energetic barrier $\Delta V_{\rm RNA}$, and  an external opposing mechanical force $F_{\rm mech}$ pushes the polymerase away from the absorbing boundary.     This process can be coarse grained into a discrete one-dimensional hopping process with waiting times $T^{(2)}_+$ and $T^{(2)}_-$.     }\label{fig:chemical2D}
\end{figure}

  The main properties of the dynamics of RNA polymerases are the following.   
 In the elongating state, RNA polymerases   can   either continue elongation (green arrows in Fig.~\ref{fig:chemical2Dxx} a) and c)), or enter a backtracking state  (blue arrows in Fig.~\ref{fig:chemical2Dxx} a) and c)).    A RNA polymerase in the backtracking state
   diffuses on the DNA template  until its active site  is realigned with the 3' end of the RNA  \cite{galburt2007backtracking, footnotesPolymerase}   This diffusive motion is often biased by either the presence of external opposing forces $F_{\rm mech}$ or by an energy barrier $\Delta V_{\rm RNA}$ related to the secondary structure of the nascent RNA \cite{zamft2012nascent, ishibashi2014transcription}, see Fig.~\ref{fig:chemical2D}.

 The dynamics of RNA polymerase can thus be described as a continuous-time Markov jump process on a two-dimensional network \cite{dangkulwanich2013complete, depken2013intermittent, ishibashi2014transcription}, see Fig.~\ref{fig:chemical2Dxx}  c).    The state of a polymerase is determined by two variables:  $X$  denotes the position of the polymerase (in nucleotides) along the DNA template, 
  and  $Y$ denotes the number of NTP molecules hydrolysed during elongation.   When $X=Y$ the polymerase is in the elongating state, and when $X<Y$ the polymerase is in the backtracked state.  The variable $N = Y-X$ denotes the depth of the backtrack.    We consider for simplicity  a stationary processs with steady-state distribution $p_{\rm ss}(X,Y)$ on a homogeneous DNA template.     Therefore, the steady-state probability to find    the polymerase  in an elongating state is $p_{\rm ss}(0) = p_{\rm ss}(X,X)$,  and  
the steady-state probability to find the polymerase backtracked by $N$ steps is $p_{\rm ss}(N) = p_{\rm ss}(X,X+N)$.

 Two relevant quantities in transcription are the pause density \cite{galburt2007backtracking, larson2012trigger, dangkulwanich2013complete, lisica2016mechanisms}, i.e., the probability per nucleotide that an elongating polymerase enters a backtracked state, and the maximum depth of a backtrack  \cite{galburt2007backtracking, depken2009origin}, i.e., the maximum number of backtracked nucleotides.      Our theory  provides expressions for the statistics of these quantities in terms of  infima and stopping-time statistics of entropy production.

If the probability flux from the elongating state into the backtracked state is smaller  than the  flux of the reverse process, which implies that entropy $S_{\rm tot}$ reduces when the polymerase enters a backtrack,  the pause density  is equal to a passage probability $\mathsf{P}^{(2)}_-$ of entropy production.   In addition we consider that an elongating polymerase only moves forward.  These conditions are necessary for backtrack entry to correspond to a first-passage process of entropy production with two absorbing boundaries of different sign.     The probability $\mathsf{P}^{(2)}_+$ then corresponds to the probability   for the polymerase to attach a new nucleotide, as illustrated in the inset in Fig.~\ref{fig:chemical2Dxx}  c).   These  probabilities  obey  Eqs.~(\ref{eq:pass1xxx}) and (\ref{eq:pass2xxx})  \cite{footnotePP},
with $-s^-_{\rm tot} = \Delta S_{{\rm be}}+F_{\rm mech}\ell/\mathsf{T}_{\rm env}$ and $s^+_{\rm tot} = (\Delta G_{\rm NTP}-F_{\rm mech}\ell)/\mathsf{T}_{\rm env}$.  Here, $\Delta S_{\rm be}  = -k_{\rm B}  \ln  \left(p_{\rm ss}(1)/p_{\rm ss}(0)\right)$ is the system entropy change when entering the backtrack.
 If $\Delta G_{\rm NTP}/k_{\rm B}\mathsf{T}_{\rm env}\gg  1$, which  holds in typical experiments \cite{dangkulwanich2013complete},    we have  $e^{-s^+_{\rm tot}/k_{\rm B}}\ll 1$, and thus we find simple expressions for the pause densities and the probability to continue elongations
\begin{eqnarray}
\mathsf{P}^{(2)}_+ &\simeq &1- e^{\left(\mathsf{T}_{\rm env}\Delta S_{\rm be}+F_{\rm mech}\ell\right)/k_{\rm B}\mathsf{T}_{\rm env}}\label{eq:pass1applx}\quad,\\ 
\mathsf{P}^{(2)}_- &\simeq &e^{\left(\mathsf{T}_{\rm env}\Delta S_{\rm be}+F_{\rm mech}\ell\right)/k_{\rm B}\mathsf{T}_{\rm env}}\quad.\label{eq:pass2applx}
\end{eqnarray} 
The backtrack entry of a polymerase, and therefore the pause density,  is  thus  determined by  a kinetic competition between an entropic term $\mathsf{T}_{\rm env}\Delta S_{\rm be}<0$ and the work done by the  force acting on the polymerase $F_{\rm mech}\ell >0$.   

We can also discuss the dynamics of backtracked RNA polymerases.  
   During a backtrack the dynamics of  RNA polymerase is captured by a one-dimensiononal 
biased diffusion process with an absorbing boundary corresponding to the transition to   elongation  \cite{depken2009origin}, see Fig.~\ref{fig:chemical2D}.   During backtracking the bias $F = \Delta V_{\rm RNA}/\ell - F_{\rm mech}$, where   $\Delta V_{\rm RNA}$ is  an energy barrier associated with  the secondary structure of the nascent RNA   \cite{zamft2012nascent, ishibashi2014transcription, lisica2016mechanisms}.     The waiting times of the corresponding Markov jump process are  equal to first-passage times of entropy production.    For a polymerase  with backtrack depth $N$,  the waiting time  to decrease $N$ by one $T^{(2)}_+(N)$, and the waiting time to increase $N$ by one  $T^{(2)}_-(N)$, are  first-passage times of entropy production with two absorbing boundaries $s^+_{\rm tot} = -k_{\rm B}\ln \left(p_{\rm ss}(N-1)/p_{\rm ss}(N)\right) + F\ell/\mathsf{T}_{\rm env}$  and $-s^-_{\rm tot} = -k_{\rm B}\ln \left(p_{\rm ss}(N+1)/p_{\rm ss}(N)\right) - F\ell/\mathsf{T}_{\rm env}$.    If the bias dominates,   the two boundaries  have opposite sign   
and we can use the theory of stopping times developed here.     The hopping probabilities for forward and backward steps in the backtrack follow then from our general expressions for passage probabilities of entropy production Eqs.~(\ref{eq:pass1xxx}) and (\ref{eq:pass2xxx}).

The maximum  backtrack depth $N_{\rm max}(t)$, at  a time $t$ after entering the backtrack state, is related to the infimum of entropy production  $S_{\rm inf}(t)$.       The infimum law~(\ref{eq:infx}) provides therefore a thermodynamic  bound on the average of the maximum extent of a backtrack.     Using  Eq.~(\ref{eq:tox}) with  $\Delta V_{\rm RNA} \simeq 0.5 k_{\rm B}\mathsf{T}_{\rm env}$ and $F_{\rm mech}\ell\simeq 0.4 k_{\rm B}\mathsf{T}_{\rm env}$  \cite{zamft2012nascent, ishibashi2014transcription, lisica2016mechanisms},  we estimate  the  upper bound for the  maximum  backtrack depth $\langle N_{\rm max}(t)\rangle \lesssim 10$ nucleotides; we have used the parameter values $F_{\rm mech} = 5 \,{\rm pN}$,  $k_{\rm B}\mathsf{T}_{\rm env} = 4\, {\rm pN}\, {\rm nm}$ and $\ell = 0.34 \,{\rm nm}$ the distance between two nucleotides. Similarily,   from Eq.~(\ref{eq:toxx}) we find that the cumulative distribution of the maximum backtrack depth is upper bounded by an exponential \cite{footnoteBoundary}.     This is consistent with 
single molecule experiments on RNA polymerase backtracking, which  have reported that the distribution of the backtrack extent is  exponential (see Fig. S3 in~\cite{galburt2007backtracking}).      
\section{Discussion}\label{sec:9}  
The second law of thermodynamics is a statement about the average entropy production when
many realizations of a mesoscopic system are considered. This fundamental law 
leaves open the question
whether fluctuations of entropy production also obey general laws   \cite{barato2015thermodynamic, gingrich2015dissipation, roldan2015decision, pietzonka2015universal, saito2015waiting, pietzonka2016affinity, polettini2016tightening}. In the present paper we 
have demonstrated
that the infimum of entropy production and  stopping times of entropy production 
exhibit statistical features that are universal and do not depend on the physical nature
of a given system.   
The simplest example that illustrates these universal features is the case where entropy production follows a stochastic drift-diffusion process  
described by the Langevin equation
\begin{eqnarray}
\frac{{\rm d}S_{\rm tot}(t)}{{\rm d}t}  = v_{\rm S} + \eta_{\rm S}(t)\quad,
\end{eqnarray}
where the  constant drift velocity $v_{\rm S}>0$ corresponds to the average entropy production rate.     The  Gaussian white noise $\eta_{\rm S}(t)$ has zero mean   and the autocorrelation $\langle \eta_{\rm S}(t) \eta_{\rm S}(0)\rangle = 2D_{\rm S}\delta(t)$.   The diffusion constant  $D_{\rm S}$   characterizes entropy production fluctuations and obeys    
 \begin{eqnarray}
 D_{\rm S} = k_{\rm B}\,v_{\rm S} \quad.
 \end{eqnarray}
These relations  can be derived from Eq.~(\ref{eq:SDD}).  These equations exhibit all the universal properties of entropy production discussed in this paper, see Section \ref{sec:7}.  

The infimum of entropy production up to a given time 
must be either zero or must be negative, even though entropy production is always positive on average. 
Here, we have shown that when 
averaging over all realizations in a nonequilibrium steady state, the average infimum of
entropy production is  always greater than or equal to $-k_{\rm B}$. 
Furthermore, the global infimum of entropy production is exponentially distributed with mean $-k_{\rm B}$.    This is an exact result for the distribution of global infima of a correlated stochastic process, which is interesting because only few similar results are known \cite{majumdar2003extreme, comtet2005precise, majumdar2014extreme}.    

 Our results are of special interest for experiments, because entropy production reaches its global infimum  in a finite time.     The exact results on entropy infima could be verified experimentally, for example  using colloidal particles trapped with optical tweezers \cite{blickle2007characterizing, gomez2009experimental, PhysRevE.90.032116}.    Other experimental systems which could be used to test our results are single molecule experiments  \cite{nishiyama2002chemomechanical, collin2005verification, toyabe2011thermodynamic} or mesoscopic electronic systems such as quantum dots  \cite{saira2012test}.

We have furthermore shown that the stopping-time statistics and the passage statistics of entropy production exhibit universality.       
We  have found a remarkable symmetry for the stopping times of entropy production.   In a network of states, this symmetry relation implies that, for each link between two states, the statistics of waiting times is the same for forward and backward jumps along this link.   Measuring statistics  of waiting times thus reveals whether two forward and backward transitions along a link in a network of states are each other's time reverse.    If the corresponding waiting-time distributions are not equal, then forward and backward transitions are not related by time reversal.

Our work is based on the finding that in  steady state  $e^{-S_{\rm tot}(t)/k_{\rm B}}$ is a martingale process.   A martingale is  a process for which the mean is unpredictable, or equivalently, represents a fair game with no net gain or loss.      
The theory of martingales is well developed because of its  importance to fields such as quantitative finance    \cite{musiela2006martingale, bjork2009arbitrage} and decision theory \cite{tartakovsky2014sequential}.   In stochastic thermodynamics  martingales have not yet
attracted much attention  \cite{chetrite2011two}.     
Our work reveals the relevance of martingales to nonequilibrium  thermodynamics.    We also show that entropy production itself is a submartingale,  and  relate this fact to the second law of thermodynamics.     Remarkably,  the universal statistics of infima and stopping times all follow from the  martingale property of $e^{-S_{\rm tot}(t)/k_{\rm B}}$.

Our results on entropy production fluctuations provide  expressions for the 
hopping probabilities, the statistics of waiting times, and  the extreme-value statistics of active molecular processes.  
We have illustrated our results on active molecular processes
described by one degree of freedom, and on the stochastic dynamics of  DNA transcription by RNA polymerases.    
Our theory provides  expressions for the  probability of molecular motors to step forwards or backwards in terms of the entropy produced during the forward and backward steps, and relates waiting-time statistics of forward and backward transitions.     Moreover, the  infimum law  provides a thermodynamic bound on the  maximal excursion of a motor against the effective force that drives its motion.       For the dynamics of RNA polymerases this implies that the statistics of the maximum backtrack depth is bounded by a limiting exponential distribution.    This provides predictions about the mean and the second moment of backtrack depths  that could be  tested  in future single-molecule experiments.

Cells and proteins often execute complex functions at random times.  
Stopping times provide a powerful tool to
characterize timing of stochastic biological processes.  
It  will be interesting to explore whether our approach to  the statistics of stopping times 
of nonequilibrium processes is also relevant to more complex systems, such as, flagellar motor switching \cite{morse2015flagellar}, sensory systems \cite{sartori2014thermodynamic, barato2014efficiency, bo2015thermodynamic, das2016lower}, self-replicating nucleic acids \cite{england2013statistical, sartori2015thermodynamics} or  cell-fate decisions \cite{PhysRevLett.104.228104, siggia2013decisions}.
\acknowledgements
We thank Heinrich Meyr, Meik D\"orpinghaus, Mostafa Khalili-Marandi, Andre Cardoso Barato, Ana Lisica  and Stephan W.~Grill for
stimulating discussions.  We thank Matteo Polettini for bringing Refs.~\cite{jia2014cycle} and \cite{bauer2014affinity} to our attention. 
E.R.~acknowledges funding from Spanish Government, Grant TerMic (FIS2014-52486-R).

\appendix

\section{MARTINGALE PROCESSES}\label{app:a}
We define martingale processes and discuss their properties using concepts and notations from   probability and martingale theory \cite{liptser2013statistics}.  
\subsection{Definition of a martingale process}
 A sequence of random variables $X_1, X_2, X_3, \ldots$ is {\it martingale} when $\mathbb{E}\left(|X_n|\right)<\infty$, for all $n\in\mathbb{N}$, and the conditional expectation $\mathbb{E}\left(X_n|X_{1}, X_2, \ldots, X_{m}\right)= X_{m}$, for all $m\leq n$.    A martingale is thus a process for which the mean is unpredictable, or equivalently, represents a fair game with no net gain or loss.    Note that here we use the standard notation $\mathbb{E}_{\mathbb{P}}\left(X_n\right) = \langle  X_n\rangle_{\mathbb{P}} $ for the expectation of a random variable $X_n$ with respect to the measure $\mathbb{P}$.
 
For our purposes, we also need  a general definition of a martingale process with respect to a filtered probability space.   We consider a stochastic process $X(\omega;t)$, with $t\in[0,\infty[$, and a filtered probability space $(\Omega, \mathcal{F}, \left\{\mathcal{F}(t)\right\}_{t\geq 0}, \mathbb{P})$.    For processes in continuous time, we consider that $X$ is right continuous.  
A process $X(\omega;t)$ is a {\it martingale} with respect to the filtered probability space $(\Omega, \mathcal{F}, \left\{\mathcal{F}(t)\right\}_{t\geq 0}, \mathbb{P})$ when: $X$ is adapted to the filtration  $\left\{\mathcal{F}(t)\right\}_{t\geq 0}$;  $X$ is integrable, i.e.,~$\mathbb{E}\left(|X(\omega, t)|\right)<\infty$; and the conditional expectation $\mathbb{E}\left(X(\omega;t)|\mathcal{F}(s)\right) = X(\omega;s)$ for all $s<t$.   The latter condition is equivalent to 
\begin{eqnarray}
\int_{\omega\in \Phi}X(\omega;s){\rm d}\mathbb{P} = \int_{\omega\in \Phi}X(\omega;t) {\rm d}\mathbb{P}\quad, \label{eq:martxFirst}
\end{eqnarray}
for  any $s<t$ and  for any $\Phi\in\mathcal{F}(s)$.   
A {\it sub-martingale} satisfies the inequality  
 \begin{eqnarray}
\int_{\omega\in \Phi}X(\omega;s){\rm d}\mathbb{P} \leq \int_{\omega\in \Phi}X(\omega;t) {\rm d}\mathbb{P}\quad,\label{eq:martprop}
\end{eqnarray}
for  any  $s<t$ and for any $\Phi\in\mathcal{F}(s)$.  The Wiener process, also called a Brownian motion, is an example of a martingale process.

\subsection{Uniform integrability} 
Uniformly integrable processes play an important role in martingale theory.  
 We call a stochastic process  $X$ uniform integrable on the probability space $(\Omega, \left\{\mathcal{F}(t)\right\}_{t\geq 0}, \mathcal{F}, \mathbb{P})$ if given $\epsilon>0$, there exists a $K\in[0,\infty)$, such that
\begin{eqnarray}
\mathbb{E}_{\mathbb{P}} \left(|X(t)|I_{|X(t)|> K}\right)< \epsilon\quad,
\end{eqnarray}
for all $t\geq 0$.
The indicator function $I_{|X(t)|> K}$ is defined as: 
\begin{eqnarray}
I_{|X(t)|> K}= \left\{\begin{array}{ccc}1 &{\rm if}& |X(t)|>K\\ 0 &{\rm if}& |X(t)|\leq K\end{array}\right. \quad.
\end{eqnarray}
A bounded random process is uniformly integrable, and a uniformly-integrable process is integrable.  Uniform integrability can thus be seen as a condition less stringent than boundedness, but more stringent than integrability.

For a uniformly integrable martingale process, the random variable  $X(\omega;\infty) = \lim_{t\rightarrow\infty}X(\omega;t)$ exists \cite{liptser2013statistics}, and it is finite  for $\mathbb{P}$-almost all $\omega\in\Omega$.   
A process is uniform integrable, if and only if  \cite{liptser2013statistics}:
\begin{eqnarray}
X(\omega;t) = \mathbb{E}\left(X(\omega;+\infty) |\mathcal{F}\left(t\right)\right)\quad, \label{eq:condmart}
\end{eqnarray}
for $t\geq 0$, and with $X(\omega;+\infty)$ integrable, i.e., $\mathbb{E}(|X(\omega;+\infty) |)<\infty$.

\subsection{Doob's maximal inequality} $\\$
 Doob's maximal inequality provides an upper bound on the cumulative distribution of the supremum of a nonnegative submartingale process  $X(\omega;t)$ in a time interval $I$ \cite{doob1953stochastic}, viz.,   
\begin{eqnarray}
\mathsf{Pr}\left({\rm sup}_{t\in I}X(\omega;t)\geq \lambda\right) \leq \frac{1}{\lambda}\mathbb{E}_{\mathbb{P}}\left(X(\omega;t)\right)\quad, \label{eq:doobopt}
\end{eqnarray}  
for any constant $\lambda> 0$.  Doob's maximal inequality, given by Eq.~(\ref{eq:doobopt}), holds for nonnegative submartingales in discrete time, and  for right-continuous nonnegative  submartingales
in continuous time  \cite{liptser2013statistics}.  

Note that Doob's maximal inequality is a unique property of martingales, and a  stronger result than 
Markov's inequality.  Markov's inequality 
provides an upper bound on the cumulative distribution 
of a nonnegative random variable $X$:
\begin{eqnarray}
\mathsf{Pr}\left(X(\omega;t)\geq \lambda\right) \leq \frac{1}{\lambda}\mathbb{E}_{\mathbb{P}}\left(X(\omega;t)\right)\quad, \label{eq:markov}
\end{eqnarray} 
with  $\lambda>0$.  Since $\mathsf{Pr}\left(X(\omega;t)\geq \lambda\right)\geq \mathsf{Pr}\left({\rm sup}_{t\in I}X(\omega;t)\geq \lambda\right)$, Markov's inequality does not imply Doob's maximal inequality, but, Doob's maximal inequality does imply Markov's inequality for  nonnegative martingales.   

\subsection{Doob's optional sampling theorem}\label{Doobmart}
We consider  a uniformly integrable martingale process $X(\omega;t)$ adapted to the filtration $\left\{\mathcal{F}(t)\right\}_{t\geq 0}$ and two stopping times $T_1(\omega)$ and $T_2(\omega)$, with the property that for each $\omega\in\Omega$, $T_1(\omega)\leq T_2(\omega)$. Additionally, in continuous time $X$ is right continuous.    Doob's optional sampling theorem states then that \cite{liptser2013statistics, williams1991probability}:
\begin{eqnarray}
\int_{\omega\in \Phi}X(\omega;T_2(\omega))\,{\rm d}\mathbb{P} = \int_{\omega\in \Phi}X(\omega;T_1(\omega))\, {\rm d}\mathbb{P} \label{eq:dooboptxx}
\end{eqnarray}
for each set $\Phi\in\mathcal{F}(T_1)$.      The sub-$\sigma$-algebra $\mathcal{F}(T_1)$, corresponding to the stopping time $T_1$, is defined as 
\begin{eqnarray}
\mathcal{F}(T_1)=\left\{\Phi\subset \Omega: \forall t \geq 0, \Phi \cap \left\{\omega:T_1(\omega)\leq t\right\} \in \mathcal{F}(t) \right\}\quad. \nonumber\\
\end{eqnarray}

\section{PROBABILITY THEORY FOR STOCHASTIC ENTROPY PRODUCTION}\label{app:c}
We define stochastic entropy production within the formalism of modern probability theory \cite{kolmogorovgrundbegriffe, royden1988real, liptser2013statistics}.  We use this definition of entropy production to  a derive general properties of stochastic entropy production.
\subsection{Filtered probability space of a  process at the mesoscopic scale}
We describe a mesoscopic system with the coarse-grained state variables $\bq(\tau)$ and $\tilde{\bq}(\tau)$ and  write for the full trajectory $\omega= \left\{\bq(\tau), \tilde{\bq}(\tau)\right\}_{\tau\in(-\infty, \infty)}$ .   The trajectory $\omega$  contains thus, for one given realization,  the full information on the dynamics of our mesoscopic system.   

In order to associate probabilities to sets of realizations  $\omega$, we consider the 
 filtered probability space $(\Omega, \mathcal{F}, \left\{\mathcal{F}(t)\right\}_{t\geq 0}, \mathbb{P})$.   The set $\Omega$ contains all possible realizations $\omega$ of the process, and $\mathcal{F}$ is the generated $\sigma$-algebra
Trajectories over a finite time interval $\omega^t_0 = \left\{\bq(\tau), \tilde{\bq}(\tau)\right\}_{0\leq \tau\leq t}$ generate the sub-$\sigma$-algebra $\mathcal{F}(t)$, and a corresponding probability space $(\Omega, \mathcal{F}(t), \left.\mathbb{P}\right|_{\mathcal{F}(t)})$ for all values $t\geq 0$.   In continuous time we consider filtrations  $\left\{\mathcal{F}(t)\right\}_{t\geq 0}$ that are  right-continuous.   
 
The time-translation map $\mathsf{T}_t$ operates on trajectories $\omega$ as $\mathsf{T}_t(\omega) = \left\{\bq(\tau+t), \tilde{\bq}(\tau+t)\right\}_{\tau\in(-\infty, \infty)}$.  The time-inversion map $\Theta$, with respect to the reference time $t=0$, operates on trajectories $\omega$ as  $\Theta(\omega) = \left\{\bq(-\tau), \tilde{\bq}(-\tau)\right\}_{\tau\in(-\infty, \infty)}$.   We also consider the time-inversion map $\Theta_t$, with respect to the reference time $t/2$, which operates on trajectories $\omega$ as $\Theta_t(\omega) = \left\{\bq(-\tau+t), \tilde{\bq}(-\tau+t)\right\}_{\tau\in(-\infty, \infty)}$.  
The map $\Theta_t$ is thus the composition $\Theta_t = \mathsf{T}_t\circ\Theta$.

\subsection{Entropy production of a filtered probability space}
Entropy production $S_{\rm tot}(\omega;t)$ is a   stochastic process 
adapted to a stationary filtered probability space $(\Omega, \mathcal{F}, \left\{\mathcal{F}(t)\right\}_{t\geq 0}, \mathbb{P})$.   We define  $S_{\rm tot}(\omega;t)$ as the Radon-Nikod\'{y}m derivative of the  measure  $\left.\mathbb{P}\right|_{\mathcal{F}(t)}$  with respect to  the time-reversed measure  $\left.\mathbb{P}\circ\Theta_t\right|_{\mathcal{F}(t)}$ \cite{maes2000definition, maes2003time}: 
\begin{eqnarray}
S_{\rm tot}(\omega;t) = k_{\rm B} \ln \frac{\left.{\rm d}\mathbb{P}\right|_{\mathcal{F}(t)}}{\left.{\rm d}\left(\mathbb{P}\circ \Theta_t\right)\right|_{\mathcal{F}(t)}}(\omega)\quad,  \label{eq:entropyDefApp}
\end{eqnarray}
for $t\geq 0$.   For stationary probability measures $\mathbb{P}$, Eq.~(\ref{eq:entropyDefApp}) is the same as Eq.~(\ref{Eq:entropDef}).  An analogous definition applies to   entropy production for negative values of time, $t\leq 0$.       
The definition of entropy production, given by Eq.~(\ref{eq:entropyDefApp}), requires thus absolute continuity of $\mathbb{P}$ with respect to $\mathbb{P}\circ\Theta_t$.  Note that it is possible to define entropy production for  processes with stritcly irreversible transitions \cite{murashita2014nonequilibrium}, but, Eq.~(\ref{eq:entropyDefApp}) does not apply to such processes.    Stochastic entropy production (\ref{eq:entropyDefApp})  is thus the stochastic process  of a filtered probability space that characterizes time's arrow in the possible outcomes of a random process \cite{roldan2015decision}.

 For mesoscopic processes in contact with an  equilibrated environment  the definition of entropy production, Eq.~(\ref{eq:entropyDefApp}),  is thermodynamically consistent \cite{bergmann1955new, crooks1998nonequilibrium, crooks1999entropy, maes2003time}.     In statistical physics this latter  condition is often called  local detailed balance or generalized detailed balance.

We now discuss the mathematical properties  of stochastic entropy production.    We  first show that entropy production is a process of odd parity  with respect to the time inversion operator $\Theta_t$.    Secondly, we show that $e^{-S_{\rm tot}(\omega, t)/k_{\rm B}}$ is a {\it uniformly integrable  martingale} with respect to the filtered probability space $(\Omega, \mathcal{F}, \left\{\mathcal{F}(t)\right\}_{t\geq 0}, \mathbb{P})$ generated by the dynamics of the mesoscopic degrees of freedom.   This is a key property of entropy production and allows us to apply the techniques of martingale processes  to entropy production.

\subsection{Entropy production under time reversal}\label{app:b3}
An interesting property  of entropy production is the change of its sign under the time-reversal operation
 $\Theta_t$, viz., 
\begin{eqnarray}
S_{\rm tot}(\Theta_t(\omega);t) &=& k_{\rm B} \ln \frac{\left.{\rm d}\mathbb{P}\right|_{\mathcal{F}(t)}}{\left.{\rm d}\left(\mathbb{P}\circ \Theta_t\right)\right|_{\mathcal{F}(t)}}(\Theta_t(\omega)) 
\nonumber \\ 
&=& k_{\rm B} \ln \frac{\left.{\rm d}\left(\mathbb{P}\circ \Theta_t\right)\right|_{\mathcal{F}(t)}}{\left.{\rm d}\left(\mathbb{P}\circ \Theta_t\circ \Theta_t\right)\right|_{\mathcal{F}(t)}}(\omega)
\nonumber \\  
&=& k_{\rm B} \ln \frac{\left.{\rm d}\left(\mathbb{P}\circ \Theta_t\right)\right|_{\mathcal{F}(t)}}{\left.{\rm d}\mathbb{P}\right|_{\mathcal{F}(t)}}(\omega)   \nonumber \\  
&=& -S_{\rm tot}(\omega;t) \quad,  \label{eq:timer}
\end{eqnarray}  
where we have used that $\Theta_t = \Theta^{-1}_t$.

\subsection{The exponential of negative entropy production   is a martingale in steady state}\label{app:enMar}
We now show that the process $e^{-S_{\rm tot}(\omega;t)/k_{\rm B}}$,  adapted to a stationary and filtered probability space $(\Omega, \mathcal{F}, \left\{\mathcal{F}(t)\right\}_{t\geq 0}, \mathbb{P})$, is a martingale.   
This follows directly from the fact that  $e^{-S_{\rm tot}(\omega;t)/k_{\rm B}}$ is a  Radon-Nikod\'{y}m density process.

The process $e^{-S_{\rm tot}(\omega;t)/k_{\rm B}}$ is the Radon-Nikod\'{y}m density process of the filtered probability space
$(\Omega, \mathcal{F}, \left\{\mathcal{F}(t)\right\}_{t\geq 0}, \mathbb{P})$  with respect to the filtered probability space
 $(\Omega, \mathcal{F}, \left\{\mathcal{F}(t)\right\}_{t\geq 0}, \mathbb{Q})$:
\begin{eqnarray}
e^{-S_{\rm tot}(\omega;t)/k_{\rm B}} = \frac{\left.{\rm d}\mathbb{Q}\right|_{\mathcal{F}(t)}}{\left.{\rm d}\mathbb{P}\right|_{\mathcal{F}(t)}}(\omega)\quad,
\end{eqnarray} 
with $\mathbb{Q} = \mathbb{P}\circ\Theta$, the time reversed measure and $t\geq 0$.
Consider  two sub-$\sigma$-algebras $\mathcal{F}(\tau)$ and $\mathcal{F}(t)$ of $\mathcal{F}$, with $\tau<t$.  
We first write the measure $\mathbb{Q}\left(\Phi\right)$ of a  set $\Phi\in\mathcal{F}(\tau)$ as an integral over the probability space  $\left(\Omega, \mathcal{F}(\tau), \left.\mathbb{Q}\right|_{\mathcal{F}(\tau)}\right)$
 \begin{eqnarray}
\mathbb{Q}\left(\Phi\right) &=& \int_{\omega\in \Phi}\left.{\rm d}\mathbb{Q}\right|_{\mathcal{F}(\tau)}
\nonumber \\ 
&=& \int_{\omega\in \Phi}e^{-S_{\rm tot}(\omega;\tau)/k_{\rm B}}\left.{\rm d}\mathbb{P}\right|_{\mathcal{F}(\tau)}\nonumber \\
&=& \int_{\omega\in \Phi}e^{-S_{\rm tot}(\omega;\tau)/k_{\rm B}}{\rm d}\mathbb{P}\quad.
\end{eqnarray}  
Alternatively, we write the measure $\mathbb{Q}\left(\Phi\right)$ as an integral over the probability space  $\left(\Omega, \mathcal{F}(t), \left.\mathbb{Q}\right|_{\mathcal{F}(t)}\right)$
 \begin{eqnarray}
\mathbb{Q}\left(\Phi\right) &=& \int_{\omega\in \Phi}\left.{\rm d}\mathbb{Q}\right|_{\mathcal{F}(t)}
\nonumber \\ 
&=& \int_{\omega\in \Phi}e^{-S_{\rm tot}(\omega;t)/k_{\rm B}}\left.{\rm d}\mathbb{P}\right|_{\mathcal{F}(t)}\nonumber \\
&=& \int_{\omega\in \Phi}e^{-S_{\rm tot}(\omega;t)/k_{\rm B}}{\rm d}\mathbb{P}\quad.
\end{eqnarray}    
We thus have the equality
\begin{eqnarray}
 \int_{\omega\in \Phi}e^{-S_{\rm tot}(\omega;\tau)/k_{\rm B}}{\rm d}\mathbb{P}   =  \int_{\omega\in \Phi}e^{-S_{\rm tot}(\omega;t)/k_{\rm B}}{\rm d} \mathbb{P}  \quad,
\end{eqnarray}
for all sets $\Phi\in\mathcal{F}(\tau)$, which is identical to the Eq.~(\ref{eq:martxFirst})
that defines a martingale process.  The process $e^{-S_{\rm tot}(\omega;t)/k_{\rm B}}$ is therefore a martingale process with respect to the measure $\mathbb{P}$.

\subsection{The exponential of negative entropy production is uniformly integrable}
We show that $e^{-S_{\rm tot}(\omega;t)/k_{\rm B}}$  is  uniformly integrable.    We use the necessary and sufficient condition (\ref{eq:condmart}) for uniform integrability.     The process  $e^{-S_{\rm tot}(\omega;t)/k_{\rm B}}$ is uniformly integrable since, by definition,  the following condition is met: 
 \begin{eqnarray}
 e^{-S_{\rm tot}(\omega;t)/k_{\rm B}}  = \mathbb{E}\left[\frac{{\rm d}\mathbb{P}\circ\Theta}{{\rm d}\mathbb{P}}(\omega)\Big|\mathcal{F}(t)\right]\quad,
\end{eqnarray}
with $\frac{{\rm d}\mathbb{P}\circ\Theta}{{\rm d}\mathbb{P}}(\omega) = e^{-S_{\rm tot}(\omega;+\infty)/k_{\rm B}}$ a positive and integrable random variable.

\section{STOCHASTIC DOMINANCE}\label{app:dom}
Consider two positive valued random variables $X\geq0$ and $Y\geq 0$. We define their cumulative distributions as: 
\begin{eqnarray}
F_{X}(x) &=& \mathsf{Pr}(X\leq x)\quad, \\ 
F_{Y}(x) &=& \mathsf{Pr}(Y\leq x) \quad.
\end{eqnarray}

We say that $X$ dominates $Y$ stochastically when the cumulative distribution functions of $X$ and $Y$ satisfy the relation $F_{X}(x)\geq F_Y(x)$. If $X$ dominates $Y$ stochastically, then the mean value of $X$ is smaller than the mean value of $Y$: $\langle X\rangle\leq \langle Y\rangle$.  This follows directly from the relation  
 $\langle X\rangle= \int^\infty_0 {\rm d}x \left(1- F_{X}\left(x\right)\right)$ between the mean and the cumulative distribution.

\section{PASSAGE PROBABILITIES FOR  ENTROPY PRODUCTION}\label{app:pass}
We generalize the relations for passage probabilities of entropy production, given by Eqs.~(\ref{eq:pass1xx})-(\ref{eq:pass2xx}), to  passage probabilities of entropy production for trajectories starting from a given macrostate I.    We define a macrostate I as a subset of the phase space $\left\{\bq, \tilde{\bq}\right\}$.  Here, we  consider macrostates defined by a given set of constraints on  the variables of even parity ${\rm I}_{\rm even}\subseteq \mathbb{R}^n$.          The initial state  $\omega(0)$ belongs to the macrostate I if  $\bq(0)\in {\rm I}_{\rm even}$.

We define the passage probability $\mathsf{P}^{(2)}_{+,{\rm I}}$ (and 
$\mathsf{P}^{(2)}_{-,{\rm I}}$) as the joint probability of  the process being initially in the macrostate I, i.e.,
  $\omega(0)\in {\rm I}$, and of entropy production  to pass the threshold 
 $s^+_{\rm tot}$ ($s^-_{\rm tot}$) before passing the threshold $s^-_{\rm tot}$ ($s^+_{\rm tot}$).    
Formally, the passage probabilities are defined by 
\begin{eqnarray}
\mathsf{P}^{(2)}_{+,{\rm I}} &\equiv& \mathbb{P}\left(\Phi_+ \cap \Gamma_{\rm I}  \right)\quad,\\ 
\mathsf{P}^{(2)}_{-,\rm I} &\equiv& \mathbb{P}\left(\Phi_- \cap \Gamma_{\rm I} \right)\quad,
\end{eqnarray}
with $\Phi_+$ and $\Phi_-$ the sets defined in Eqs.~(\ref{eq:Phi+}) and  (\ref{eq:Phi-}) respectively,   
and  $\Gamma_{\rm I}$ the set of trajectories starting from the macrostate I: 
\begin{eqnarray}
\Gamma_{\rm I} = \left\{\omega\in \Omega: \omega(0)\in {\rm I}\right\}\quad.
\end{eqnarray}
We also define the conjugate  probabilities  $\tilde{\mathsf{P}}^{(2)}_{+, {\rm I}}$  and  $\tilde{\mathsf{P}}^{(2)}_{-, {\rm I}}$: 
\begin{eqnarray}
\tilde{\mathsf{P}}^{(2)}_{+, \rm I} &\equiv& \left(\mathbb{P}\circ \Theta\right)\left(\Phi_+ \cap \Gamma_{\rm I}  \right)\quad,\label{eq:E14App}\\ 
\tilde{\mathsf{P}}^{(2)}_{-, \rm I}   &\equiv& \left(\mathbb{P}\circ \Theta\right)\left(\Phi_- \cap \Gamma_{\rm I}   \right)\quad.
\end{eqnarray}
Here, we have used that ${\rm I} = \tilde{\rm I}$, since we have defined the  macrostate using constraints on variables of even parity only.  

For a steady-state process nonequilibrium, i.e., $\langle S_{\rm tot}(t)\rangle>0$, $S_{\rm tot}(t)$  passes in a finite time one of the two boundaries with probability  one. We thus have:
\begin{eqnarray}
\mathsf{P}^{(2)}_{+,{\rm I}}  + \mathsf{P}^{(2)}_{-,{\rm I}}   &=& \mathbb{P}\left(\Gamma_{\rm I}\right)  \quad, \label{eq:E16App}\\ 
\tilde{\mathsf{P}}^{(2)}_{+, \rm I}  +\tilde{\mathsf{P}}^{(2)}_{-, \rm I}&=& \mathbb{P}\left(\Gamma_{\rm I}\right) \label{eq:E17App}\quad.
\end{eqnarray}

In addition, we can derive, using Doob's optional sampling theorem, the following  two identities: 
\begin{eqnarray}
\frac{\mathsf{P}^{(2)}_{+,\rm I}}{\tilde{\mathsf{P}}^{(2)}_{+,\rm I}} &=& e^{s^+_{\rm tot}/k_{\rm B}}\quad,\label{eq:E16App} \\ 
\frac{\mathsf{P}^{(2)}_{-,\rm I}}{\tilde{\mathsf{P}}^{(2)}_{-,\rm I}} &=& e^{-s^-_{\rm tot}/k_{\rm B}}\quad. \label{eq:E17App}
\end{eqnarray}
For instance, Eq.~(\ref{eq:E16App}) follows  immediately from the Eqs.~(\ref{eq:E11})-(\ref{eq:E15}), if we replace $\Phi_+$ by $\Phi_{+,\rm I}$.  Note that in Eq. (\ref{eq:E12}) we can still apply   Doob's optional sampling theorem, since the set  $\Phi_{+, {\rm I}}\in \mathcal{F}(T^{(2)})$, with $T^{(2)}$ the  stopping time when entropy reaches first one of the two thresholds $s^+_{\rm tot}$  or $s^-_{\rm tot}$.   From Eqs.~(\ref{eq:E14App}-\ref{eq:E17App}) we find the following explicit expressions for  the probabilities  $\mathsf{P}^{(2)}_{+, \rm I}$ and $\mathsf{P}^{(2)}_{+, \rm I}$:
\begin{eqnarray}
\mathsf{P}^{(2)}_{+, \rm I} &=&\mathbb{P}\left(\Gamma_{\rm I}\right)\frac{e^{s^-_{\rm tot}/k_{\rm B}}-1}{e^{s^-_{\rm tot}/k_{\rm B}}-e^{-s^+_{\rm tot}/k_{\rm B}}}  \label{eq:pass1xxApp}\quad,\\ 
\mathsf{P}^{(2)}_{-, \rm I} &=&\mathbb{P}\left(\Gamma_{\rm I}\right)\frac{1-e^{-s^+_{\rm tot}/k_{\rm B}}}{e^{s^-_{\rm tot}/k_{\rm B}}-e^{-s^+_{\rm tot}/k_{\rm B}}}\quad.\label{eq:pass2xxApp}
\end{eqnarray}
The conditional entropy production passage probabilities, for trajectories that are initially in the macrostate ${\rm I}$, are:
\begin{eqnarray}
\frac{\mathsf{P}^{(2)}_{+, \rm I}}{\mathbb{P}\left(\Gamma_{\rm I}\right)} &=&\frac{e^{s^-_{\rm tot}/k_{\rm B}}-1}{e^{s^-_{\rm tot}/k_{\rm B}}-e^{-s^+_{\rm tot}/k_{\rm B}}}  \label{eq:pass1xxAppx}\quad,\\ 
\frac{\mathsf{P}^{(2)}_{-, \rm I}}{\mathbb{P}\left(\Gamma_{\rm I}\right)}&=&\frac{1-e^{-s^+_{\rm tot}/k_{\rm B}}}{e^{s^-_{\rm tot}/k_{\rm B}}-e^{-s^+_{\rm tot}/k_{\rm B}}}\quad.\label{eq:pass2xxAppx}
\end{eqnarray}
The entropy-production passage probabilities conditioned on the initial state are thus the same 
as the unconditioned entropy-production passage probabilities, given by Eqs.~(\ref{eq:pass1xx}) and (\ref{eq:pass2xx}).

 \section{STOPPING-TIME FLUCTUATION THEOREMS FOR ENTROPY PRODUCTION}\label{sec:proofApp} 
In this Appendix we derive the entropy stopping-time fluctuation relations.   We consider the definition of entropy production 
 $S_{\rm tot}(\omega;t)$, given by Eq.~(\ref{Eq:entropDef}).    An $s_{\rm tot}$-stopping time $T_+$ is a stopping time for which  $S_{\rm tot}(\omega;T_+(\omega)) = s_{\rm tot}$.
 
We first derive the fluctuation relation for entropy stopping times in the first subsection, and we consequently apply this fluctuation relation to first-passage times in the second subsection.  In the third subsection we consider a fluctuation relation for entropy stopping times of trajectories starting in a macrostate I.  In the last subsection we use this generalized fluctuation relation to derive a fluctuation relation for waiting times of stochastic processes.

\subsection{Fluctuation theorem for entropy stopping times}
The fluctuation theorem for entropy stopping times, given by Eq.~(\ref{eq:stopp}), in Section \ref{sec:6}, follows from the following identities:
\begin{eqnarray}
\lefteqn{\frac{\mathbb{P}\left[\Theta_{T_+}\left(\Phi_{T_+\leq t}\right)\right]}{\mathbb{P}\left[\Phi_{T_+\leq t}\right]}} \label{eq:D5}
 \\ 
 &=& \frac{\displaystyle\int_{\omega\in\Theta_{T_+}\left(\Phi_{T_+\leq t}\right)}\left.{\rm d}\mathbb{P}\right|_{\mathcal{F}(t)}}{\displaystyle\int_{\omega\in\Phi_{T_+\leq t}}\left.{\rm d}\mathbb{P}\right|_{\mathcal{F}(t)}}  \label{eq:D6}
 \\ 
  &=& \frac{\displaystyle\int_{\omega\in\Phi_{T_+\leq t}}\left.{\rm d}\left(\mathbb{P}\circ \Theta_{T_+}\right)\right|_{\mathcal{F}(t)}}{\displaystyle\int_{\omega\in\Phi_{T_+\leq t}}\left.{\rm d}\mathbb{P}\right|_{\mathcal{F}(t)}}  \label{eq:D7}
 \\ 
  &=& \frac{\displaystyle\int_{\omega\in\Phi_{T_+\leq t}}\left.{\rm d}\left(\mathbb{P}\circ \mathsf{T}_{T_+} \circ\Theta \right)\right|_{\mathcal{F}(t)}}{\displaystyle\int_{\omega\in\Phi_{T_+\leq t}}\left.{\rm d}\mathbb{P}\right|_{\mathcal{F}(t)}}  \label{eq:D8}
 \\ 
  &=& \frac{\displaystyle\int_{\omega\in\Phi_{T_+\leq t}}\left.{\rm d}\left(\mathbb{P}\circ \Theta \right)\right|_{\mathcal{F}(t)}}{\displaystyle\int_{\omega\in\Phi_{T_+\leq t}}\left.{\rm d}\mathbb{P}\right|_{\mathcal{F}(t)}}  \label{eq:D9}
 \\ 
  &=& \frac{\displaystyle\int_{\omega\in\Phi_{T_+\leq t}}e^{-S_{\rm tot}(t;\;\omega)/k_{\rm B}}\left.{\rm d}\mathbb{P}\right|_{\mathcal{F}(t)}}{\displaystyle\int_{\omega\in\Phi_{T_+\leq t}}\left.{\rm d}\mathbb{P}\right|_{\mathcal{F}(t)}}  \label{eq:D10}
 \\  
  &=& \frac{\displaystyle\int_{\omega\in\Phi_{T_+\leq t}}e^{-S_{\rm tot}(T_+;\;\omega)/k_{\rm B}}\left.{\rm d}\mathbb{P}\right|_{\mathcal{F}(t)}}{\displaystyle\int_{\omega\in\Phi_{T_+\leq t}}\left.{\rm d}\mathbb{P}\right|_{\mathcal{F}(t)}}  \label{eq:D11}
 \\ 
  &=& e^{-s_{\rm tot}/k_{\rm B}}\frac{\displaystyle\int_{\omega\in\Phi_{T_+\leq t}}\left.{\rm d}\mathbb{P}\right|_{\mathcal{F}(t)}}{\displaystyle\int_{\omega\in\Phi_{T_+\leq t}}\left.{\rm d}\mathbb{P}\right|_{\mathcal{F}(t)}}  \label{eq:D12}
 \\ 
 &=&e^{-s_{\rm tot}/k_{\rm B}}\quad. \label{eq:D13}
\end{eqnarray}
In Eq.~(\ref{eq:D6}) we write the measures of the $\mathcal{F}(t)$-measurable sets $\Theta_{T_+}\left(\Phi_{T_+\leq t}\right)$ and $\Phi_{T_+\leq t}$    in terms of an integral over a probability space.   In Eq.~(\ref{eq:D7}) we  have transformed the variables in the integral  using  the measurable morphism $\Theta_{T_+}$ and  the {\it change of variables formula}.  The change of variables formula relates two integrals under a change of variables, viz.,
\begin{eqnarray}
\int_{\omega\in\Phi}X(\omega) \: {\rm d} \mathbb{P} = \int_{\omega\in\phi(\Phi)}(X\circ\phi^{-1})\left(\omega\right) {\rm d}( \mathbb{P}\circ \phi^{-1})  \label{eq:change} \nonumber \\
\end{eqnarray}  
with $X$ a random variable, measurable in a probability space $(\Omega, \mathcal{F}, \mathbb{P})$;  
$\Phi$ a measurable set in this probability space, i.e., $\Phi\in\mathcal{F}$; and 
 $\phi:\mathcal{F}\rightarrow\mathcal{F}'$ a measurable morphism   from one probability space $(\Omega, \mathcal{F}, \mathbb{P})$ to another probability space  $(\Omega', \mathcal{F}', \mathbb{P}\circ\phi)$, with the property that $\phi^{-1}(\Xi)\in\mathcal{F}$ for each $\Xi\in\mathcal{F}'$  (see excercise 1.4.38 in \cite{tao2011introduction}).    In Eq.~(\ref{eq:D8}) we use the definition of the composition $\Theta_{T_+} = \mathsf{T}_{T_+}\circ\Theta$.    In Eq.~(\ref{eq:D9}) we use that  $\mathbb{P}$ is a stationary measure and thus  $\mathbb{P} = \mathbb{P}\circ\mathsf{T}_{T_+}$. 
 In Eq.~(\ref{eq:D10}) we use the Radon-Nikod\'{y}m theorem, given by Eq.(\ref{eq:ro}), in order to change the  integral over the measure  $\mathbb{P}$ into an integral over the measure $\mathbb{P}\circ\Theta$, and then use our measure-theoretic definition of entropy production, given by Eq.~(\ref{Eq:entropDef}), to write the Radon-Nikod\'{y}m derivative in terms of entropy production $S_{\rm tot}(t)$. 
  Since $e^{-S_{\rm tot}(t)/k_{\rm B}}$ is a uniformly integrable martingale with respect to $\mathbb{P}$, we use in Eq.~(\ref{eq:D11}) Doob's optional sampling theorem Eq.~(\ref{eq:dooboptxx}).    In order to apply this theorem we additionally need  that
  $e^{-S_{\rm tot}(t)/k_{\rm B}}$  is right continuous.
    In Eq.~(\ref{eq:D12}) we use the fact that $T_+(\omega)$ is an $s_{\rm tot}$-stopping time such that $s_{\rm tot} = S_{\rm tot}(\omega;T_+(\omega))$.  In the last Eq.~(\ref{eq:D13}) we use that $\Phi_{T_+\leq t}$   has non-zero measure and $a/a = 1$ for $a\neq 0$.  

The fluctuation theorem Eq.~(\ref{eq:stopp}) can be written as a ratio between the cumulative distribution of the $s_{\rm tot}$-stopping time $T_+$ and a conjugate $(-s_{\rm tot})$-stopping time $T_-$.
We define the $(-s_{\rm tot})$-stopping time $T_-$ associated to the stopping time $T_+$ by
\begin{eqnarray}
T_-\left[ \Theta_{T_+(\omega)}\left(\omega\right)\right]  = T_+(\omega)
\end{eqnarray} 
We thus have 
\begin{eqnarray}
\Phi_{T_-\leq t} = \Theta_{T_+}\left(\Phi_{T_+\leq t}\right)\quad,   \label{eq:1x}
\end{eqnarray} 
and the fluctuation relation, given by Eq.~(\ref{eq:stopp}), reads:
\begin{eqnarray}
\frac{\mathbb{P}\left[\Phi_{T_-\leq t}\right]}{\mathbb{P}\left[\Phi_{T_+\leq t}\right]} = e^{s_{\rm tot}/k_{\rm B}}\quad.  \label{eq:1xx}
\end{eqnarray}
Since the probability-density functions of $T_+$ and $T_-$ are given by
\begin{eqnarray}
p_{T_+}\left(t\right)&=& \frac{{\rm d}}{{\rm d}t}\mathbb{P}\left[\Phi_{T_+\leq t}\right]\quad,\\ 
p_{T_-}\left(t\right)&=& \frac{{\rm d}}{{\rm d}t}\mathbb{P}\left[\Phi_{T_-\leq t}\right]\quad,
\end{eqnarray}
we find the following fluctuation theorem for the distributions of entropy stopping times:
\begin{eqnarray}
\frac{p_{T_+}\left(t\right)}{p_{T_-}\left(t\right)} = e^{s_{\rm tot}/k_{\rm B}}\quad, \label{eq:fptapp}
\end{eqnarray}
which equals to Eq.~(\ref{eq:FPTime}).

\subsection{First-passage time fluctuation theorems for entropy production}
We apply the fluctuation theorem, given by Eq.~(\ref{eq:fptapp}), to first-passage times of entropy production.  We consider here first-passage times with one absorbing boundary $T^{(1)}_{\pm}$ and first-passage times with two absorbing boundaries  $T^{(2)}_{\pm}$ (see subsection \ref{sec:IVA}).   

The first-passage time $T^{(1)}_{\pm}$  denotes the time when the process $S_{\rm tot}(\omega;t)$ passes first the threshold $\pm s_{\rm tot}$.  If  $S_{\rm tot}(\omega;t)$ never passes the threshold $\pm s_{\rm tot}$, then $T^{(1)}_{\pm}=+\infty$.

The first-passage time $T^{(2)}_{+}$ 
denotes the time at which the process $S_{\rm tot}(\omega;t)$ passes first the threshold $s_{\rm tot}$, given that it has not reached $-s_{\rm tot}$ before: 
\begin{eqnarray}
T^{(2)}_{+}  = \left\{\begin{array}{ccc} T^{(1)}_{+},  &&T^{(1)}_{+}  < T^{(1)}_{-}\\ +\infty,   && T^{(1)}_{+}>T^{(1)}_{-}\end{array}\right. \quad.
\end{eqnarray}
Analogously, we define $T^{(2)}_{-}$:
  \begin{eqnarray}
T^{(2)}_{-}  = \left\{\begin{array}{ccc} +\infty,    &&T^{(1)}_{+} < T^{(1)}_{-}\\ T^{(1)}_{-},   && T^{(1)}_{+}>T^{(1)}_{-}\end{array}\right. \quad. 
\end{eqnarray}
Note that other   $s_{\rm tot}$-stopping times can be defined such as the times of second passage.

Since entropy production is a process of odd parity with respect to the time-reversal map $\Theta_t$, see Eq.~(\ref{eq:timer}), we have the relations: 
\begin{eqnarray}
\Theta_{T^{(1)}_+}\left(\Phi_{T^{(1)}_+ \leq t}\right) = \Phi_{T^{(1)}_- \leq t}\quad,
\end{eqnarray}
and 
\begin{eqnarray}
\Theta_{T^{(2)}_+}\left(\Phi_{T^{(2)}_+ \leq t}\right) = \Phi_{T^{(2)}_- \leq t}\quad.
\end{eqnarray}
Using Eqs.~(\ref{eq:1x}) and (\ref{eq:1xx}) we thus find the first-passage-time fluctuation relations: 
\begin{eqnarray}
\frac{p_{T^{(1)}_+}\left(t\right)}{p_{T^{(1)}_-}\left(t\right)} = e^{s_{\rm tot}/k_{\rm B}}\quad, \label{eq:fptappfpt1}
\end{eqnarray}
and 
\begin{eqnarray}
\frac{p_{T^{(2)}_+}\left(t\right)}{p_{T^{(2)}_-}\left(t\right)} = e^{s_{\rm tot}/k_{\rm B}}\quad.\label{eq:fptappfpt2}
\end{eqnarray}
\subsection{Generalized  fluctuation theorem for stopping times of entropy production}\label{app:D3}
We define macrostates as subsets ${\rm I}$  $({\rm II})$ of the phase space of configurations $\left\{\bq, \tilde{\bq}\right\}$.   We also consider the subsets $\tilde{\rm I}$ $\left(\tilde{{\rm II}}\right)$ of the corresponding time-reversed states $\tilde{\rm I} = \left\{ (\bq, \tilde{\bq}): (\bq, -\tilde{\bq}) \in {\rm I}   \right\}$  $(\tilde{{\rm II}} = \left\{ (\bq, \tilde{\bq}): (\bq, -\tilde{\bq}) \in {\rm II}  \right\})$.   We associate to each macrostate ${\rm I}$ a  set  of trajectories $\Gamma_{\rm I}$: 
\begin{eqnarray}
\Gamma_{\rm I}&=& \left\{\omega\in\Omega: \omega(0)\in  {\rm I}\right\} \quad,
\end{eqnarray}  
Note that Doob's optional sampling theorem, given by Eq.~(\ref{eq:dooboptxx}),  applies to the set $\Phi_{T_+\leq t}\cap \Gamma_{\rm I}$, since $\Phi_{T_+\leq t}\cap \Gamma_{\rm I}\in\mathcal{F}\left(T_+\right)$.    
We can  therefore follow the steps in Eqs.~(\ref{eq:D5})-(\ref{eq:D12})  to find the following generalized fluctuation relation
\begin{eqnarray}
\frac{\mathbb{P}\left[\Theta_{T_+}\left(\Phi_{T_+\leq t}\cap\Gamma_{\rm I}\right)\right]}{\mathbb{P}\left[\Phi_{T_+\leq t}\cap \Gamma_{\rm I}\right]}  = e^{-s_{\rm tot}/k_{\rm B}}\quad. \label{eq:fluctIntermediate}
\end{eqnarray} 
\subsection{Fluctuation relations for waiting times}\label{app:D4}
Waiting times  $T^{{\rm I}\rightarrow {\rm {\rm II}}}$ denote  the time a process takes to travel between two macrostates ${\rm I}$ and ${\rm II}$ (the time it takes for the system to change its macrostate from I to II).   Here we derive a fluctuation theorem for waiting times along tajectories that produce a given amount of entropy production.  The  entropy waiting time $T^{{\rm I}\rightarrow {\rm {\rm II}}}_+$ denotes the  time  a process  needs to travel from the macrostate ${\rm I}$ into the macrostate ${\rm II}$ while producing a positive entropy $s_{\rm tot}$, and given that the process has not returned to the macrostate ${\rm I}$ before.  Analogously,  the entropy  waiting time   $T^{{\rm \tilde{II}}\rightarrow {\rm \tilde{I}}}_-$ denotes the time  a process needs to   travel from the macrostate  ${\rm \tilde{II}}$ into the macrostate ${\rm \tilde{I}}$ while producing a negative  entropy $-s_{\rm tot}$, and  given that the process has not returned to the macrostate $\tilde{{\rm II}}$ before.   

 In order to define waiting times, we first define  stopping times $T^{{\rm I}}$ and $T^{{\rm II}}$ that denote the time when a process reaches a given macrostate I or II, respectively:
\begin{eqnarray}
T^{{\rm I}}(\omega) &=& {\rm inf}\Big \{ t>0\cup\left\{+\infty\right\}: \omega(t)\in {\rm I}  \Big \}\,,   \\
T^{{\rm II}}(\omega) &=& {\rm inf}\Big \{ t>0\cup\left\{+\infty\right\}: \omega(t)\in {\rm II}  \Big \}\,.
\end{eqnarray} 
The waiting time $T^{{\rm I}\rightarrow {\rm {\rm II}}}$  denotes the time a process takes to travel from I to II 
\begin{eqnarray}
T^{{\rm I}\rightarrow {\rm {\rm II}}}(\omega) = \left\{\begin{array}{ccc} T^{{\rm II}}(\omega)  && T^{{\rm II}}(\omega) <   T^{{\rm I}}(\omega)\quad, \\ +\infty  && T^{{\rm II}}(\omega) >  T^{{\rm I}}(\omega) \quad, \end{array}\right.
\end{eqnarray} 
for all trajectories $\omega$ for which $\omega(0)\in {\rm I}$, 
and the associated $s_{\rm tot}$-waiting time $T^{ {\rm I} \rightarrow { {\rm II}}}_+$ is defined as
\begin{eqnarray}
T^{ {\rm I} \rightarrow { {\rm II}}}_+(\omega) &=&  \left\{\begin{array}{ccc} T^{{\rm I}\rightarrow {\rm II}}(\omega)  && S_{\rm tot}(T^{{\rm I}\rightarrow {\rm II}}(\omega);\omega) = s_{\rm tot}  \quad, \\ +\infty &&  S_{\rm tot}(T^{{\rm I}\rightarrow {\rm II}}(\omega);\omega)\neq s_{\rm tot}\quad,\end{array}\right. \nonumber \\
 \end{eqnarray}
 for all trajectories $\omega$ for which $\omega(0)\in {\rm I}$.
We also define a  $(-s_{\rm tot})$-stopping time $T^{ {\rm \tilde{II}}\rightarrow {\rm \tilde{I}}}_-$, denoting the time when a  trajectory  reaches   the macro state ${\rm \tilde{I}}$, and produces a negative total entropy $-s_{\rm tot}$:
\begin{eqnarray}
T^{ {\rm \tilde{II}} \rightarrow { {\rm \tilde{I}}}}_-(\omega) &=&  \left\{\begin{array}{ccc}T^{ {\rm \tilde{II}} \rightarrow { {\rm \tilde{I}}}}(\omega)   && S_{\rm tot}(T^{ {\rm \tilde{II}} \rightarrow { {\rm \tilde{I}}}}(\omega);\omega) = -s_{\rm tot}  \quad, \\ +\infty &&  S_{\rm tot}(T^{ {\rm \tilde{II}} \rightarrow { {\rm \tilde{I}}}}(\omega);\omega)\neq -s_{\rm tot}\quad.\end{array}\right. \nonumber \\
 \end{eqnarray}

 We apply the generalized fluctuation theorem, given by  Eq.~(\ref{eq:fluctIntermediate}), on the stopping time  $T^{{\rm I}\rightarrow {\rm {\rm II}}}_+$ and the set $\Gamma_{\rm I}$, and find
\begin{eqnarray}
\frac{\mathbb{P}\left[\Phi_{T^{{ {\rm I}}\rightarrow { {\rm II}}}_+\leq t}\cap\Gamma_{\rm I}\right] }{  \mathbb{P}\left[\Theta_{T^{{ {\rm I}}\rightarrow { {\rm II}}}_+}\left(\Phi_{T^{{ {\rm I}}\rightarrow { {\rm II}}}_+\leq t}\cap \Gamma_{\rm I}\right)\right]}  = e^{s_{\rm tot}/k_{\rm B}}\quad. \label{eq:fluctIntermediatexxx}
\end{eqnarray}
  Note that the waiting time $T^{ {\rm I} \rightarrow { {\rm II}}}_+(\omega)$ is defined on all trajectories $\omega\in \Omega$, but we are interested in those trajectories for which $\omega(0)\in {\rm I}$, and thus set $\omega\in\Gamma_{\rm I}$.  
Since
\begin{eqnarray}
\Theta_{T^{{ {\rm I}}\rightarrow { {\rm II}}}_+}\left(\Phi_{T^{{ {\rm I}}\rightarrow {{\rm II}}}_+\leq t}\cap \Gamma_{\rm I} \right)= \Phi_{T^{{ {\rm \tilde{II}}}\rightarrow {\tilde{\rm I}}}_-\leq t}\cap \Gamma_{\tilde{{\rm II}}}\quad,\nonumber \\
\end{eqnarray}
we have \begin{eqnarray}
\frac{\mathbb{P}\left[\Phi_{T^{{\rm {\rm I}}\rightarrow {\rm {\rm II}}}_+\leq t}\cap \Gamma_{\rm I}\right]}{\mathbb{P}\left[\Phi_{T^{{\rm {\rm \tilde{II}}}\rightarrow { \tilde{\rm I}}}_-\leq t}\cap \Gamma_{\tilde{{\rm II}}}\right]        }  = e^{s_{\rm tot}/k_{\rm B}}\quad. \label{eq:fluctIntermediatexxxx}
\end{eqnarray}
We can write the  probability densities of the entropy waiting times $T^{ {\rm I}\rightarrow {\rm {\rm II}}}_+$ and $T^{\tilde{{\rm {\rm II}}}\rightarrow \tilde{{\rm I}}}_-$ in term of the measures on the right-hand side of the previous equations:
 \begin{eqnarray}
p_{T^{ {\rm I}\rightarrow {\rm {\rm II}}}_+}\left(t\right)&=& \frac{1}{\mathbb{P}\left[\Gamma_{\rm I}\right]}\frac{{\rm d}}{{\rm d}t}\mathbb{P}\left[\Phi_{T^{{\rm I} \rightarrow { {\rm {\rm II}}}}_+\leq t}\cap\Gamma_{\rm I}\right]\quad, \\
p_{T^{\tilde{{\rm II}}\rightarrow  \tilde{\rm I}}_-}\left(t\right)&=&  \frac{1}{\mathbb{P}\left[\Gamma_{\tilde{{\rm II}}}\right]}\frac{{\rm d}}{{\rm d}t}\mathbb{P}\left[\Phi_{T^{\tilde{{\rm II}} \rightarrow { \tilde{\rm I}}}_-\leq t}\cap\Gamma_{\tilde{{\rm II}}}\right]\,.  \nonumber \\
\end{eqnarray} 
We thus find the following fluctuation theorem for waiting times between macrostates:
\begin{eqnarray}
\frac{p_{T^{{ \rm I}\rightarrow { {\rm II}}}_+}\left(t\right)}{p_{T^{{ \tilde{{\rm II}}}\rightarrow {\tilde{\rm I}}}_-}\left(t\right)} = e^{s_{\rm tot}/k_{\rm B} + \log \frac{\mathbb{P}\left[\Gamma_{\tilde{{\rm II}}}\right]}{\mathbb{P}\left[\Gamma_{\rm I}\right]}}\quad. \label{eq:fptt}
\end{eqnarray}
  If we set $\Gamma_{\rm I} = \Gamma_{{\rm II}} = \Omega$, Eq.~(\ref{eq:fptt}) is equal to the entropy stopping-time fluctuation theorem, given by Eq.~(\ref{eq:fptapp}).

The quantity in the exponential of   the right hand side of (\ref{eq:fptt}) has in general  no particular meaning.  For macrostates defined by variables of even parity, i.e., $\tilde{\rm I} ={\rm I}$ and   $\tilde{{\rm II}} = {\rm II}$, 
we have  the relation $\Gamma_{\tilde{{\rm II}}} = \Gamma_{{\rm II}}$.    We recognize then in  Eq.~(\ref{eq:fptt}) the system entropy change: 
\begin{eqnarray}
\Delta s_{\rm sys} = -k_{\rm B}\: \log \frac{\mathbb{P}\left[\Gamma_{ {\rm II}}\right]}{\mathbb{P}\left[\Gamma_{\rm I}\right]} \quad,
\end{eqnarray}
between the two macrostates ${\rm I}$ and ${\rm II}$.  

A particular important example of the waiting-time fluctuation theorem, given by Eq.~(\ref{eq:fptt}), is for macrostates ${\rm I}$ and ${\rm II}$ that correspond to  one single point in phase space, i.e.,  ${\rm I} = \left\{\bq_{\rm I}\right\}$ and ${\rm II} = \left\{\bq_{{\rm II}}\right\}$.  We find then the relation:
\begin{eqnarray}
\frac{p_{T^{{\rm  I}\rightarrow { {\rm II}}}_+}\left(t\right)}{p_{T^{{ {\rm II}}\rightarrow {\rm  I}}_-}\left(t\right)} = e^{s_{\rm env}/k_{\rm B}}\quad,
\end{eqnarray}
with $s_{\rm env} = s_{\rm tot}-\Delta s_{\rm sys}$ the change in the environment entropy, and $\Delta s_{\rm sys}$ the system-entropy change, which here is given by
\begin{eqnarray}
\Delta s_{\rm sys} = -k_{\rm B}\:\log \frac{p_{\rm ss}(\bq_{ {\rm II}})}{p_{\rm ss}(\bq_{\rm I})}\quad.
\end{eqnarray}

\section{INFIMUM STATISTICS  FOR A DRIFT-DIFFUSION PROCESS}
\label{app:BPI}
In this appendix we derive explicit expressions for the distribution and the mean of the infimum of entropy production in the drift-diffusion process $X(t)$.    Since in a drift-diffusion process entropy production is $S_{\rm tot}(t)/k_{\rm B}= (v/D) \left(X(t)-X(0)\right)$, the infimum statistics of entropy production follow from the infimum statistics of $X(t)$.

\subsection{Cumulative distribution of the infimum of entropy production}
The cumulative distribution $\mathsf{Pr}\left(\sup X(t) \leq L\right)$ of the supremum  $\sup X(t)$,  of  a stochastic process over a time interval $[0,t]$, equals  the survival probability $Q_X(L,t)$ of the process  in the interval $(-\infty,L)$ at time $t$~\cite{marzuoli2015extreme,majumdar2014extreme}:
\begin{eqnarray}
\mathsf{Pr}\left(\sup X(t) \leq L\right) = \mathsf{Pr}(X(s) \leq L ; s\leq t)=Q_X(L,t),\nonumber\\
\label{eq:CDFQ}
\end{eqnarray}
with $L>0$.
For a general stochastic process, the  survival probability in an interval can be calculated from the  
distribution  $p_T(\tau;L)$ of first-passage times to reach an absorbing boundary located at $L$:\begin{equation}
Q_X(L,t) =1- \int_0^t \text{d}\tau \,p_T(\tau;L) \quad.
\label{eq:QXdef}
\end{equation}
We use this relation between  to determine the cumulative distribution of extreme values of $X(t)$.

The infimum of a drift-diffusion process with positive drift is equal to the supremum of a drift-diffusion process with the same but negative drift.  
We therefore  consider the following two conjugate drift-diffusion processes: 
\begin{enumerate}
\item  $X_{+}(t)$  with velocity $v$, diffusion $D$, and initial condition $X_{+}(0)=0$, 
\item  $X_{-}(t)$  with velocity $-v$ and diffusion $D$, and initial condition $X_{-}(0)=0$,
\end{enumerate}
and $v>0$ in both processes.  The dynamics of both processes follows from Eq.~(\ref{eq:LESF}) for $V(x)=0$ and for, respectively, a positive and negative velocity.
   The infimum value of $X_{+}(t)$ equals to minus the supremum of the conjugate process $X_{-}(t)$. In the following we derive analytical expressions for the statistics of the supremum of $X_{-}(t)$, and use these   to obtain the statistics of the infimum of $X_{+}(t)$.

The survival  probability of $X_{-}(t)$  can be obtained from the distribution $p_T$ of  first-passage times to first reach the threshold $L$, see Eq.~(\ref{eq:QXdef}).   The first-passage time distribution  is given by
\begin{equation}
p_T(t;L)=\frac{L}{\sqrt{4\pi Dt^3}}e^{-(L+vt)^2/4Dt}\quad.
\label{eq:Waldneg}
\end{equation}
Substituting Eq.~\eqref{eq:Waldneg} in~\eqref{eq:QXdef} results in the following expression for the survival probability of $X_-$
\begin{equation}
Q_{X_{-}}(L,t) =1- \frac{1}{2}\left[ \text{erfc}\left(\frac{L+vt}{\sqrt{4Dt}}  \right) + e^{-vL/D}\text{erfc}\left(\frac{L-vt}{\sqrt{4Dt}}  \right)   \right]\,
\label{eq:Q-}
\end{equation}
where erfc is the complementary error function. Equation~\eqref{eq:Q-}  yields  the cumulative density function of the supremum of $X_{-}$, as follows from Eq.~\eqref{eq:CDFQ}.

 From the relation between the conjugate processes, we relate the cumulative distribution of  the infimum of $X_+$ over the interval $[0,t]$, ${\rm inf}\, X_+(t)$,  to the survival probability of $X_-(t)$:
\begin{eqnarray}
\lefteqn{\mathsf{Pr}\left(-\inf X_{+}(t) \leq L\right) }&&\nonumber \\
&=&\mathsf{Pr}\left(\inf X_{+}(t) \geq -L\right)  \nonumber\\
&=&  \mathsf{Pr}\left(\sup X_{-}(t) \leq L\right)\nonumber\\
&=& Q_{X_{-}}(L,t)\quad. \label{eq:chainSP}
\end{eqnarray}
Using Eq.~\eqref{eq:Q-} and the property $\text{erfc}(x)+\text{erfc}(-x)=2$, we obtain an analytical expression for the  cumulative distribution of the infimum of the position of a drift-diffusion process with positive velocity:
\begin{eqnarray}
\lefteqn{\mathsf{Pr}\left(-\inf X_{+}(t) \leq L\right)}&& \label{eq:CDFINFX}\\
&=& \frac{1}{2}\left[ \text{erfc}\left(\frac{-L-vt}{\sqrt{4Dt}}  \right) - e^{-vL/D}\text{erfc}\left(\frac{L-vt}{\sqrt{4Dt}}  \right)   \right].\nonumber
\end{eqnarray}

Finally, for the stochastic process $ S_{\rm tot}(t)/k_{\rm B}= (v/D) X_{+}(t)$ the infimum distribution can be obtained by replacing in~\eqref{eq:CDFINFX} $v$ and $D$ by its effective values for the process $S_{\rm tot}(t)/k_{\rm B}$, given by $v_{\rm eff}=v^2/D$ and $D_{\rm eff}=v^2/D$.   Defining  $ \sigma (t)  = \langle S_{\rm tot} (t)\rangle / k_{\rm B} = (v^2/D) t$ we obtain:
\begin{eqnarray}
\lefteqn{\mathsf{Pr}\left(- \frac{S_{\rm inf}(t)}{k} \leq s\right) }&&\\
&=& \frac{1}{2}\left[ \text{erfc}\left(\frac{-s-\sigma (t)}{2\sqrt{\sigma (t)}}  \right) - e^{-s}\text{erfc}\left(\frac{s-\sigma (t)}{2\sqrt{\sigma (t)}}  \right)   \right]\;,\nonumber
\end{eqnarray}
 which equals to Eq.~\eqref{eq:infCDFS}.

\subsection{Mean infimum of entropy production}
We first determine the distribution of the  the mean infimum of $X_{+}$, and then compute its mean value.   
Note that the infimum of  $X_{+}$  equals  minus the supremum of the conjugate process $X_{-}$:
\begin{equation}
\left\langle \inf X_{+}(t) \right\rangle = - \left\langle \sup X_{-}(t) \right\rangle\;.
\label{eq:X+X-}
\end{equation}
The cumulative distribution of the supremum of  $X_{-}$ is given by
\begin{eqnarray}
\lefteqn{\mathsf{Pr}\left(\sup X_{-}(t) \leq L\right)} &&\nonumber\\
& =& Q_{X_{-}}(L,t) \nonumber \\
&=& \frac{1}{2}\left[ \text{erfc}\left(\frac{-L-vt}{\sqrt{4Dt}}  \right) - e^{-vL/D}\text{erfc}\left(\frac{L-vt}{\sqrt{4Dt}}  \right)   \right]\, ,\nonumber \\ 
 \label{eq:CDFX-}
\end{eqnarray}
where we have used Eq.~\eqref{eq:Q-} and the property $\text{erfc}(x)+\text{erfc}(-x)=2$. The distribution of the supremum of $X_{-}$ can be found deriving Eq.~\eqref{eq:CDFX-}  with respect to $L$, which yields:
\begin{eqnarray}
\lefteqn{\mathsf{Pr}\left(\sup  X_{-}(t) = L\right) } && \nonumber \\
&=& \frac{1}{\sqrt{\pi Dt}}e^{-v^2 t/4D}+\frac{v}{2D}e^{-vL/D} \text{erfc}\left[ \frac{v\sqrt{t}}{2\sqrt{D}}   \right]\,.\nonumber \\ 
 \label{eq:PDFX-}
\end{eqnarray}
The mean of the supremum of $X_{-}$ can be calculated integrating its probability distribution  
\begin{equation}
 \left\langle \sup X_{-}(t) \right\rangle=\int_0^{\infty} \text{d}L\, \mathsf{Pr}\left(\sup X_{-}(t) = L\right) L,
\end{equation}
which after some algebra yields
\begin{eqnarray}
\lefteqn{ \left\langle \sup X_{-}(t) \right\rangle} && \label{eq:meansup-DD}\\
&=&\frac{D}{v}\text{erf}\left[\frac{v\sqrt{t}}{2\sqrt{D}}\right] - \frac{vt}{2} \text{erfc}\left[ \frac{v\sqrt{t}}{2\sqrt{D}}   \right] +\sqrt{\frac{Dt}{\pi}}e^{-v^2 t/4D} \;. \nonumber
\end{eqnarray}
From Eqs.~\eqref{eq:X+X-} and~\eqref{eq:meansup-DD} we find an exact expression for the mean infimum of a drift-diffusion process with positive velocity:
\begin{eqnarray}
\lefteqn{\left\langle \inf X_{+}(t) \right\rangle} && \\
&=&-\frac{D}{v}\text{erf}\left[\frac{v\sqrt{t}}{2\sqrt{D}}\right] + \frac{vt}{2} \text{erfc}\left[ \frac{v\sqrt{t}}{2\sqrt{D}}   \right] -\sqrt{\frac{Dt}{\pi}}e^{-v^2 t/4D} \;. \nonumber
\label{eq:meaninfDD}
\end{eqnarray}
Using $ \sigma (t)  = \langle S_{\rm tot} (t)\rangle / k_{\rm B} = (v^2/D) t$ and $S_{\rm tot} (t) = (v/D)X_+(t)$, we obtain, from the epxression (\ref{eq:meaninfDD}), an analytical expression for the mean infimum of entropy production in a drift-diffusion process at time $t$:
\begin{eqnarray}
\lefteqn{\left\langle  \frac{S_{\rm inf}(t)}{k_{\rm B}}\right\rangle }&& \\
&=&- \text{erf}\left[\frac{\sqrt{\sigma(t)}}{2}\right] +\frac{\sigma(t)}{2}\text{erfc}\left[\frac{\sqrt{\sigma(t)}}{2}\right]- \sqrt{\frac{\sigma(t)}{\pi}}e^{-\sigma(t)/4} .\nonumber
\end{eqnarray}
The above result equals to the Eq.~\eqref{eq:INFSDD} in the main text.

\section{PASSAGE PROBABILITIES AND FIRST-PASSAGE TIMES  FOR A DRIFT-DIFFUSION PROCESS}\label{app:F}
We determine the passage statistics of entropy production  of a drifted Brownian particle  with diffusion coefficient $D$ and drift velocity $v$, as given by the Langevin  Eq.~(\ref{eq:LESF}) (with $V(x)=0$).    
Since in a drift-diffusion process entropy production is $S_{\rm tot}(t)/k_{\rm B}= (v/D) \left(X(t)-X(0)\right)$, the infimum statistics of entropy production follow from the infimum statistics of $X(t)$.
 \subsection{First-passage-time statistics for one boundary, or two symmetric boundaries}
The first-passage-time distribution for  $X$  to pass at time $T =t$ for the first time the threshold $L>0$, starting from the initial condition $X(0)=0$,  is given by Wald's distribution~\cite{redner2001guide,sato1994inverse}
\begin{equation}
p_{T}(t;L)=\frac{|L|}{\sqrt{4\pi Dt^3}}e^{-(L-vt)^2/4Dt}\quad.  
\label{eq:pft}
\end{equation}
Equation~\eqref{eq:pft}  implies:
\begin{equation}
\frac{p_{T}(t;L)}{p_{T}(t;-L)} = e^{vL/D}\quad.
\label{eq:1BF}
\end{equation}
Note that the argument of the exponential equals to the  P\'eclet number, $\text{Pe}=vL/D$.

The distribution of  entropy first-passage times $T^{(1)}_{+}$ with one absorbing boundary located at $s_{\rm tot}$ is equal to  the first-passage-time distribution for the position of the particle,  given by Eq.~(\ref{eq:pft}), with one absorbing boundary located at  $L = s_{\rm tot}D/(vk_{\rm B})$. We thus replace in Eq.~\eqref{eq:1BF} $L$ by $s_{\rm tot}D/(vk_{\rm B})$,  and find the first-passage-time distribution of entropy production given by Eq.~(\ref{eq:FPTime}).
An analogous relation to Eq.~(\ref{eq:1BF}) holds for  two-boundary first-passage times  $T^{(2)}_{+}$ for entropy production  in the drift-diffusion process, and can be derived using the results in Sec.~2.2.2.2 in~\cite{redner2001guide} (see also~\cite{roldan2015decision}). 
\subsection{First-passage-time statistics for two asymmetric boundaries}
We consider the drift-diffusion process with two absorbing boundaries located at $L_+\geq 0$ and $-L_- \leq  0$, and with the initial position $X(0)=0$.  The passage probabilities $\mathsf{P}^{(2)}_+$ and $\mathsf{P}^{(2)}_-$, to first reach $L_+$ and $-L_-$, respectively, are \cite{redner2001guide}
\begin{eqnarray}
\mathsf{P}^{(2)}_+ &=& \frac{e^{vL_-/D}-1}{e^{vL_-/D}-e^{-vL_+/D}} \label{eq:p+x}\quad,\\ 
\mathsf{P}^{(2)}_- &=& \frac{1-e^{-vL_+/D}}{e^{vL_-/D}-e^{-vL_+/D}}\label{eq:p-x}\quad.
\end{eqnarray}
The corresponding entropy-production passage probabilities follow from  the expressions (\ref{eq:p+x}) and (\ref{eq:p-x}) using the threshold values $s^-_{\rm tot}/k_{\rm B} = vL_-/D$ and $s^+_{\rm tot}/k_{\rm B} = vL_+/D$.  Equations  (\ref{eq:p+x}) and (\ref{eq:p-x})  are in correspondence with the Eqs.~(\ref{eq:pass1xxx}) and (\ref{eq:pass2xxx}) for passage probabilities of entropy production.

Notably the first-passage-time fluctuation theorem, given by Eq.~(\ref{eq:FPTime}), does not simply extend to asymmetric boundaries.  From the expression (\ref{eq:pft}) for the first-passage-time distribution of the position with one absorbing boundary, we find
\begin{eqnarray}
\frac{p_{T}(t;L_+)}{p_{T}(t;-L_-)
} = \frac{L_+}{L_-}\:e^{v(L_+-L_-)/(2D)}\:e^{-\left(L^2_+-L^2_-\right)/(4Dt)}\quad.
\end{eqnarray}
Using $s^-_{\rm tot}/k_{\rm B} = vL_-/D$ and $s^+_{\rm tot}/k_{\rm B} = vL_+/D$, we find for the ratio of the first-passage-times of entropy production
\begin{eqnarray}
\frac{p_{T^{(1)}_+}(t;s^+_{\rm tot})}{p_{T^{(1)}_-}(t;-s^-_{\rm tot})
} =  \frac{s^+_{\rm tot}}{s^-_{\rm tot}}\:e^{\left(s^+_{\rm tot}+s^-_{\rm tot}\right)\left(1-\frac{s^+_{\rm tot}-s^-_{\rm tot}}{2k_{\rm B}\sigma(t)}\right)/(2k_{\rm B})}\quad.
\end{eqnarray}
For asymmetric boundaries, the ratio of the first-passage-time distributions is time dependent, and converges to a finite value in the limit $t\rightarrow \infty$~\cite{saito2015waiting}.  
Consequently, for asymmetric boundaries the mean first-passage time for entropy production to reach the positive threshold $s^+_{\rm tot}$  is  different than the mean first-passage time for entropy production to reach the negative threshold $s^-_{\rm tot}$,  $\langle T^{(1)}_+ \rangle\neq  \langle T^{(1)}_-\rangle$. When $s^+_{\rm tot} = s^-_{\rm{ tot}}=s_{\rm tot}$ we recover the time independent ratio, $e^{s_{\rm tot}/k_{\rm B}}$, as given by the first-passage-time fluctuation theorem.  

\section*{References}
\bibliographystyle{ieeetr} 
\bibliography{bibliography}

\begin{thebibliography}{100}
\bibitem{parity}
Variables that are even under time reversal do not change sign under time
  reversal (e.g.~position variables) whereas variables that are odd under time
  reversal change sign under time reversal (e.g.~momenta or current).
\bibitem{footnotePP}
Note that here we use passage probabilities of entropy production with a given
  initial state as discussed in Appendix D.
\bibitem{footnoteBoundary}
The bound holds despite the presence of an absorbing boundary, since the effect
  of this boundary is to reduce the maximum depth of the backtrack.
\end{thebibliography}

\end{document}